\documentclass[12pt]{article}
\usepackage{fullpage}
\usepackage{cite}
\usepackage{amstext,amsfonts,amsbsy,eucal,amssymb}
\usepackage{epsfig}
\usepackage{psfrag}

\newcounter{definition}

\def\thefigure{\arabic{section}.\arabic{figure}}
\def\theequation{\thesection.\arabic{equation}}
\def\appendix{
  \setcounter{section}{0}
  \setcounter{subsection}{0}
  \par
  \def\thesection{Appendix \Alph{section}}
  \def\theequation{\Alph{section}.\arabic{equation}}
  \def\thefigure{\Alph{section}.\arabic{figure}}
}
\makeatletter
\def\fnum@figure{Fig. \thefigure}
\@addtoreset{equation}{section}
\@addtoreset{figure}{section}
\makeatother

\def\thesubsection{Step \arabic{subsection}}
  
\date{
Novenber 18,2001
}
\title{
A direct calculation of the free energy\\
 from the Bethe ansatz equation\\
 for the Heisenberg model.
} 
\author{
  Go {\sc Kato}\thanks{kato@monet.phys.s.u-tokyo.ac.jp}\, and  
  Miki {\sc Wadati}\thanks{wadati@phys.s.u-tokyo.ac.jp} \\
  Department of Physics, Graduate School of Science,\\
  University of Tokyo,\\
  Hongo 7-3-1, Bunkyo-ku, Tokyo 113-0033, Japan.
}

\begin{document}
\maketitle \setlength{\baselineskip}{1.8em}
\begin{abstract}
  Thermodynamics of the XXX Heisenberg model is studied. The trace of
  the Boltzmann weight with respect to the Hilbert space is taken in the
  thermodynamic limit with the number of up-spins being fixed.  The
  expression of the trace gives an explanation why the correct thermodynamic
  quantities are derived from the string hypothesis.
  Combining this with the previous result, we conclude that the free
  energy can be calculated only by assuming the Bethe ansatz equation.
  The method is more direct than other known methods which were used
  to derive the free energy.
\end{abstract}

\newpage
\section{Introduction}
There are two well-known methods to calculate the free energy for
quantum integrable systems. One is the Thermodynamic Bethe Ansatz
(TBA) method~\cite{yang}, and the other is the Quantum Transfer Matrix
(QTM)
method~\cite{koma,inoue_suzuki,wadati_akutsu,suzuki_akutsu_wadati,suzuki_nagao_wadati}.
Both methods, however, are still unsatisfactory. In TBA case, we assume the form of the entropy. And we use
the string hypothesis~\cite{takahashi_a}
for some models,
 whose validity is not yet proved.  While QTM is a general formulation, it is difficult to analyze the resultant equations. We solve
them asymptotically or numerically.
 In this paper, we present a direct method whereby the free energy
is derived without logical jumps, assumptions and numerical supports.

We treat the spin-half XXX Heisenberg model. The Hamiltonian of the
system is
\begin{eqnarray}
  H
  &=&
  -J\sum_{j=1}^{L}\left(
    S^x_{j}S^x_{j+1}+S^y_{j}S^y_{j+1}+S^z_{j}S^z_{j+1}
  \right)
  -h\sum_{j=1}^L S^z_{j}
  +{\rm constant},
\label{eq:Hamiltonian}
\end{eqnarray}
where $L$ is the number of sites, $J$ is the coupling constant and $h$
expresses the external field. 
  This model is interesting since it has
bound states, and correspondingly the Bethe equations have complex
solutions.  Throughout the paper, we assume
periodic boundary condition, and use a unit which makes $J=1$.

We briefly summarize our previous work~\cite{go_xxx} on this system.
We suppose the expression of $Z_M$, the trace of the Boltzmann weight
with respect to the Hilbert space in which the number of up-spins is
fixed to be $M$. An infinite sum $\sum_MZ_M$ defined by the expression is
analyzed by use of combinatorial relations.  Then, it is proved that
the free energy $-\beta\log\sum_MZ_M$ can be expressed in term of the
function which satisfies an integral equation.  The result perfectly
agrees with the free energy derived from a different
method~\cite{takahashi_c}.

Our purpose is to derive directly the expression;
\begin{eqnarray}
  Z_{M}
&\equiv&
  {\rm Tr}e^{-\beta H_M}\makebox[2cm]{}  \beta=1/k_BT
\label{eq:Z_LM_Tr_formulation(formal)}\\
&=&
\frac{e^{hM}}{M!}
  \sum_{\theta\in\Theta_M}
  \left[
    \prod_{\sigma\in\theta}
    N_{\sigma}!
  \right]
  \sum_{\zeta\in\bar\Theta\left(\theta\right)}
  \mu\left(\hat0_\theta,\zeta\right)
  \left[
    \prod_{\theta'\in\zeta}
    \int_{-\infty}^{ \infty}
    dx_{\theta'}
  \right]
  \left|\frac{\partial I}{\partial x}\right|_{L,\zeta}
    e^{-\beta E\left(\zeta\right)}
\label{eq:result_of_this_paper}
\end{eqnarray}
in the thermodynamic limit $L\rightarrow \infty$. All the notation in
(\ref{eq:result_of_this_paper}) are explained in the following
sections. Here, $H_M$ denotes the Hamiltonian (\ref{eq:Hamiltonian})
acting on the Hilbert space in which the number of up-spins is
restricted to a constant $M$.  In this paper, we use only the Bethe
Ansatz; the eigen energy $E$ of $H_M$ is given by
\begin{eqnarray}
  E+hM=\sum_{m=1}^{M}\frac2{x_m^2+1},
\label{eq:eigen_energy}
\end{eqnarray}
where $\{x_m\}$ satisfies the Bethe equations,
\begin{eqnarray}
  \left[\frac{x_m+i}{x_m-i}\right]^{L}
&=&
  \prod_{m'\neq m}
  \frac{x_m-x_{m'}+2i}{x_m-x_{m'}-2i},
\end{eqnarray}
which are due to periodic boundary condition.  Note that
(\ref{eq:result_of_this_paper}) is the expression of $Z_M$ supposed in
the previous paper~\cite{go_xxx}. We thus complete a new method to derive
the free energy of this system independent of TBA and QTM.

The derivation goes as follows.  First, we express the trace in a
series with respect to solutions of the Bethe equations, which
constitute a complete set of the system.  Next, the series are
replaced by integrals over pseudo-momenta.  This replacement is
justified in the same way as that in the calculation of the free
energy for a $\delta$-function bose system \cite{go_b,go_d,go_e}.  The
replacement, however, has a difficulty, because the Bethe equations
have complex solutions.  The difficulty is resolved by taking a
``good'' integral path. Finally, using combinatorial relations, $Z_M$ (\ref{eq:result_of_this_paper}) is obtained.

We also show that the expression of $Z_M$ gives a reason why the
string hypothesis, that is, the substitution of the Bethe equations
for string center equations, is appropriate.  The substitution is a critical 
procedure when we apply the thermodynamic Bethe Ansatz to an integrable system
which has bound states.

The outline of this paper is the following. In \S2, we derive the
expression (\ref{eq:result_of_this_paper}) of $Z_M$ only from the
Bethe ansatz.  In \S3, we examine the string hypothesis by use of
thus proved expression of $Z_M$.  The last section is devoted to the
concluding remarks. Technical details of calculations and proofs are summarized
in Appendices A-H.

\section{Direct derivation of $Z_{M}$ from Bethe equations}
In order to explain our analysis of
(\ref{eq:Z_LM_Tr_formulation(formal)}), we begin with several
symbols and notations. We divide the procedure into 9 steps;
definitions of symbols are placed on the top of each step where
they first appear.

\subsection{}
With the  eigen energies $E$ (\ref{eq:eigen_energy}), $Z_M$ becomes
\begin{eqnarray}
  M!  Z_{M}e^{-\beta hM}
&=&
  \sum
  \exp\left(-\beta{\sum_{m=1}^{M}\frac2{x_m^2+1}}\right).
\label{eq:Z_LM_summation_formulation(formal)}
\end{eqnarray}
Here, $\sum$ without subscript means a summation over all the
``different'' ``physical solutions'' of the Bethe equations. It
is interpreted as a summation with respect to a set of integers
$\{I_m\}$ corresponding to a ``physical solution'', where $\{I_m\}$ is
related to $\{x_m\}$ by
\begin{eqnarray}
    \left[\frac{x_m+i}{x_m-i}\right]^{L}
&=&
  e^{-2\pi i I_m}
  \prod_{m'\in\theta,\neq m}
  \left[\frac{x_m-x_{m'}+2i}{x_m-x_{m'}-2i}\right]^{N_{m'}}.
\label{eq:modified_Bethe}
\end{eqnarray}
The ``physical solution'' means a solution of the Bethe equations
which contains some pseudo-momenta $x_m$ of the ``same value''.
The symbol $\theta$ and the words ``different'', ``same value'' will be explained later.

\subsection{}
\begin{itemize}
\item definition: $\Theta(\sigma)$

  Let $\sigma$ be a set which has a finite number of elements.
  $\Theta\left(\sigma\right)$ denotes all the patterns of division of
  a set $\sigma$.  A pattern of division $\theta$ is a set with
  elements each of which is a cluster.  The cluster $\sigma'$ is one
  of the pieces into which a set $\sigma$ is divided, and the cluster
  is also regarded as a set;
  \begin{eqnarray}
    \Theta(\sigma)&=&\left\{\;\theta\;\left|\;
      \sigma=\bigoplus_{\sigma'\in\theta}\sigma'\;
    \right.\right\}.
  \end{eqnarray}
  In case of $\sigma=\{1,2,\cdots,n\}$, we write
  $\Theta_n$ for $\Theta(\{1,2,\cdots,n\})$ for simplicity.

\item definition: $N_\sigma$

  Let $\sigma$ be a set or a sequence which has a finite number of
  elements.  The number of elements of a set or a sequence $\sigma$ is
  denoted by $N_\sigma$.

\end{itemize}

 Using the above symbols,
(\ref{eq:Z_LM_summation_formulation(formal)}) can be rewritten in the thermodynamic limit as
\begin{eqnarray}
  M!  Z_{M}e^{-\beta hM}
&=&
  \sum_{\theta\in\Theta_M}
  \mu\left(\hat 0_M,\theta\right)
  \int
  e^{-\beta E_\theta}
  \prod_{\sigma\in \theta}dI_\sigma,
\label{eq:Z_LM_integral_formulation(formal)}
\end{eqnarray}
where $\mu(\hat0_M,\theta)$ and $E_\theta$ are 
respectively defined as
\begin{eqnarray}
  \mu\left(\hat0_M,\theta\right)
&\equiv&
  \prod_{\sigma\in\theta}\left(-\right)^{N_\sigma-1}\left(N_\sigma-1\right)!,
\\
  E_\theta
&\equiv&
  \beta\sum_{\sigma\in \theta}\frac{2N_\sigma}{x_\sigma^2+1}
,\quad\theta\in\Theta_M.
\label{eq:statistical_energy}
\end{eqnarray}
This rewriting is justified in the same way as we have done for the
$\delta$-function boson case~\cite{go_b,go_d,go_e}.  Roughly speaking, the
sum can be replaced with a multiple integral in taking the
thermodynamic limit, and extinction of non-physical states from the
sum is realized by the coefficients
$\frac1{M!}\sum_{\theta\in\Theta_M}\mu\left(\hat 0_M,\theta\right)$.
The relation between $\{x_\sigma\}$ and $\{I_\sigma\}$ is defined by
\begin{eqnarray}
  \left[\frac{x_\sigma+i}{x_\sigma-i}\right]^{L}
&=&
  e^{-2\pi i I_\sigma}
  \prod_{\sigma'\in\theta,\neq\sigma}
  \left[\frac{x_\sigma-x_{\sigma'}+2i}{x_\sigma-x_{\sigma'}-2i}\right]^{N_{\sigma'}},
  \quad\quad\sigma\in\theta.
\label{eq:statistical_Bethe}
\end{eqnarray}
We note that eqs.(\ref{eq:statistical_Bethe}) are
eqs.(\ref{eq:modified_Bethe}) restricted by the conditions;
\begin{eqnarray}
  x_m\;=\;x_\sigma\quad
  I_m\;=\;I_\sigma\quad\quad m\in\sigma.
\end{eqnarray}
The function $\mu(\hat0_M,\theta)$ is a special case of the M\"obius
function $\mu(\theta',\theta)$ which comes from a natural
definition of partial order among elements in $\Theta_M$.
The most important thing in
(\ref{eq:Z_LM_integral_formulation(formal)}) is the definition of the
paths of the multiple integrals. We define the path of the
multiple integral corresponding to $\theta$. The integral path is a
manifold on which $\{ I_\sigma\}$ is a set of real numbers and
$\{x_\sigma\}$ are continuously distributed. Concretely, the integral
path is an $N_\theta$-dimensional surface defined by
conditions,
\begin{eqnarray}
  \left|
    \left(\frac{x_\sigma-i}{x_\sigma+i}\right)^{L}
    \prod_{\sigma'\in\theta,\neq\sigma}
    \left(
      \frac{x_\sigma-x_{\sigma'}+i}{x_\sigma-x_{\sigma'}-2i}
    \right)^{N_{\sigma'}}
  \right|
&=&
  1.
\label{eq:integral_path_orijinal}
\end{eqnarray}
The orientation of the integral path is chosen so that the value of an integral $\int
\prod dI_\sigma$ for any part of the integral path is positive.  Here
and hereafter, we consider conditions such as
(\ref{eq:integral_path_orijinal}) for an integral path.

For rewriting (\ref{eq:Z_LM_summation_formulation(formal)}) into
(\ref{eq:Z_LM_integral_formulation(formal)}), it is necessary to
define the words ``different'' and ``same value'' used in step 1. We
regard $\{x_m\}$ in eq.(\ref{eq:modified_Bethe}) as a multivalued vector function
of $\{\exp(2\Pi i I_m)\neq 0, \infty\}$.  Whether
$x_{m'}$ and $x_{m''}$ take the ``same value'' or not depends on
neighborhoods of the point $\{x_m\}$ on the Riemann surface defined by
the function.  We say that $x_{m'}$ and $x_{m''}$ take the ``same
value'' if and only if there is a point $\{x'_m\}$ in any
neighborhood where $x'_{m'}=x'_{m''}$ and $x_{m'}\neq x'_{m'}$.  In
case $x_{m'}=x_{m''}\neq \infty,\pm i$, this definition indicates that
$x_{m'}$ and $x_{m''}$ take the same value.  Therefore, in almost
all the region of variables $\{x_m\}$, this definition leads to a
valid definition of the ``physical solution''.  On the other hand, we
have to prove that the ``physical solution'' is a real physical
solution when some variables $x_m$ are equal to $\infty$ or $\pm i$,
where a real physical solution means that the Bethe vector
corresponding to the solution expresses a non-trivial eigen vector.
Next, ``different'' is defined as follows. Solutions of
(\ref{eq:integral_path_orijinal}) corresponding to different points on
the Riemann surface are ``different'', where we regard
(\ref{eq:integral_path_orijinal}) as an equation for $\{x_m\}$ fixing $\{\exp(2\Pi i I_m)\}$ to be constants. At any point on the
Riemann surface, the number of ``different'' solutions is
\begin{eqnarray}
  \lim_{\epsilon\rightarrow0} \max
  \quad 
  \makebox{
    number of points on $R_\epsilon$
    corresponding to the same $\{\exp(2\Pi i I_m) \}$
  },
\end{eqnarray}
where $R_\epsilon$ is an $\epsilon$-neighborhood of the point on the
Riemann surface.  For example, in case $M=2$, the point on the Riemann
surface $x_1,x_2=\infty$, $I_1,I_2=$integer
corresponds to two ``different'' solutions, and to two ``physical
solutions''. But, there is only one eigen state corresponding to the
point. We interpret this as follows. The ``values'' $x_1$ and
$x_2$ are not the ``same value'', and one of the ``physical
solutions'' is given by exchanging the ``values'' $x_1,x_2$ of the other
``physical solution''.

\subsection{}
In \ref{app:independency_of_integral_path}, we prove that the value of
the integral
\begin{eqnarray}
  \int
  e^{-\beta E_\theta}
  \prod_{\sigma\in \theta}dI_\sigma,
\label{eq:tmp_app_b_4}
\end{eqnarray}
along the integral path
\begin{eqnarray}
    \left|
      \left(\frac{x_\sigma-i}{x_\sigma+i}\right)^L
      \prod_{\sigma'\in\theta,\neq\sigma}
      \left(\frac{x_\sigma-x_{\sigma'}+2i}{x_\sigma-x_{\sigma'}-2i}
        \right)^{N_{\sigma'}}
    \right|
&=&
  A_{\sigma},\quad \sigma\in\theta,
\label{eq:tmp_app_b_03}
\label{eq:integral_path_modified}
\label{eq:integral_path_Bethe}
\end{eqnarray}
does not depend on $A_\sigma$ in case that all $A_\sigma$ are finite and
positive. The relation between $x_\sigma$ and $I_\sigma$ is defined by
(\ref{eq:statistical_Bethe}) and $E_\theta$ is defined by
(\ref{eq:statistical_energy}). Note that we define the orientation of
the path so that an integral $\int \prod
dI_\sigma$ for any part of the path is positive.

In the multiple integral (\ref{eq:Z_LM_integral_formulation(formal)}), we thus change the integral
path (\ref{eq:integral_path_orijinal}) into
(\ref{eq:integral_path_modified}).
We prove in \ref{app:definition_of_integral_path} that the expression
(\ref{eq:Z_LM_integral_formulation(formal)}) can be rewritten as
\begin{eqnarray}
  (\ref{eq:Z_LM_integral_formulation(formal)})
&=&
  \sum_{\theta\in\Theta_M}\mu\left(\hat 0_M,\theta\right)
  \left[
    \prod_{\sigma\in \theta}
    \int_{\left|x_\sigma-i\right|=0+}
    dx_\sigma
  \right]
  \left|
    \frac{\partial I_\sigma}{\partial x_\sigma}
  \right|_{L,\theta}
  e^{-\beta E_\theta},
\label{eq:Z_LM_integral_formulation_1}
\end{eqnarray}
by defining $A_\sigma$ as a set of sufficiently small real number set,
where $\left|x_\sigma-i\right|=0+$ indicates the integral path that
$x_\sigma$ turns around a point $i$ anticlockwise.

\subsection{}
\begin{itemize}
\item definition: $\Lambda(\theta)$

  Let $\theta$ be a set with a finite number of elements.  We denote
  by $\Lambda\left(\theta\right)$ all the patterns of connection of a
  set $\theta$.  What we call a pattern of connection satisfies the
  following two conditions.  1) Any two elements of $\theta$ are
  connected or not. Simply, there is no multiple connection.  2) There
  is no closed path in the connections.  Then, a pattern of connection $\lambda$
  is a set of elements each of which corresponds to a
  connection $\eta$.  To
  summarize, $\Lambda(\theta)$ satisfies conditions,
  \begin{eqnarray}
    &{\rm if} \quad \eta\in\lambda\in\Lambda(\theta),
  \quad {\rm then}\quad
    \eta=\left\{\sigma,\sigma'\right\},\quad \sigma,\sigma'\in\theta,
  \nonumber\\
  &{\rm if}\quad
    \left\{\sigma_1,\sigma_2\right\},
    \left\{\sigma_2,\sigma_3\right\},
    \cdots,
    \left\{\sigma_{m-1},\sigma_m\right\}
    \in\lambda\in\Lambda\left(\theta\right),
\quad {\rm then}\quad\{\sigma_1, \sigma_m\}\not \in \lambda,
  \end{eqnarray}
  and has the most elements of all sets which satisfy the above conditions.

\item definition: $G_\theta(\lambda)$

  Let $\theta$ be a set with a finite number of elements, and
  $\lambda\in\Lambda(\theta)$.  $G_\theta(\lambda)$ is an element of
  $\Theta(\theta)$.  In other words, $G_\theta(\lambda)$ is a pattern
  of division of $\theta$. We call that
  $\sigma$ and $\sigma'$ in $\theta$ are indirectly connected by
  $\lambda$ when $\sigma$ is linked to $\sigma'$ through one or
  some connections in $\lambda$. Two elements in $\theta$ are indirectly
  connected by $\lambda$ if and only if there is a cluster $\theta'\in
  G_\theta(\lambda)$ containing the two elements.   Precisely, $G_\theta(\lambda)$
  satisfies the conditions,
  \begin{eqnarray}
    G_\theta\left(\lambda\right)
  &&\in\quad
    \Theta\left(\theta\right)
  \nonumber\\ 
   {\rm if} \quad \left\{\sigma,\sigma'\right\}\in\lambda,
  &&{\rm then}\quad
    \sigma,\sigma'\in\theta'\in G_\theta\left(\lambda\right),
  \end{eqnarray}
  and has the most elements of all sets which satisfy the conditions.

\item definition: $\Theta(\theta|\lambda)$

  Let $\theta$ be a set with a finite number of elements, and
  $\lambda\in\Lambda(\theta)$.  $\Theta(\theta|\lambda)$ consists of
  all elements $\zeta$ satisfying the following conditions: $\zeta$ is
  in $\Theta(\theta)$, and any $\theta'\in G_\theta(\lambda)$ is a
  union of some sets in $\zeta$. Then, $\Theta(\theta|\lambda)$ is
  \begin{eqnarray}
   \Theta\left(\theta|\lambda\right)&=&\left\{\;\zeta\;|\;
   \zeta= G_\theta\left(\lambda'\right)
,\quad
   \lambda'\subseteq\lambda
\right\}.
  \end{eqnarray}

\item definition: $\stackrel{\zeta}{\sim}$

  Let $\zeta\in\Theta(\theta)$, and $\theta$ be a set with a finite
  number of elements. We denote by $\stackrel{\zeta}{\sim}$ a $\zeta$-dependent equivalent relation
  with respect to elements in $\theta$,
  \begin{eqnarray}
    \sigma\stackrel{\zeta}{\sim}\sigma'&&{\rm when}\quad\sigma,\sigma'\in\theta'\in\zeta.
  \end{eqnarray}

\item definition: $l_\lambda\left(\sigma,\sigma'\right)$

  Let $\sigma,\sigma'\in\theta$, and $\lambda\in\Lambda(\theta)$.  We
  define $l_\lambda\left(\sigma,\sigma'\right)$ in case that $\sigma$
  and $\sigma'$ are indirectly connected by $\lambda$, or
  $\sigma=\sigma'$.  In case $\sigma=\sigma'$,
  $l_\lambda\left(\sigma,\sigma\right)=0$. In the other case,
  $l_\lambda\left(\sigma,\sigma\right)$ is the number of connections
  from $\sigma$ to $\sigma'$.

\item definition: $\theta[\zeta,\sigma]$

  Let $\sigma$ be an element of a set which belongs to $\zeta$.
  We denote by $\theta[\zeta,\sigma]$ a set which is in $\zeta$ and
  includes $\sigma$.  That is, $\theta\left[\zeta,\sigma\right]$ is
  the set which satisfies a condition,
  \begin{eqnarray}
    \sigma\in\theta\left[\zeta,\sigma\right]\in\zeta.
  \end{eqnarray}

\item definition: $\Theta\left(\theta|\lambda|\sigma_1,\cdots,\sigma_m\right)$

  Let $\theta=\{\sigma_1,\cdots,\sigma_m,\cdots\}$ be a finite set and
  $\lambda$ be in $\Lambda\left(\theta\right)$.  The
  symbol $\Theta\left(\theta|\lambda|\sigma_1,\cdots,\sigma_m\right)$
  is defined as follows; it consists of all the elements $\zeta$ in
  $\Theta\left(\theta|\lambda\right)$ which satisfy the condition
  $\theta[\zeta,\sigma_k]\neq\theta[\zeta,\sigma_{k'}]$ when $k\neq
  k'$ and $k,k'\leq m$.  Then,
  $\Theta\left(\theta|\lambda|\sigma_1,\cdots,\sigma_m\right)$
  satisfies the condition,
  \begin{eqnarray}
    {\rm if}\quad \zeta \;\in\;
    \Theta\left(\theta|\lambda|\sigma_1,\cdots,\sigma_m\right), \quad
    k,k'\leq m, \quad
    {\rm then}\quad\zeta\;\in\;\Theta\left(\theta|\lambda\right), \quad
    \sigma_k\not\stackrel{\zeta}{\sim} \sigma_{k'},
  \end{eqnarray}
  and has the most elements of all sets satisfying the condition.

\end{itemize}

Using these symbols, we can express the change of the integral path from $|x_\sigma-i|=0+$
into $(-\infty,\infty)$.  Then,
(\ref{eq:Z_LM_integral_formulation_1}) becomes
\begin{eqnarray}
  (\ref{eq:Z_LM_integral_formulation(formal)})
&=&
  \sum_{\theta\in\Theta_M}
  \mu\left(\hat0_M,\theta\right)
  \sum_{\lambda\in\Lambda\left(\theta\right)}
  \left[\prod_{\sigma\in\theta}N_\sigma\right]^{-1}
  \left[
    \prod_{\{\sigma,\sigma'\}\in\lambda}
    N_{\sigma} N_{\sigma'}
  \right]
  \sum_{\zeta\in\Theta\left(\theta|\lambda\right)}
  \sum_{\{\sigma^{\left(\theta_k\right)}\}\in\{\theta_k\in\zeta\}}
\nonumber\\&&
  \left[
    \prod_{\theta'\in\zeta}
    \int_{-\infty+i\min\left(\sigma^{(\theta')}\right)\delta}
        ^{+\infty+i\min\left(\sigma^{(\theta')}\right)\delta}
    \frac{dx_{\theta'}}{2\pi}
  \right]
  \left[
    \prod_{\theta'\in G_\theta\left(\lambda\right)}
    \left(
      \sum_{\sigma\in\theta'}
      \frac
        {2N_{\sigma}L}
        {
          \left(
            x_{\theta[\zeta,\sigma ]}
            +2l_\lambda\left(\sigma,\sigma^{(\theta[\zeta,\sigma])}\right)i
          \right)^2
          +1
        }
    \right)
  \right]
\nonumber\\&&
  \left[
    \prod_{\{\sigma,\sigma'\}\in\lambda,\;\sigma\not\stackrel{\zeta}{\sim}\sigma'}
    \frac
      {-4}{
        \left(
           x_{\theta[\zeta,\sigma ]}
          -x_{\theta[\zeta,\sigma']}
          +2l_\lambda\left(\sigma ,\sigma^{(\theta[\zeta,\sigma ])}\right)i
          -2l_\lambda\left(\sigma',\sigma^{(\theta[\zeta,\sigma'])}\right)i
        \right)^2
        +4
      }
  \right]
\nonumber\\&&
  e^{
    -\beta\sum_{\sigma\in\theta}
    \frac
      {2N_{\sigma}}
      {
        \left(
          x_{\theta[\zeta,\sigma]}
          +2l_\lambda\left(\sigma,\sigma^{(\theta[\zeta,\sigma])}\right)i
        \right)^2
        +1
      }
  }.
\label{eq:Z_LM_integral_formulation_2}
\end{eqnarray}
We have used two simplified notations:
\begin{eqnarray}
  \sum_{\{\sigma^{\left(\theta_k\right)}\}\in\{\theta_k\in\zeta\}}
&\equiv&
  \sum_{\sigma^{(\theta_1)}\in\theta_1}
  \sum_{\sigma^{(\theta_2)}\in\theta_2}
  \cdots
  \sum_{\sigma^{(\theta_{N_\zeta})}\in\theta_{N_\zeta}},
\quad  \theta_1,\theta_2\cdots\theta_{N_\zeta}\in\zeta,
\end{eqnarray}
and $\min(\sigma)\equiv\min_{n\in\sigma}n$ where
$\sigma\subset\mathbb{N}$.

A remark is in order. In (\ref{eq:Z_LM_integral_formulation_2}), the
series $\sum_{\zeta\cdots}$ with respect to $\zeta$ can
be divided into two parts. One is a set in which all elements have
only one element, and the other is a set which consists of the other
elements.  The former is a set of terms corresponding to
(\ref{eq:Z_LM_integral_formulation_1}) with an integral path
$(-\infty,\infty)$. The latter is a set of terms corresponding to
residues which the modification of the integral path generates.

From now on, we prove the equivalence of
(\ref{eq:Z_LM_integral_formulation_1}) and
(\ref{eq:Z_LM_integral_formulation_2}).  Equivalently, we prove that
\begin{eqnarray}
  &&
    \left[
    \prod_{\sigma\in \theta}
    \int_{\left|x_\sigma-i\right|=0+}
    dx_\sigma
  \right]
  \left[\prod_{\sigma\in\theta}N_\sigma\right]^{-1}
    \left[
\prod_{\theta'\in G_\theta\left(\lambda\right)}
\left(
      \sum_{\sigma\in\theta}
      \frac{2N_\sigma L}{x^2_\sigma+1}
\right)
    \right]
\nonumber\\&&
    \left[
      \prod_{\{\sigma,\sigma'\}\in\lambda}
      \frac{-4N_\sigma N_{\sigma'}}{\left(x_\sigma-x_\sigma'\right)^2+4}
    \right]
    e^{-\beta\sum_{\sigma\in\theta}\frac{2JN_\sigma}{x_\sigma^2+1}}
\nonumber\\&=&
  \left[\prod_{\sigma\in\theta}N_\sigma\right]^{-1}
  \left[
    \prod_{\{\sigma,\sigma'\}\in\lambda}
    N_{\sigma} N_{\sigma'}
  \right]
  \sum_{\zeta\in\Theta\left(\theta|\lambda\right)}
  \sum_{\{\sigma^{\left(\theta_k\right)}\}\in\{\theta_k\in\zeta\}}
\nonumber\\&&
  \left[
    \prod_{\theta'\in\zeta}
    \int_{-\infty+i\min\left(\sigma^{(\theta')}\right)\delta}
        ^{+\infty+i\min\left(\sigma^{(\theta')}\right)\delta}
    \frac{dx_{\theta'}}{2\pi}
  \right]
  \left[
    \prod_{\theta'\in G_\theta\left(\lambda\right)}
    \left(
      \sum_{\sigma\in\theta'}
      \frac
        {2N_{\sigma} L}
        {
          \left(
            x_{\theta[\zeta,\sigma]}
            +2l_\lambda\left(\sigma,\sigma^{(\theta[\zeta,\sigma])}\right)i
          \right)^2
          +1
        }
    \right)
  \right]
\nonumber\\&&
  \left[
    \prod_{\{\sigma,\sigma'\}\in\lambda,\;\sigma\not\stackrel{\zeta}{\sim}\sigma'}
    \frac
      {-4}{
        \left(
           x_{\theta[\zeta,\sigma ]}
          -x_{\theta[\zeta,\sigma']}
          +2l_\lambda\left(\sigma ,\sigma^{(\theta[\zeta,\sigma ])}\right)i
          -2l_\lambda\left(\sigma',\sigma^{(\theta[\zeta,\sigma'])}\right)i
        \right)^2
        +4
      }
  \right]
\nonumber\\&&
  \exp\left[
    -\beta\sum_{\sigma\in\theta}
    \frac
      {2N_{\sigma}}
      {
        \left(
          x_{\theta[\zeta,\sigma]}
          +2l_\lambda\left(\sigma,\sigma^{(\theta[\zeta,\sigma])}\right)i
        \right)^2
        +1
      }
  \right],
\label{eq:modification_of_integral_path_1_tmp_0}
\end{eqnarray}
where $\theta\in\Theta_M$ and $\zeta\in\Lambda\left(\theta\right)$,
since both sides of this equation become
(\ref{eq:Z_LM_integral_formulation_1}) and
(\ref{eq:Z_LM_integral_formulation_2}) respectively when we apply
$\sum_{\theta\in\Theta_M} \mu\left(\hat0_M,\theta\right)
\sum_{\lambda\in\Lambda\left(\theta\right)}$ to both sides.  Here, we
have substituted an explicit expression of the Jacobian,
\begin{eqnarray}
  \left(2\pi\right)^{N_{\theta}}
  \left|\frac{\partial I_\sigma}{\partial x_\sigma'}\right|_{L,\theta}
&\equiv&
  \left[\prod_{\sigma\in\theta}N_\sigma\right]^{-1}
  \sum_{\lambda\in\Lambda\left(\theta\right)}
  \left[\prod_{\left\{\sigma,\sigma'\right\}\in\lambda}
    -\frac{4N_{\sigma}N_{\sigma'}}{\left(x_{\sigma}-x_{\sigma'}\right)^2+4}
  \right]
\nonumber\\&&{}\times
  \prod_{\theta'\in G_\theta\left(\lambda\right)}
  \left[
    \sum_{\sigma\in\theta'}
    \frac{2N_\sigma L}{x^2_\sigma+1}
  \right].
\label{eq:def_jacobian_n}
\end{eqnarray}
A sufficient condition of
(\ref{eq:modification_of_integral_path_1_tmp_0}) is that an expression
\begin{eqnarray}
&&  \left[\prod_{\sigma\in\theta}N_\sigma\right]^{-1}
  \left[
    \prod_{\{\sigma,\sigma'\}\in\lambda}
    N_{\sigma} N_{\sigma'}
  \right]
  \sum_{\zeta\in\Theta\left(\theta|\lambda|\sigma_1,\cdots,\sigma_m\right)}
  \sum_{\{\sigma^{(\theta_k)}\}\in\{\theta_k\in\zeta\}_{k>m}}
  \left[
    \prod_{k=1}^m
    \int_{\left|i-x_{\theta_k}\right|_+=\bar\delta_{\sigma_k}}
    \frac{dx_{\theta_k}}{2\pi}
  \right]
\nonumber\\&&
  \left[
    \prod_{k=m+1}^{N_\zeta}
    \int_{-\infty+i\delta_{\sigma^{(\theta_k)}}}
        ^{+\infty+i\delta_{\sigma^{(\theta_k)}}}
    \frac{dx_{\theta_k}}{2\pi}
  \right]
    \left[
      \prod_{\theta'\in G_\theta\left(\lambda\right)}
      \left(
      \sum_{\sigma\in\theta'}
      \frac
        {2N_{\sigma} L}
        {
          \left(
            x_{\theta[\zeta,\sigma]}
            +2l_\lambda\left(\sigma,\sigma^{(\theta[\zeta,\sigma])}\right)i\right)^2+1}
\right)
    \right]
    \nonumber\\&&
    \left[
      \prod_{\{\sigma,\sigma'\}\in\lambda,\;\sigma\not\stackrel{\zeta}{\sim}\sigma'}
        \frac
          {-4}{
            \left(
               x_{\theta[\zeta,\sigma ]}
              -x_{\theta[\zeta,\sigma']}
              +2l_\lambda\left(\sigma ,\sigma^{(\theta[\zeta,\sigma ])}\right)i
              -2l_\lambda\left(\sigma',\sigma^{(\theta[\zeta,\sigma'])}\right)i\right)^2+4}
    \right]
    \nonumber\\&&
    \exp
    \left[-\beta
      \sum_{\sigma\in\theta}
      \frac
        {2N_{\sigma}}
        {
          \left(
            x_{\theta[\zeta,\sigma]}
            +2l_\lambda\left(\sigma,\sigma^{(\theta[\zeta,\sigma])}\right)i\right)^2+1}
    \right]
\label{eq:modification_of_integral_path_1_tmp_1}
\end{eqnarray}
dose not depend on $m$, where we demand that the constants
$\delta_{\sigma_k}$ and $\bar\delta_{\sigma_k}$ satisfy
\begin{eqnarray}
  1>> \delta_{\sigma_1}> \delta_{\sigma_{2}}>\cdots >0,
\quad\quad
 0< \bar\delta_{\sigma_1}<\bar\delta_{\sigma_{2}}<\cdots<<1.
\label{eq:large_small_relation}
\end{eqnarray}
A proof of this fact is shown in
\ref{app:modification_of_integral_path_1}.  This fact is a sufficient
condition, because (\ref{eq:modification_of_integral_path_1_tmp_1})
becomes the l.h.s.(r.h.s.) of
eq.(\ref{eq:modification_of_integral_path_1_tmp_0}) in case $m$ is
$N_\theta$($0$). In (\ref{eq:modification_of_integral_path_1_tmp_1}),
we have used a simplified notation
$\sum_{\{\sigma^{\left(\theta_k\right)}\}\in\{\theta_k\in\zeta\}_{k>m}}$
where $\zeta$ is in
$\Theta\left(\theta|\lambda|\sigma_1,\cdots,\sigma_{m'}\right)$: it is
defined as
\begin{eqnarray}
\!\!\!\!\!\!\!  \sum_{\{\sigma^{\left(\theta_k\right)}\}\in\{\theta_k\in\zeta\}_{k>m}}
\!\!\!\!\!\!\!
F\left(\{\sigma^{\left(\theta_k\right)}\}\right)
&\equiv&\!\!\!
  \sum_{\sigma^{(\theta_{m+1})}\in\theta_{m+1}}
  \sum_{\sigma^{(\theta_{m+2})}\in\theta_{m+2}}
  \!\!\!\cdots\!\!\!
  \sum_{\sigma^{(\theta_{N_\zeta})}\in\theta_{N_\zeta}}\!\!\!\!\!
\left.F\left(\{\sigma^{\left(\theta_k\right)}\}\right)\right|_{\sigma^{(\theta_k)}=\sigma_{k\leq m}},
\\&&
  \theta_1,\theta_2\cdots\theta_{N_\zeta}\in\zeta.
\nonumber
\end{eqnarray}
We have numbered the elements $\sigma_1,\cdots\sigma_{N_\theta}$ in
$\theta\in\Theta_M$ so that $\{\sigma_m\}$ satisfy the condition
$\min(\sigma_1)>\min(\sigma_2)>\cdots$. And when $\zeta$ is in
$\Theta\left(\theta|\lambda|\sigma_1,\cdots,\sigma_m\right)$ and
$0<k\leq N_\zeta$,  elements $\theta'\in\zeta$ have been named  $\theta_k$
so that $\theta_k$ is equal to $\theta[\zeta,\sigma_k]$ in case $k\leq
m$. These notations are used only in
(\ref{eq:modification_of_integral_path_1_tmp_1}) and
\ref{app:modification_of_integral_path_1}.

\subsection{}
\begin{itemize}
\item definition: $\tilde\Theta(\theta)$

  Let $\theta$ be a set with a finite number of elements.  A set
  $\tilde\Theta\left(\theta\right)$ consists of all elements
  $\tilde\zeta$ satisfying the following conditions.  First,
  $\tilde\zeta$ is a set of sequences as elements.  Second, all the
  elements in the sequences are in $\theta$.  Third, a set of sets
  $\zeta$ derived from $\tilde\zeta$ is in $\Theta(\theta)$.  Here,
  $\zeta$ is derived from $\tilde\zeta$ when we replace sequences in
  $\tilde\zeta$ with sets by ignoring the order of the sequences.
  Note that the number of $\tilde\zeta$'s which become a $\zeta$ by the
  above procedure is $\prod_{\theta'\in\zeta}N_{\theta'}!$.
\end{itemize}

Using these definitions, we change an expression of summations in
(\ref{eq:Z_LM_integral_formulation_2}) with respect to permutations.
Then, (\ref{eq:Z_LM_integral_formulation_2}) becomes
\begin{eqnarray}
&&  (\ref{eq:Z_LM_integral_formulation(formal)})
\nonumber\\
&=&
  \sum_{\theta\in\Theta_M}
  \sum_{\tilde\zeta\in\tilde\Theta\left(\theta\right)}
  \sum_{\lambda\in\Lambda\left(\tilde\zeta\right)}
  \left[
    \prod_{\left(\sigma_1,\sigma_2,\cdots\right)=\tilde\theta\in\tilde\zeta}
    \int_{-\infty+i\min\left(\sigma_1\right)\delta}
        ^{+\infty+i\min\left(\sigma_1\right)\delta}
    \frac{dx_{\tilde\theta}}{2\pi}
  \right]
\nonumber\\&&
  \left[
    \prod_{\tilde\zeta'\in G_{\tilde\zeta}\left(\lambda\right)}
    \left(
      \sum_{\tilde\theta\in\tilde\zeta'}
      \sum_{\sigma _{m }\in \tilde\theta =\left(\sigma _1,\cdots\right)}
      \frac
        {2N_{\sigma_m} L}
        {
          \left(
            x_{\tilde\theta}
            +2\left(m-1\right)i
          \right)^2
          +1
        }
    \right)
  \right]
\nonumber\\&&
  \left[
    \prod_{\{\tilde\theta,\tilde\theta'\}\in\lambda}
    \left(
      \sum_{\sigma _{m }\in \tilde\theta =\left(\sigma _1,\cdots\right)}
      \sum_{\sigma'_{m'}\in \tilde\theta'=\left(\sigma'_1,\cdots\right)}
      \frac
        {-4 N_{\sigma_m}N_{\sigma'_{m'}}}
        {
          \left(
             x_{\tilde\theta}
            -x_{\tilde\theta'}
            +2\left(m-m'\right)i
          \right)^2
          +4
        }
    \right)
  \right]
\nonumber\\&&
  e^{
    -\beta
    \sum_{\tilde\theta\in\tilde\zeta}
    \sum_{\sigma _{m }\in \tilde\theta =\left(\sigma _1,\cdots\right)}
    \frac
      {2N_{\sigma_m}}
      {
        \left(
           x_{\tilde\theta}
          +2\left(m-1\right)i
        \right)^2
        +1
      }
  }
\nonumber\\&&
{}\makebox[-0.5cm]{}  \prod_{\left(\sigma_1,\cdots\right)=\tilde\theta\in\tilde\zeta}\!\!
  \left[
   \! \left(-\right)^{N_{\sigma_1}-1}\!\left(N_{\sigma_1}-1\right)!
    N_{\sigma_1}^{-1}
    \!\!\!\prod_{\sigma_{m>1}\in\tilde\theta}\!\!
    \left[
      \sum_{\theta'\in\Theta\left(\sigma_m\right)}\!\!\!\!
      N_{\sigma_{m-1}}^{N_{\theta'}}
      \!\!\prod_{\sigma'\in\theta'}
      \left(-\right)^{N_{\sigma'}-1}\!\left(N_{\sigma'}-1\right)!
    \right]\!
  \right]\!\!.
\label{eq:Z_LM_integral_formulation_3}
\end{eqnarray}

The ``change of an expression of summations with respect to
permutations'' means a change from a series $A$ to a series $B$ satisfying the
following two conditions: First, the value of $A$ and $B$ are the
same.  Second, the equivalence can be shown by making a correspondence
between some elements in $A$ and in $B$.

For example, in case of the simple relations,
\begin{eqnarray}
  \sum_{m\geq0}\sum_{n\geq0}a^mb^n&=&
  \sum_{m\geq0}\sum_{n=0}^{m}a^{m-n}b^n,
\label{eq:example_sum_1}\\
  \sum_{m\geq0}\sum_{n\geq0}a^mb^n&=&
  \sum_{m\geq0}
a^m\frac{1-\left(\frac b a\right)^{m+1}}{1-\frac b a},
\label{eq:example_sum_2}
\end{eqnarray}
we say that the r.h.s. is made by means of changing an expression of
summation $\sum_{m\geq0}\sum_{n\geq0}$ in the l.h.s. with respect to
permutations. In case of (\ref{eq:example_sum_1}), when we write terms in $\sum_{m\geq0}\sum_{n\geq0}$ and
$\sum_{m\geq0}\sum_{n=0}^{m}$ as $(m,n)_l$ and $(m,n)_r$, $(m,n)_l$ and $(m+n,n)_r$ are the same. And, in case of
(\ref{eq:example_sum_2}), when we write terms in
$\sum_{m\geq0}\sum_{n\geq0}$ and $\sum_{m\geq0}$ as $(m,n)_l$ and $(m)_r$, a sum
of terms $(m,0)_l,\cdots (m,m)_l$ is equal to $(m)_r$.

We introduce the words ``correspond'' and ``correspondence'' as
follows.  When we say ``an element $a$ corresponds to $b$'' where $a$
and $b$ are elements under $\sum$, e.g. $(m,n)_l$ and
$(m,n)_r$, we demand the following two facts.  Values of
both $a$ and $b$ are the same, and the correspondence makes a
one-to-one correspondence between $A$ and $B$ where $a\in A$ and $b\in
B$. In such case, we say ``there is a one-to-one correspondence
between $A$ and $B$''.  Remark that we have given the extra meaning to
the word ``correspond'' and ``correspondence''.  In case of
(\ref{eq:example_sum_1}) there is a one-to-one correspondence between
$\{(m,n)_l\}$ and $\{(m,n)_r\}$, and $(m,n)_l$ corresponds to
$(m+n,n)_r$.  And, when we say ``a subset $\{a\}$ corresponds to an
element $b$'' where $a\in\{a\}$ and $b$ are elements under
$\sum$, e.g. $(m,n)_l$ and $(m)_r$, we demand the following two facts.
The summation of values over $a\in\{a\}$ is equal to a value
of $b$ and the correspondence makes a many-to-one
correspondence between $A$ and $B$ where $a\in A$ and $b\in B$.  In
such case, we say that there is a many-to-one correspondence between
$A$ and $B$.  Therefore, in case of (\ref{eq:example_sum_2}) there is
a many-to-one correspondence between $\{(m,n)_l\}$ and $\{(m)_r\}$,
and $(m,0)_l,\cdots ,(m,m)_l$ corresponds to $(m)_r$.  ``One-to-many
correspondence'', ``an element corresponds to a set'', ``many-to-many
correspondence'' and ``a set corresponds to a set'' should be understood in the
same way.

We note that when we change an expression of summations in
(\ref{eq:Z_LM_integral_formulation_2}) into
(\ref{eq:Z_LM_integral_formulation_3}), there is a many-to-one
correspondence between $\{\lambda,\zeta,\{\sigma^{(\theta_k)}\}\}$ and
$\{\tilde\zeta,\lambda\}$ associated with
\begin{eqnarray}
  \sum_{\theta\in\Theta_M}
  \sum_{\lambda\in\Lambda\left(\theta\right)}
  \sum_{\zeta\in\Theta\left(\theta|\lambda\right)}
  \sum_{\{\sigma^{\left(\theta_k\right)}\}\in\{\theta_k\in\zeta\}}
&\leftrightarrow&
  \sum_{\theta\in\Theta_M}
  \sum_{\tilde\zeta\in\tilde\Theta\left(\theta\right)}
  \sum_{\lambda\in\Lambda\left(\tilde\zeta\right)}.
\end{eqnarray}
A subset $A$ of $\{\lambda,\zeta,\{\sigma^{(\theta_k)}\}\}$
corresponds to $\tilde\zeta',\lambda'$ in $\{\tilde\zeta,\lambda\}$,
where $\lambda'',\zeta'',\{\sigma^{(\theta_k)}\}$ in $A$ and
$\tilde\zeta',\lambda'$ satisfy the following conditions.  There is a
bijection $f$ from the set $\zeta''$ to the set $\tilde\zeta'$ which
satisfies
\begin{eqnarray}
  \bigoplus_{\sigma\in\theta'',l_{\lambda''}\left(\sigma,\sigma^{(\theta'')}\right)=m-1}\sigma
&=&
  \sigma'_m
,
\end{eqnarray}
where $f\left(\theta''\right)=\tilde\theta'$ and
$\left(\sigma'_1,\cdots\right)=\tilde\theta'$.  And, $\lambda'$ can be
written as
\begin{eqnarray}
  \lambda'
&=&
  \left\{
    \left\{\tilde\theta',\tilde\theta''\right\}
    \left|
    \left\{\sigma',\sigma''\right\}\in\zeta'',\;
    \sigma'\in f^{-1}\left(\tilde\theta'\right),\;
    \sigma''\in f^{-1}\left(\tilde\theta''\right)
  \right.\right\}.
\end{eqnarray}
It is clear that this correspondence is a ``many-to-one correspondence''.

\subsection{}
\begin{itemize}
\item definition: $\theta_>(\tilde\theta)$ 

  Let $\tilde\theta$ be a sequence of sets with a finite
  number of elements. $\theta_>(\tilde\theta)$  is defined by
  \begin{eqnarray}
    \theta_>\left(\tilde\theta\right)&\equiv&
    \prod_{\sigma_{m>1}\in\tilde\theta=(\sigma_1,\cdots)}
    \theta\left(N_{\sigma_{m-1}}- N_{\sigma_{m}}+\frac12\right),
  \end{eqnarray}
  where the $\theta$-function is defined as
  \begin{eqnarray}
    \theta\left(n\right)
    &\equiv&
    \left\{
      \begin{array}{lll}
        1        &\makebox{in case of}&n>0\\
        \frac12  &\makebox{in case of}&n=0\\
        0        &\makebox{in case of}&n<0.\\
      \end{array}
    \right.
  \end{eqnarray}

\item definition: $\Lambda_c(\theta)$

  Let $\theta$ be a set with a finite number of elements.
  $\Lambda_c(\theta)$ consists of all elements $\lambda$ satisfying
  following two condition. First, $\lambda$ is in
  $\Lambda(\theta)$. Second, any two of elements in $\theta$ are
  indirectly connected by $\lambda$, that is to say,
  $N_{G_\theta(\lambda)}=1$.  Then,
  \begin{eqnarray}
    \Lambda_c\left(\theta\right)
  &=&
    \left\{
      \lambda|\lambda\in\Lambda\left(\theta\right),N_{G_\theta(\lambda)}=1
    \right\}.
  \end{eqnarray}

\item definition: $D\left(\tilde\theta\right)$

  Let $\tilde\theta$ be a sequence of sets, where the number of
  elements in each set decreases in order of the sequence.
  $D\left(\tilde\theta\right)$ consists of elements $\tilde\zeta$
  satisfying the following conditions. First, $\tilde\zeta$ is a set
  of sequences $\tilde\theta'$ which consist of sets $\sigma$.  Second, all
  sets $\sigma$ in any sequence $\tilde\theta'$ have the same number of
  elements. Third,
  \begin{eqnarray}
    \sigma_k
    &=&
    \bigoplus_{(\sigma'_1,\cdots)=\tilde\theta'\in\tilde\zeta}\sigma'_k,
  \end{eqnarray}
  where $(\sigma_1,\sigma_2,\cdots)=\tilde\theta$.

\item definition: $\delta\left(\tilde\zeta,\lambda,n\right)$

  Let $\tilde\zeta$ be in $D\left(\tilde\theta\right)$, $\lambda$ be
  in $\Lambda_c\left(\tilde\zeta\right)$, $n$ be in a set $\sigma$ and $\tilde\theta$ be in $\tilde\Theta(\sigma)$.
  $\delta\left(\tilde\zeta,\lambda,n\right)$ is a function whose value
  is $1$ in case $\tilde\zeta,\lambda,n$ satisfy the following
  conditions and is $0$ otherwise.  First, there is only one sequence
  $\tilde\theta^{(0)}$ in $\tilde\zeta$ which has the smallest number
  of elements in $\tilde\zeta$.  Second, when we write the sequence
  $\tilde\theta^{(0)}$ as $(\sigma^{(0)}_1,\cdots)$, $n$ is in
  $\sigma^{(0)}_1$. Third,
  \begin{eqnarray}
    {\rm if}\quad l_\lambda\left(\tilde\theta^{(0)},\tilde\theta' \right)+1=
    l_\lambda\left(\tilde\theta^{(0)},\tilde\theta''\right),
    \;
    l_\lambda\left(\tilde\theta',\tilde\theta''\right)=1
    &{\rm then}\quad
    N_{\tilde\theta'}< N_{\tilde\theta''}
  \end{eqnarray}
  for any $\tilde\theta',\tilde\theta''\in\tilde\zeta$.
  This function is sometimes referred to as $\delta$-function, but is rather close to  the Kronecker's $\delta$.

\item definition: $M_\theta$

  Let $\theta$ be a set or sequence of sets which have the same number
  of elements.  Then, $M_\theta$ means $N_\sigma$ where
  $\sigma\in\theta$.
\end{itemize}
 
In terms of those notation, we change the expression
(\ref{eq:Z_LM_integral_formulation_3}) into
\begin{eqnarray}
&&  (\ref{eq:Z_LM_integral_formulation(formal)})
\nonumber\\
&=&
  \sum_{\theta\in\Theta_M}
  \sum_{\tilde\zeta\in\tilde\Theta\left(\theta\right)}
  \left[
    \prod_{\tilde\theta \in\tilde\zeta}\delta_>\left(\tilde\theta\right)
  \right]
  \sum_{\lambda\in\Lambda\left(\tilde\zeta\right)}
  \left[
    \prod_{\left(\sigma_1,\sigma_2,\cdots\right)=\tilde\theta\in\tilde\zeta}
    \int_{-\infty+i\min\left(\sigma_1\right)\delta}
        ^{+\infty+i\min\left(\sigma_1\right)\delta}
    \frac{dx_{\tilde\theta}}{2\pi}
  \right]
\nonumber\\&&
    \left[
      \prod_{\tilde\zeta'\in G_{\tilde\zeta}\left(\lambda\right)}
      \left(
        \sum_{\tilde\theta\in\tilde\zeta'}
        \sum_{\sigma _{m }\in \tilde\theta =\left(\sigma _1,\cdots\right)}
        \frac
          {2\left(N_{\sigma_{m}}-N_{\sigma_{m+1}}\right)m L}
          {
            \left(
              x_{\tilde\theta}
              +\left(m-1\right)i
            \right)^2+m^2
          }
      \right)
    \right]
\nonumber\\&&
  \left[
    \prod_{\{\tilde\theta,\tilde\theta'\}\in\lambda}
    \left(
      \sum_{\sigma _{m }\in \tilde\theta =\left(\sigma _1,\cdots\right)}
      \sum_{\sigma'_{m'}\in \tilde\theta'=\left(\sigma'_1,\cdots\right)}
      \left(
        -\left(N_{\sigma _{m }}-N_{\sigma _{m +1}}\right)
         \left(N_{\sigma'_{m'}}-N_{\sigma'_{m'+1}}\right)
      \right)
\right.\right.
\nonumber\\&&
\left.\left.
      K_{m,m'}\left(
        x_{\tilde\theta}-x_{\tilde\theta'}+\left(m-m'\right)i
      \right)
    \right)
  \right]
\nonumber\\&&
    \exp\left[
      -\beta
      \sum_{\tilde\theta\in\tilde\zeta}
      \sum_{\sigma _{m }\in \tilde\theta =\left(\sigma _1,\cdots\right)}
      \frac
        {2\left(N_{\sigma_{m}}-N_{\sigma_{m+1}}\right)m}
        {
          \left(
            x_{\tilde\theta}
            +\left(m-1\right)i
          \right)^2
          +m^2
        }
    \right]
\nonumber\\&&
  \prod_{\left(\sigma_1,\cdots\right)=\tilde\theta\in\tilde\zeta}
  \left[
    \sum_{\tilde\zeta'\in D\left(\tilde\theta\right)}
    \sum_{\lambda'\in\Lambda_c\left(\tilde\zeta'\right)}
    \delta\left(\tilde\zeta',\lambda',\min\left(\sigma_1\right)\right)
    \left[
      \prod_{\tilde\theta'\in\tilde\zeta'}
      \left(
        \left(-\right)^{M_{\tilde\theta'}-1}\left(M_{\tilde\theta'}-1\right)!
        M_{\tilde\theta'}^{-1}
        \left(
          M_{\tilde\theta'}!
        \right)^{N_{\tilde\theta'}-1}
      \right)
    \right]
\right.\nonumber\\&&\left.
    \left[
      \prod_{\{\tilde\theta',\tilde\theta''\}\in\lambda'}
      \left(-M_{\tilde\theta'}M_{\tilde\theta''}\right)
    \right]
  \right],
\label{eq:moveing_of_integral_path_tmp_1}
\end{eqnarray}
where the function $K_{nm}(x)$ is defined as
\begin{eqnarray}
{}\makebox[-0.5cm]{}  K_{n,m}\left(x\right)
&\equiv&
  \left\{
\begin{array}{ll}
\kappa_{\left|n-m\right|}\left(x\right)
+2\kappa_{\left|n-m\right|+2}\left(x\right)
+\cdots
+2\kappa_{n+m-2}\left(x\right)
+\kappa_{n+m}\left(x\right)
& n\neq m\\
2\kappa_{2}\left(x\right)
+\cdots
+2\kappa_{2n-2}\left(x\right)
+\kappa_{2n}\left(x\right)
& n=m
\end{array}
\right.
\label{eq:define_K_n_m}
\end{eqnarray}
and $\kappa_{n}(x)$ is defined as
\begin{eqnarray}
  \kappa_{n}\left(x\right)&\equiv&\frac{2n}{x^2+n^2}.
\end{eqnarray}
In \ref{sec:moveing_of_integral_path}, we give a proof of a relation,
\begin{eqnarray}
&&    \left(-\right)^{N_{\sigma_1}-1}\left(N_{\sigma_1}-1\right)!
    N_{\sigma_1}^{-1}
    \prod_{\sigma_{m>1}\in\tilde\theta}
    \left[
      \sum_{\theta'\in\Theta\left(\sigma_m\right)}
      N_{\sigma_{m-1}}^{N_{\theta'}}
      \prod_{\sigma'\in\theta'}
      \left(-\right)^{N_{\sigma'}-1}\left(N_{\sigma'}-1\right)!
    \right]
\nonumber\\&=& 
    \sum_{\tilde\zeta'\in D\left(\tilde\theta\right)}
    \sum_{\lambda'\in\Lambda_c\left(\tilde\zeta'\right)}
    \delta\left(\tilde\zeta',\lambda',\min\left(\sigma_1\right)\right)
    \left[
      \prod_{\tilde\theta'\in\tilde\zeta'}
      \left(
        \left(-\right)^{M_{\tilde\theta'}-1}\left(M_{\tilde\theta'}-1\right)!
        M_{\tilde\theta'}^{-1}
        \left(
          M_{\tilde\theta'}!
        \right)^{N_{\tilde\theta'}-1}
      \right)
    \right]
\nonumber\\&&
    \left[
      \prod_{\{\tilde\theta',\tilde\theta''\}\in\lambda'}
      \left(-M_{\tilde\theta'}M_{\tilde\theta''}\right)
    \right],
\label{eq:moveing_of_integral_path_coodinate_relation}
\end{eqnarray}
where
$\left(\sigma_1,\cdots\right)=\tilde\theta\in\tilde\zeta\in\tilde\Theta\left(\theta\in\Theta_M\right)$.
The equality of (\ref{eq:Z_LM_integral_formulation_3}) and
(\ref{eq:moveing_of_integral_path_tmp_1}) is proved by using this
relation and executing some elementary calculations.

\subsection{}
\begin{itemize}
\item definition: $\tilde{\bar\Theta}(\theta)$

  Let $\theta$ be a set which has a finite number of sets as elements.
  $\tilde{\bar\Theta}(\theta)$ consists of all elements $\tilde\zeta$
  satisfying the following two conditions. First, $\tilde\zeta$ is in
  $\tilde\Theta(\theta)$. Second, any sequence in $\tilde\zeta$
  satisfies the condition that all sets as elements in the sequence
  have the same number of elements. Then,
  \begin{eqnarray}
    \tilde{\bar\Theta}(\theta)&=&\left\{\;
      \tilde\zeta\in\tilde\Theta(\zeta)\;\left|\;
        N_\sigma=N_{\sigma'}, \quad\sigma,\sigma'\in\tilde\theta\in\tilde\zeta
      \right.\right\}.
  \end{eqnarray}
  Here and hereafter, we use the symbol $\in$ for an inclusion
  relation between a sequence and its element like a set and its
  element.

\item definition: $\lambda[\lambda,\tilde\zeta']$ 

  Let $\tilde\zeta'$ be a subset of a set $\tilde\zeta$ and
  $\lambda$ be an element of $\Lambda(\tilde\zeta)$.
  We write $\lambda[\lambda,\tilde\zeta']$ as
 \begin{eqnarray}
   \lambda[\lambda,\tilde\zeta']
   &\equiv&
   \left\{
   \left\{
     \tilde\theta,\tilde\theta'
   \right\}\in\lambda
     \left|     \tilde\theta,\tilde\theta'\in\tilde\zeta'
   \right.\right\}.
 \end{eqnarray}

\end{itemize}

By using those symbols, we change an expression of summations in
(\ref{eq:moveing_of_integral_path_tmp_1}) with respect to permutation.
Then, (\ref{eq:moveing_of_integral_path_tmp_1}) becomes
\begin{eqnarray}
&&  (\ref{eq:Z_LM_integral_formulation(formal)})
\nonumber\\
&=&
  \sum_{\theta\in\Theta_M}
  \sum_{\tilde\zeta\in\tilde{\bar\Theta}\left(\theta\right)}
  \sum_{\lambda\in\Lambda\left(\tilde\zeta\right)}
  \sum_{\lambda'\subseteq\lambda}
  \left[
    \prod_{\tilde\zeta'\in G_{\tilde\zeta}\left(\lambda'\right)}
    \delta\left(
      \tilde\zeta',
      \lambda\left[\lambda',\tilde\zeta'\right],
      {\min}_1\left(\tilde\zeta'\right)
    \right)
  \int_{-\infty+i\min_1\left(\tilde\zeta'\right)\delta}
      ^{-\infty-i\min_1\left(\tilde\zeta'\right)\delta}
    \frac{dx_{\tilde\zeta'}}{2\pi}
    \right]
\nonumber\\&&
  \left[
    \prod_{\tilde\zeta' \in G_{\tilde\zeta }\left(\lambda \right)}
    \left(
      \sum _{\tilde\zeta''\in G_{\tilde\zeta'}\left(\lambda\left[\lambda',\tilde\zeta'\right]\right)}
      \sum_{\tilde\theta''\in\tilde\zeta''}
      \frac
        {2M_{\tilde\theta''}N_{\tilde\theta''}L}
        {
          \left(x_{\tilde\zeta''}+\left(N_{\tilde\theta''}-1\right)i\right)^2
          +N_{\tilde\theta''}^2
        }
    \right)
  \right]
  \left[
    \prod_{\{\tilde\theta,\tilde\theta'\}\in\lambda'}
  \left(-    M_{\tilde\theta}M_{\tilde\theta'}
\right)
  \right]
\nonumber\\&&
  \left[
    \prod_{
      \{\tilde\theta,\tilde\theta'\}\in\lambda-\lambda',\;
      \tilde\zeta' \ni\tilde\theta ,\;
      \tilde\zeta''\ni\tilde\theta'
    }
    \left(-M_{\tilde\theta}M_{\tilde\theta'}\right)
    K_{N_{\tilde\theta},N_{\tilde\theta'}}\left(
      x_{\tilde\zeta'}
      -x_{\tilde\zeta''}
      +\left(N_{\tilde\theta}-N_{\tilde\theta'}\right)i
    \right)
  \right]
\nonumber\\&&
  e^{-\beta
      \sum _{\tilde\zeta'\in G_{\tilde\zeta}\left(\lambda'\right)}
      \sum_{\tilde\theta'\in\tilde\zeta'}
      \frac
        {2M_{\tilde\theta'}N_{\tilde\theta'}}
        {
          \left(x_{\tilde\zeta'}+\left(N_{\tilde\theta'}-1\right)i\right)^2
          +N_{\tilde\theta'}^2
        }
    }
\prod_{\tilde\theta\in\tilde\zeta}
\left[
  \left(-\right)^{M_{\bar\theta}-1}\left(M_{\bar\theta}-1\right)!
  M_{\bar\theta}^{-1}
  \left(
    M_{\bar\theta}!
  \right)^{N_{\bar\theta}-1}
\right].
\label{eq:moveing_of_integral_path_tmp_2}
\end{eqnarray}
Here, we have simplified the function
$\min_{(\sigma_1,\cdots)=\tilde\theta\in\tilde\zeta}
\left(\min\left(\sigma_1\right)\right)$ as $
{\min}_1(\tilde\zeta)$ where $\tilde\zeta$ is in
$\tilde\Theta(\theta)$ and $\theta$ is in $\Theta_M$.  In the change
from (\ref{eq:moveing_of_integral_path_tmp_1}) into
(\ref{eq:moveing_of_integral_path_tmp_2}), there is a one-to-many
correspondence between $\{\tilde\zeta,\lambda,\{\lambda'\}\}$ and
$\{\tilde\zeta,\lambda,\lambda'\}$ associated with
\begin{eqnarray}
&&
  \sum_{\theta\in\Theta_M}
  \sum_{\tilde\zeta\in\tilde\Theta\left(\theta\right)}
  \left[
    \prod_{\tilde\theta \in\tilde\zeta}\delta_>\left(\tilde\theta\right)
  \right]
  \sum_{\lambda\in\Lambda\left(\tilde\zeta\right)}
  \prod_{\left(\sigma_1,\cdots\right)=\tilde\theta\in\tilde\zeta}
  \left[
    \sum_{\tilde\zeta'\in D\left(\tilde\theta\right)}
    \sum_{\lambda'\in\Lambda_c\left(\tilde\zeta'\right)}
  \right]
\nonumber\\
&\leftrightarrow&
  \sum_{\theta\in\Theta_M}
  \sum_{\tilde\zeta\in\tilde{\bar\Theta}\left(\theta\right)}
  \sum_{\lambda\in\Lambda\left(\tilde\zeta\right)}
  \sum_{\lambda'\subseteq\lambda}.
\end{eqnarray}
Each $\tilde\zeta',\lambda_0,\{\lambda'\}$ in $\{\tilde\zeta,\lambda,\{\lambda'\}\}$ corresponds to a subset $A$ of
$\{\tilde\zeta,\lambda,\lambda'\}$, where
$\tilde\zeta',\lambda_0,\{\lambda'\}$ and
$\tilde\zeta'',\lambda'',\lambda'''$ in $A$ satisfy the following two
conditions.  First, by use of $\tilde\zeta''$ and $\lambda'''$,
$\tilde\zeta'$ and $\{\lambda'\}$ can be written as
\begin{eqnarray}
    \tilde\zeta'
&=&
\left\{\;\tilde\theta\;\left|\;\left(\sigma_k\right)=\left(\oplus_{
  (\sigma'_1,\cdots)=\tilde\theta'\in\tilde\zeta'''}\sigma'_k\right)
,\quad\tilde\zeta'''\in G_{\tilde\zeta''}\left(\lambda'''\right)
\;\right.\right\}
\nonumber\\
  \{\lambda'\}
&=&
  \left\{
    \lambda
  \left|
    \lambda=\lambda\left[\lambda''',\tilde\zeta'''\right],
     \tilde\zeta'''\in G_{\tilde\zeta''}\left(\lambda'''\right)
  \right.\right\}.
\end{eqnarray}
Second, $\lambda_0$ can be written as 
\begin{eqnarray}
  \lambda_0&=&
  \left\{
    \left\{\tilde\theta,\tilde\theta'\right\}
    \left|
    \tilde\theta =f\left(\tilde\theta'' \right),\;
    \tilde\theta'=f\left(\tilde\theta'''\right),\;
    \left\{\tilde\theta'',\tilde\theta'''\right\}\in\lambda''-\lambda'''    
  \right.\right\},
\end{eqnarray}
where $f$ is a onto-mapping from the set $\tilde\zeta''$ to the set
$\tilde\zeta'$ satisfying a condition: For any $\tilde\theta''\in
\tilde\zeta''$ there is a $\tilde\zeta\in
D\left(f(\tilde\theta'')\right)$ in which $\tilde\theta''$ exists.  It
is clear that this correspondence is a ``one-to-many correspondence''.

\subsection{}
We change the integral path of a variable $x_{\tilde\theta}$ in
(\ref{eq:moveing_of_integral_path_tmp_2})from
$(-\infty+i\min(\sigma_1)\delta,+\infty+i\min(\sigma_1)\delta)$ to
$(-\infty-(N_{\tilde\theta}-1)i,+\infty-(N_{\tilde\theta}-1)i)$. Then,
we get
\begin{eqnarray}
&&  (\ref{eq:Z_LM_integral_formulation(formal)})
\nonumber\\
&=&
  \sum_{\theta\in\Theta_M}
  \sum_{\tilde\zeta\in\tilde{\bar\Theta}\left(\theta\right)}
  \sum_{\lambda\in\Lambda\left(\tilde\zeta\right)}
  \left[
    \prod_{\tilde\theta\in\tilde\zeta}
    \int_{-\infty-iN_{\tilde\theta}\delta}
        ^{+\infty-iN_{\tilde\theta}\delta}
    \frac{dx_{\tilde\theta}}{2\pi}
  \right]
  \left[
    \prod_{\tilde\zeta'\in G_\lambda\left(\tilde\zeta\right)}
    \left(
      \sum_{\tilde\theta\in\tilde\zeta}
      \frac
        {2M_{\tilde\theta}N_{\tilde\theta}L}
        {
          \left(x_{\tilde\theta}+\left(N_{\tilde\theta}-1\right)i\right)^2
          +N_{\tilde\theta}^2
        }
    \right)
  \right]
\nonumber\\&&
  \left[
    \prod_{\{\tilde\theta,\tilde\theta'\}\in\lambda}
    \left(-M_{\tilde\theta}M_{\tilde\theta'}\right)
    K_{N_{\tilde\theta},N_{\tilde\theta'}}\left(
      x_{\tilde\theta}
      -x_{\tilde\theta}
      +\left(N_{\tilde\theta}-N_{\tilde\theta'}\right)i
    \right)
  \right]
\nonumber\\&&
  e^{
    -\beta
    \sum_{\tilde\theta\in\tilde\zeta}
    \frac
      {2M_{\tilde\theta}N_{\tilde\theta}}
      {
        \left(
          x_{\tilde\theta}+\left(N_{\tilde\theta}-1\right)i
        \right)^2
        +N_{\bar\theta}^2
      }
    }
\prod_{\tilde\theta\in\tilde\zeta}
\left[
  \left(-\right)^{M_{\tilde\theta}-1}\left(M_{\tilde\theta}-1\right)!
  M_{\tilde\theta}^{-1}
  \left(
    M_{\tilde\theta}!
  \right)^{N_{\tilde\theta}-1}
\right].
\label{eq:moveing_of_integral_path_tmp_3}
\end{eqnarray}
There are cancellations of terms corresponding to
$\tilde\zeta\not\in\tilde{\bar\Theta}(\theta)$ and all effects of
residues.  The equality of (\ref{eq:moveing_of_integral_path_tmp_2})
and (\ref{eq:moveing_of_integral_path_tmp_3}) is proved in
\ref{sec:proof_of_integral_path},

By further changing variables $x_{\tilde\theta}$ into
$x_{\tilde\theta}+(N_{\tilde\theta}-1)i$,
(\ref{eq:moveing_of_integral_path_tmp_3}) becomes
\begin{eqnarray}
&&  (\ref{eq:Z_LM_integral_formulation(formal)})
\nonumber\\
&=&
  \sum_{\theta\in\Theta_M}
  \sum_{\tilde\zeta\in\tilde{\bar\Theta}\left(\theta\right)}
  \sum_{\lambda\in\Lambda\left(\tilde\zeta\right)}
  \left[
    \prod_{\tilde\theta\in\tilde\zeta}
    \int_{-\infty}^{ \infty}
    \frac{dx_{\tilde\theta}}{2\pi}
  \right]
\nonumber\\&&
  \left[
    \prod_{\tilde\zeta'\in G_{\tilde\zeta}\left(\lambda\right)}
    \left(
      \sum_{\tilde\theta\in\tilde\zeta'}
      \frac
        {2M_{\tilde\theta}N_{\tilde\theta}L}
        {x_{\tilde\theta}^2+N_{\tilde\theta}^2}
    \right)
  \right]
  \left[
    \prod_{\{\tilde\theta,\tilde\theta'\}\in\lambda}
    \left(-M_{\tilde\theta}M_{\tilde\theta'}\right)
    K_{N_{\tilde\theta},N_{\tilde\theta'}}\left(
      x_{\tilde\theta}-x_{\tilde\theta'}
    \right)
  \right]
\nonumber\\&&
  e^{
    -\beta\sum_{\tilde\theta\in\tilde\zeta}
    \frac
      {2M_{\tilde\theta}N_{\tilde\theta}}
      {x_{\tilde\theta}^2+N_{\tilde\theta}^2}
  }
  \prod_{\tilde\theta\in\tilde\zeta}
  \left[
    \left(-\right)^{M_{\tilde\theta}-1}\left(M_{\tilde\theta}-1\right)!
    M_{\tilde\theta}^{-1}
    \left(
      M_{\tilde\theta}!
    \right)^{N_{\tilde\theta}-1}
  \right].
\label{eq:Z_LM_integral_formulation_4}
\end{eqnarray}

\subsection{}
\begin{itemize}
\item definition: $\bar\Theta(\theta)$

  Let $\theta$ be a set which has a finite number of sets as
  elements.  $\bar\Theta(\theta)$ consists of all elements
  $\zeta$ satisfying the following two conditions. First,
  $\zeta$ is in $\Theta(\theta)$. Second, any set $\theta'$
  in $\zeta$ satisfies the condition that all sets as elements
  in the set $\theta'$ have the same number of elements. Then,
  \begin{eqnarray}
    {\bar\Theta}(\theta)&=&\left\{\;
      \zeta\in\Theta(\zeta)\;|\;
      N_\sigma=N_{\sigma'}, \quad\sigma,\sigma'\in\theta'\in\zeta
    \right\}.
  \end{eqnarray}
\end{itemize}

We rewrite the expression of summations in
(\ref{eq:Z_LM_integral_formulation_4}) with respect to permutations.
Then, eq.(\ref{eq:Z_LM_integral_formulation_4}) becomes
\begin{eqnarray}
&&  (\ref{eq:Z_LM_integral_formulation(formal)})
\nonumber\\
&=&
  \sum_{\theta\in\Theta_M}
  \left[
    \prod_{\sigma\in\theta}
    N_{\sigma}!
  \right]
  \sum_{\zeta\in\bar\Theta\left(\theta\right)}
  \left[
    \prod_{\theta'\in\zeta}
    \left(-\right)^{N_{\theta'}-1}\left(N_{\theta'}-1\right)!
  \right]
  \left[
    \prod_{\theta'\in\zeta}
    \int_{-\infty}^{ \infty}
    \frac{dx_{\theta'}}{2\pi}
  \right]
  \left[
    \prod_{\theta'\in\zeta}
    N_{\theta'}^{-1}
  \right]
\nonumber\\&&
  \sum_{\lambda\in\Lambda\left(\zeta\right)}
  \left[
    \prod_{\zeta'\in G_{\zeta}\left(\lambda\right)}
    \left(
      \sum_{\theta'\in\zeta'}
      \frac
        {2N_{\theta'}M_{\theta'}L}
        {x_{\theta'}^2+M_{\theta'}^2}
    \right)
  \right]
  \left[
    \prod_{\{\theta',\theta''\}\in\lambda}
    \left(-N_{\theta'}N_{\theta''}\right)
    K_{M_{\theta'},M_{\theta''}}\left(
      x_{\theta'}-x_{\theta''}
    \right)
  \right]
\nonumber\\&&
    e^{
      -\beta\sum_{\theta'\in\zeta}
      \frac
        {2N_{\theta'}M_{\theta'}}
        {   x_{\theta'}^2+M_{\theta'}^2}
    }.
\label{eq:Z_LM_integral_formulation_5}
\end{eqnarray}

In this rewriting, we observe a many-to-many correspondence between
$\{\tilde\zeta,\lambda\}$ and $\{\zeta,\lambda\}$ associated with
\begin{eqnarray}
  \sum_{\theta\in\Theta_M}
  \sum_{\tilde\zeta\in\tilde{\bar\Theta}\left(\theta\right)}
  \sum_{\lambda\in\Lambda\left(\tilde\zeta\right)}
&\leftrightarrow&
  \sum_{\theta\in\Theta_M}
  \sum_{\zeta\in\bar\Theta\left(\theta\right)}
  \sum_{\lambda\in\Lambda\left(\zeta\right)}.
\end{eqnarray}
A subset $A$ of $\{\tilde\zeta,\lambda\}$ corresponds to a subset $B$
of $\{\zeta,\lambda\}$, where $\tilde\zeta,\lambda$ in $A$ and
$\zeta',\lambda'$ in $B$ satisfy the following conditions: There is a
bijection $f$ from the set $\tilde\zeta$ to the set $\zeta'$ which
satisfies
\begin{eqnarray}
  \bigoplus_{\sigma\in\tilde\theta}\sigma
\;=\;
  \bigoplus_{\sigma'\in\theta'}\sigma',
\quad
  N_{\tilde\theta}
\;=\;
  M_{\theta'}, 
\end{eqnarray}
where $f(\tilde\theta)=\theta'$.  And, $\{ \tilde\theta' ,
\tilde\theta'' \}$ is in $\lambda $, if and only if
$\{f(\tilde\theta'),f(\tilde\theta'')\}$ is in $\lambda'$.  It is
clear that this correspondence is a ``many-to-many correspondence''.

We newly introduce three functions,
\begin{eqnarray}
  \mu\left(\hat0_\theta,\zeta\right)
&\equiv&
  \prod_{\theta'\in\zeta}
  \left(-\right)^{N_{\theta'}-1}\left(N_{\theta'}-1\right)!,
\\
  E\left(\zeta\right)
&\equiv&
  \sum_{\theta'\in\zeta}
  \frac{2N_{\theta'}M_{\theta'}}{x_{\theta'}^2+M_{\theta'}^2}
\label{eq:energy_of_string_center}
\end{eqnarray}
and
\begin{eqnarray}
\left(2\pi\right)^{N_\zeta}
 \left|\frac{\partial I}{\partial x}\right|_{L,\zeta}
&\equiv&
  \left[
    \prod_{\theta'\in\zeta}
    N_{\theta'}^{-1}
  \right]
  \sum_{\lambda\in\Lambda\left(\zeta\right)}
  \left[
    \prod_{\zeta'\in G_{\zeta}\left(\lambda\right)}
    \left(
      \sum_{\theta'\in\zeta'}
      \frac
        {2N_{\theta'}M_{\theta'}L}
        {x_{\theta'}^2+M_{\theta'}^2}
    \right)
  \right]
\nonumber\\&&
  \left[
    \prod_{\{\theta',\theta''\}\in\lambda}
    \left(-N_{\theta'}N_{\theta''}\right)
    K_{M_{\theta'},M_{\theta''}}\left(
      x_{\theta'}-x_{\theta''}
    \right)
  \right],
\label{eq:d_m_string_center_equation}
\end{eqnarray}
where $\theta\in\Theta_M$ and $\zeta\in\bar\Theta(\theta)$. In terms of 
these functions, the expression (\ref{eq:Z_LM_integral_formulation_5})
can be simplified as
\begin{eqnarray}
M!Z_{M}e^{-hM}&=&
  \sum_{\theta\in\Theta_M}
  \left[
    \prod_{\sigma\in\theta}
    N_{\sigma}!
  \right]
  \sum_{\zeta\in\bar\Theta\left(\theta\right)}
  \mu\left(\hat0_\theta,\zeta\right)
  \left[
    \prod_{\theta'\in\zeta}
    \int_{-\infty}^{ \infty}
    dx_{\theta'}
  \right]
  \left|\frac{\partial I}{\partial x}\right|_{L,\zeta}
    e^{-\beta E\left(\zeta\right)}.
\label{eq:Z_LM_integral_formulation_final}
\end{eqnarray}
In this way, we have proved the equivalence of
(\ref{eq:Z_LM_integral_formulation(formal)}) and
(\ref{eq:Z_LM_integral_formulation_final}).  Remark that
(\ref{eq:Z_LM_integral_formulation_final}) is the expression which we
suppose $Z_M$ to be in the previous paper~\cite{go_xxx}.

\section{A relation between $Z_M$ and the string hypothesis}
The function (\ref{eq:d_m_string_center_equation}) is the Jacobian
 between $\{x_\theta\}$ and $\{I_\theta\}$
defined by relations,
\begin{eqnarray}
  \left[
    \frac{x_{\theta'}+M_{\theta'} i}{x_{\theta'}-M_{\theta'} i}
  \right]^{L}
&=&
  e^{-2\pi i I_{\theta'}}
  \prod_{{\theta''}\in\zeta,\neq{\theta'}}
  E_{M_{\theta'},M_{\theta''}}\left(x_{\theta'}-x_{\theta''}\right)^{N_{\theta''}},
\label{eq:m_string_center_eqation}
\end{eqnarray}
where $\theta\in\Theta_M$, $\zeta\in\bar\Theta(\theta)$ and
\begin{eqnarray}
  E_{n,m}\left(x\right)
&\equiv&
  \frac
    {\left[x-\left(n+m\right)i\right]\left[x-\left|n-m\right|i\right]}
    {\left[x+\left(n+m\right)i\right]\left[x+\left|n-m\right|i\right]}
  \prod_{k=1}^{\min\left(n,m\right)}
  \left[
  \frac
    {x+\left(n+m-2k\right)i}
    {x-\left(n+m-2k\right)i}
  \right]^2.
\end{eqnarray}
We emphasize the following:  We regard all the variable
$\{I_\theta\}$ as integer and interpret the value $x_{\theta'}$ as an
$M_{\theta'}$-string center. The relations
(\ref{eq:m_string_center_eqation}) are the equations which are
introduced by the string hypothesis in case of any
$N_{\theta'\in\zeta}=1$. And, each term of 
(\ref{eq:energy_of_string_center}) is the energy
corresponding to $M_{\theta'}$-string in the string hypothesis.

We study eq. (\ref{eq:Z_LM_integral_formulation_final}) again. As we
have shown in \cite{go_xxx}, the expression
(\ref{eq:Z_LM_integral_formulation_final}) can be derived by summing
up the string center equation formally. The derivation is summarized
as follows.  From the string hypothesis, the free energy is written as
\begin{eqnarray}
M!Z_{M}e^{-hM}&=&
  \sum_{\theta\in\Theta_M}
  \left[
    \prod_{\sigma\in\theta}
    N_{\sigma}!
  \right]
  \sum_{\zeta\in\bar\Theta\left(\theta\right)}
  \mu\left(\hat0_\theta,\zeta\right)
  \sum_{\{I_\theta\}}
    e^{-\beta E\left(\zeta\right)},\label{eq:tmp_2002_10_31_1}
\end{eqnarray}
where $\sum_{\{I_\theta\}}$ means a summation over all the
real number solutions of (\ref{eq:m_string_center_eqation}) on
condition that $\{ I_{\theta}\}$ are integers.  Note that the
coefficients are due to the symmetry of the quasi-particles. Using a
relation
\begin{eqnarray}
  \sum e^{-\beta E\left(\zeta\right)}
&=&
  \prod_{\theta'\in\zeta}\left(\int_{-\infty}^\infty dx_{\theta'}\right) e^{-\beta E\left(\zeta\right)}
\label{eq:replace_keishiki_st}
\end{eqnarray}
in the thermodynamic limit, (\ref{eq:Z_LM_integral_formulation_final})
can be shown from (\ref{eq:tmp_2002_10_31_1}). This explains how the string hypothesis leads to the
correct result.

In this derivation, we use
\begin{eqnarray}
  \lim\sum_{\{n_i\}} f\left(\{x_j\left(\{n_i\}\right)\}\right)
&=&
\int \left|\frac{dn}{dx}\right|f\left(\{x_j\}\right)\prod dx_j
\label{eq:replace_keishiki}
\end{eqnarray}
where the limit means the thermodynamic limit, and the integral path
in the r.h.s. is the region containing points $\{x_j\}$ summed up in
the l.h.s.  A simple example of  (\ref{eq:replace_keishiki}) is
\begin{eqnarray}
  \lim\sum_{n_1=-\infty}^\infty
\cdots  \sum_{n_1=-\infty}^\infty
 \exp\left(-\beta \sum_{i=1}^N k_i ^2\right)
&=&
\prod_{i=1}^N \left(\int_{-\infty}^{\infty}
 \frac{Ldk_i}{2\pi}\right)
\exp\left(-\beta \sum_{i=1}^N k_i ^2\right)
\\
&&2\pi n_i=k_iL,
\nonumber
\end{eqnarray}
which is a partition function of the free particles. We have used this
modification from (\ref{eq:Z_LM_summation_formulation(formal)}) to
(\ref{eq:Z_LM_integral_formulation(formal)}).

There remains a problem with this derivation of
(\ref{eq:Z_LM_integral_formulation_final}) from the string hypothesis.
When we use eq.(\ref{eq:replace_keishiki}), a condition is required.
That is, the orientation of the integral path in the r.h.s. of
(\ref{eq:replace_keishiki}) has the following property: the value of an
integral $\int \left|\frac{dn}{dx}\right|\prod dx_j$ for any part of
the integral path is positive.  Note that we have defined the
orientation of the integral path
(\ref{eq:Z_LM_integral_formulation(formal)}) by this condition.  The
integral path in (\ref{eq:replace_keishiki_st}), however, does not
have the property.  The value of the integral is $O(L^{M})$ and the
difference between the r.h.s. of (\ref{eq:replace_keishiki_st}) and the
value of the integral using the truly oriented integral path is
$O(L^{M-1.5})$ where $M$ is the number of up-spins. A further
research is required to conclude whether this causes the difference in
the thermodynamic quantities or not.

Note that in \cite{go_xxx}, we do not claim that the derivation of
(\ref{eq:Z_LM_integral_formulation_final}) using the string hypothesis
is complete, but claim that (\ref{eq:Z_LM_integral_formulation_final})
is related to the string hypothesis and the free energy is properly
derived when we assume the relation
(\ref{eq:Z_LM_integral_formulation_final}). On the contrary, in this
paper, we have derived (\ref{eq:Z_LM_integral_formulation_final}) step
by step only assuming the Bethe ansatz equations.

\section{Conclusion}
In this paper, we have calculated ${\rm Tr}e^{-\beta H_M}$ for the XXX Heisenberg model, which is the
trace of the Boltzmann weight under the restriction that the number of
up-spin $M$ is fixed. This method relies only on the Bethe ansatz
equations.  Using this method and the result in \cite{go_xxx}, we have
obtained the free energy, whose expression is perfectly agree with TBA.
In a sense, we have generalized the direct method or the Bethe ansatz cluster expansion method~\cite{go_b,go_d,go_e} into models with bound states. We emphasize that this derivation of the free energy is independent of
QTM, TBA and is free from the string hypothesis. We can replace the Boltzmann
weight with some other functions in case that the Boltzmann weight and
the functions have the same analyticity.  Therefore, it may be
possible to calculate some other thermodynamic quantities $<A>$ in the
same way by replacing the Boltzmann weight $e^{-\beta H}$ with $Ae^{-\beta H}$.

\appendix

\def\thesubsection{\Alph{section}. \arabic{subsection}}

\section{Path-independence of  the  integrals}
\label{app:independency_of_integral_path}
We prove that the value of the integral (\ref{eq:tmp_app_b_4}) along
the path (\ref{eq:tmp_app_b_03}) does not depend on $A_\sigma$ in case
all $A_\sigma$ are finite positive numbers, where $E_\theta$ is
defined by (\ref{eq:statistical_energy}) and the relation between
$x_\sigma$ and $I_\sigma$ is defined by (\ref{eq:statistical_Bethe}).
In other words, when we suppose that $\{A_\sigma\}$ are continuous
functions with respect to $0\leq t\leq1$ and none of
$\{A_\sigma(t)\in\mathbb{R}_{>0}\}$ is equal to $\infty$, the
integrals for the paths specified by $\{A_\sigma(0)\}$ and
$\{A_\sigma(1)\}$ have the same value.

To prove this, we define the integral (\ref{eq:tmp_app_b_4}) more
precisely.  The reason is that, there is some ambiguity in the
treatment of the region where the relation between $\{x_\sigma\}$ and
$\{I_\sigma\}$ is not analytic.  We define the integral path
(\ref{eq:tmp_app_b_03}) and the integrand of (\ref{eq:tmp_app_b_4}) as
follows.  We denote by $\Omega$ a metric space
$(y_1\in\mathbb{C},y_2,\cdots)$ which is topologically equivalent to a
direct product of one-point compactified complex plane $\otimes
S^2$. We need not specify the distance function definitely.  Let
$\Omega_I$ be a subspace of $\Omega$.  The points on $\Omega_I$ are
written as $(\{x_\sigma\},\{e^{2\pi I_\sigma i}\})$, where they
satisfy the relations (\ref{eq:statistical_Bethe}), $x_\sigma\neq\pm
i,\infty$ and $x_{\sigma'}-x_{\sigma''}\neq\pm2i$.  Here, we regard
$\{x_\sigma\}$ as $x_\sigma,x_{\sigma'},x_{\sigma''},\cdots$ where
${\sigma},{\sigma'},{\sigma''},\cdots\in\theta $. We define
$\bar\Omega_I$ as a subspace of the closure of $\Omega_I$ in $\Omega$
where $ |\Im I_\sigma|$ is not equal to $\infty$.  The topological
space $\partial \Omega_I $ is defined to be a subset of $\bar\Omega_I$
where $\{x_\sigma\}$ as a function of $\{I_\sigma\}$ is
non-analytic or the integrand is non-analytic with respect to
$\{I_\sigma\}$. The integral path (\ref{eq:tmp_app_b_03}) represents
the space $\bar\Omega_I$ under the restriction $|e^{2\pi I_\sigma
  i}|=A_\sigma$.  The orientation of the integral path is defined so
that the value of an integral $\int \prod dI_\sigma$ for any part of
the integral path is positive.  We shall 
prove in \ref{sec:dim_of_space} the the following two facts : 1) The dimension of $\partial
\Omega_I$ under the restriction $|e^{2\pi I_\sigma i}|=A_\sigma$ is
not more than $N_\theta-1$. 2) The dimension of $\partial \Omega_I $
is not more than $2N_\theta-2$. Here, $2N_\theta$ is the dimension of
the space $\Omega_I$.  The first indicates that the orientation is
defined for all the region on the space $\bar\Omega_I$ under the
restriction $|e^{2\pi I_\sigma i}|=A_\sigma$, Therefore, the space can
be considered as an oriented manifold or $N_\theta$-chain. Here and
hereafter, we identify the symbol $\{A_\sigma\}$ with the
$N_\theta$-chain.  And, the second indicates that the boundary
$\partial\{A_\sigma\}$ is equal to zero as $N_\theta-1$-chain.  In
\ref{sec:bounded} we shall prove that 3) the integrand of
(\ref{eq:tmp_app_b_4}) converges to a finite value in the limit on any
sequence of points on $\Omega_I$ which approaches to a point on
$\bar\Omega_I$.  Then, we define the value of the integrand of
(\ref{eq:tmp_app_b_4}) as the limit at any point on $\bar\Omega_I$.
Thus, the integral (\ref{eq:tmp_app_b_4}) has been defined.

Next, we define three $N_\theta$-chains
$\{A''_\sigma(t,\delta,\delta')\}$, $\{A'_\sigma(t,\delta)\}$ and
$\Delta A''(\delta,\delta')$ embedded in $\Omega_I$ as integral paths
of (\ref{eq:tmp_app_b_4}).  The $\{A'_\sigma(t,\delta)\}$ is
continuously changed by $0\leq t\leq1$ and $0\leq\delta<<1$.  We may
choose so that $\{A_\sigma(t)\}-\{A'_\sigma(t,0)\}$ is equal to zero,
where we regard $\{A_\sigma(t)\}$ as an $N_\theta$-chain.  The
dimension of the region $\{A'_\sigma(t,\delta)\}\cap \partial \Omega_I
$ is not more than $N_\theta-2$ for any $t$, in case of
$\delta\neq0$. And, $\partial\{A'_\sigma(t,\delta)\}$ is equal to
zero.  The existence of the $N_\theta$-chain $\{A'_\sigma(t,\delta)\}$
is assured by 1) and the fact $\partial\{A_\sigma(t)\}=0$.  Note that
the expression $\{A'_\sigma(t,\delta)\}$ is merely a symbol of
$N_\theta$-chain and does not indicate that the integral path
satisfies the condition (\ref{eq:tmp_app_b_03}).
$\{A''_\sigma(t,\delta,\delta')\}$ is a part of
$\{A'_\sigma(t,\delta)\}$ where the distance between all of the points
and $\partial \Omega_I $ is more than $\delta'$.  We can also regard
$\{A''_\sigma(t,\delta,\delta')\}$ as an $N_\theta+1$-chain depending
on $\delta$ and $\delta'$ by regarding $0\leq t\leq1$ as a variable.
And, we define the orientation of the $N_\theta+1$-chain so that
$\partial\{A''_\sigma(t,\delta,0)\}$ is equal to
$\{A''_\sigma(1,\delta,0)\}-\{A''_\sigma(0,\delta,0)\}$.  Then,
$\Delta A''(\delta,\delta')$ is defined to be
$\partial\{A''_\sigma(t,\delta,\delta')\}-\{A''_\sigma(1,\delta,\delta')\}+\{A''_\sigma(0,\delta,\delta')\}$.

Finally, we evaluate the difference between the values of the integral
(\ref{eq:tmp_app_b_4}) for the integral paths, $\{A_\sigma(0)\}$ and
$\{A_\sigma(1)\}$.  The fact 3) indicates that the
integrals along the paths $\{A'_\sigma(t,+0)\}$ and $\{A_\sigma(t)\}$
are the same, and  the integrals along the paths
$\{A''_\sigma(t,\delta,+0)\}$ and $\{A'_\sigma(1,\delta)\}$ are the
same.  The difference between the values of the integral for the paths
$\{A''_\sigma(0,\delta,\delta')\}$ and
$\{A''_\sigma(1,\delta,\delta')\}$ is the integral along
the path $\Delta A''(\delta,\delta')$ when $\delta'>0$, because
(\ref{eq:tmp_app_b_4}) is a multiple complex integral and all the
point on $\{A_\sigma(t,\delta,\delta')\}$ is regular.  The integral
for the path $\Delta A''(\delta,+0)$ is equal to zero when $\delta>0$,
because the dimension of $\Delta A''(\delta,+0)$ is $N_\sigma-1$ and
the integrand is finite.  Thus, the values of the integral
(\ref{eq:tmp_app_b_4}) do not depend on the integral paths,
$\{A_\sigma(0)\}$ and $\{A_\sigma(1)\}$.

\subsection{}
\label{sec:dim_of_space}
We prove that the dimension of $\partial \Omega_I$, which is defined
in \ref{app:independency_of_integral_path}, is not more than
$2N_\theta-2$ where $2N_\theta$ is the dimension of $\Omega_I$.  As a
corollary, the number of solution $\{x_\sigma\}$ of eqs.
(\ref{eq:statistical_Bethe}) does not depend on $\{e^{I_\theta}\}$
when $\{e^{I_\theta}\}$ is in $\bar\Omega_I-\partial\Omega_I$. We also
prove that the dimension of the non-analytic region is not more than
$N_\theta-1$ in any subspace $|e^{I_\theta}|=$constant. As a
corollary, the dimension of $\bar\Omega_I$ under the restriction
$|e^{I_\theta}|=$constant is $N_\theta$.

The first is proved by mathematical induction as follows. 

In case of $N_\theta=1$, $\partial\Omega_I$ contains only one point
$(x_\sigma=\infty,e^{I_\sigma}=1)$.  Then, the dimension of
$\partial\Omega_I$ is zero.

There are two sufficient conditions when any point is in
$\partial\Omega_I$.  One is that $\{x_\sigma\}$ regarded as
a function of $\{I_\sigma\}$ is not analytic. The other is
that the integrand is not analytic when we regard the argument of the
function as $\{x_\sigma\}$.  The inverse mapping theorem indicates
that one of the following three relations holds when the first condition
holds. One is that the Jacobian (\ref{eq:def_jacobian_n}) is equal to
$0$, and the other are $x_\sigma=\pm i, \infty$ or
$x_{\sigma'}-x_{\sigma''}=\pm 2i$.  The second relation is equivalent
to the second condition.  Therefore, we shall show that the dimension
of the space which satisfies the previous three relations is not more
than $2N_\theta-2$.  Note that, in \ref{sec:singurality} we show the
following fact.  In case $\{x_{\sigma}\},\{e^{I_\sigma}\}$ are in
$\partial\Omega_I$, the condition that some variables satisfy
$x_\sigma=\pm i$ or $x_{\sigma'}-x_{\sigma''}=\pm 2i$ is equivalent to
the condition that there are subsets $\theta_\pm\subseteq\theta$ which
satisfy $x_{\sigma_\pm}=\pm i$, $x_{\sigma_\pm}-x_{\sigma'}\neq\pm
2i$, $x_{\sigma'}\neq\pm i,\infty$ and
$x_{\sigma'}-x_{\sigma''}\neq\pm 2i$ where $\sigma',\sigma''\in
\theta-\theta_+-\theta_-$ and $\sigma_\pm\in \theta_\pm$.

We divide the space $\partial\Omega_I$ into three subspaces: The first
subspace is $\partial\Omega_I$ under the restrictions $x_{\sigma\in
  \theta'}= \infty$ and $x_{\sigma\not\in \theta'}\neq\infty$ where
$\theta'\subseteq\theta$, the second under the restriction,
$x_{\sigma}\neq\infty$, $x_{\sigma_\pm\in\theta\pm}=\pm i$,
$x_{\sigma\not\in\theta\pm}\neq\pm i$ where
$\theta_\pm\subseteq\theta$, and  the third under the restriction
$x_\sigma\neq\pm i, \infty$ and $x_{\sigma'}-x_{\sigma''}\neq\pm 2i$.

In case that $\{x_{\sigma}\},\{e^{I_\sigma}\}$ are in the first
subspace, eq.(\ref{eq:statistical_Bethe}) are reduced to
\begin{eqnarray}
  \prod_{\sigma'\in\theta',\neq\sigma}
  \left[\frac{x_\sigma-x_{\sigma'}+2i}{x_\sigma-x_{\sigma'}-2i}\right]^{N_{\sigma'}}
&=&
  e^{2\pi i I_\sigma}
,
  \quad\quad\sigma\in\theta',
\label{eq:tmp021024-01}\\
  \left[\frac{x_\sigma-i}{x_\sigma+i}
\right]^{L}
  \prod_{\sigma'\in\theta-\theta',\neq\sigma}
  \left[\frac{x_\sigma-x_{\sigma'}+2i}{x_\sigma-x_{\sigma'}-2i}\right]^{N_{\sigma'}}
&=&
  e^{2\pi i I_\sigma}
,
  \quad\quad\sigma\in\theta-\theta'.
\label{eq:tmp021024-02}
\end{eqnarray}
Eq.(\ref{eq:tmp021024-01}) requires a condition
$\sum_{\sigma\in\theta'}N_\sigma I_\sigma\in \mathbb{Z}$.  And, eq.(\ref{eq:tmp021024-02}) is the Bethe equation with respect to $N_\sigma-1$
variables.  These indicate that the dimension of the first
subspace is not more than than $2N_\theta-2$.

In case that $\{x_{\sigma}\},\{e^{I_\sigma}\}$ are in the second
subspace, eq.(\ref{eq:statistical_Bethe}) becomes
\begin{eqnarray}
&  \prod_{\sigma\in\theta-\theta_+-\theta_-}
  \left[\frac{
      \left(x_\sigma+3i\right)
      \left(x_\sigma+ i\right)
      }{
      \left(x_\sigma-3i\right)
      \left(x_\sigma-i\right)
      }\right]^{N_\sigma\sum_{\sigma'\subseteq\theta_+}N_\sigma}
\quad=\quad
e^{
2\pi i \sum_{\sigma\in\theta_++\theta_-}N_\sigma I_\sigma
}
\label{eq:tmp021024-03}\\
&  \left[\frac{x_\sigma-i}{x_\sigma+i}
\right]^{L}
  \prod_{\sigma'\in\theta,\neq\sigma}
  \left[\frac{x_\sigma-x_{\sigma'}+2i}{x_\sigma-x_{\sigma'}-2i}\right]^{N_{\sigma'}}
  \prod_{\sigma'\in\theta_+}
  \left[\frac{x_\sigma+i}{x_\sigma-3i}\right]^{N_{\sigma'}}
  \prod_{\sigma'\in\theta_-}
  \left[\frac{x_\sigma+3i}{x_\sigma-i}\right]^{N_{\sigma'}}
\quad=\quad
  e^{2\pi i I_\sigma}
,\label{eq:tmp021024-04}\\
&  \makebox[4cm]{}\sigma\in\theta-\theta_+-\theta_-
\nonumber
\end{eqnarray}
The first equation is obtained as follows.  We raise both sides of
eq.(\ref{eq:statistical_Bethe}) with respect to
$\sigma\in\theta_++\theta_-$ to the power of $N_\sigma$.  Then, the
product of these equations is the first equation in case of
\begin{eqnarray}
  \lim\frac
{\prod_{\sigma\in\theta_+}\left(x_\sigma-i\right)^{N_\sigma}}
{\prod_{\sigma\in\theta_-}\left(x_\sigma+i\right)^{N_\sigma}}
&=&1,
\label{eq:tmp_app_b_100}
\end{eqnarray}
where $\lim$ means a limit on a sequence of points on $\Omega_I$
converging to a point where $x_{\sigma{\pm} \in\theta_{\pm}}=\pm i$,
$x_{\sigma\not\in\theta_{\pm}}\neq\pm i$ and $|\Im
I_\sigma|\neq\infty$.  (\ref{eq:tmp_app_b_100}) is proved in
\ref{sec:limit_relation}. We define a projection from $\bar\Omega_I$
onto the $4N_{\theta-\theta_+-\theta_-}$-dimensional space
$(\{x_{\sigma\in \theta-\theta_+-\theta_-}\},\{e^{I_{\sigma\in
    \theta-\theta_+-\theta_-}}\})$. The projection is the elimination
of the $2N_{\theta_++\theta_-}$ variables $\{x_{\sigma\in
  \theta_++\theta_-}\},\{e^{I_{\sigma\in \theta_++\theta_-}}\}$.
Then, using the projection, we define two spaces; 1) the union of the second subspace and the
inverse image of a point, 2) the direct image of the second subspace.
Eq.(\ref{eq:tmp021024-03}) indicates that the dimension of the union
is not more than $2N_{\theta_++\theta_-}-2$. Eq.(\ref{eq:tmp021024-04})
indicates that the dimension of the direct image is not more than
$2N_{\theta-\theta_+-\theta_-}$.

The dimension of the third subspace is not more than the dimension of
$\{x_\sigma\}$ in $\Omega_I$ where the Jacobian is equal to $0$, that
is $2N_\theta-2$, because $\{e^{I_\sigma}\}$ regarded as a function of
$\{x_{\sigma'}\}$ are injections in this subspace.

Thus, we have proved that the dimension of $\partial\Omega_I$ is not
more than $2N_\theta-2$.

The second fact for the non-analytic region is proved by re-evaluating the dimension of the three
subspace under the restriction $|e^{I_\sigma}|$=constant.

\subsection{}
\label{sec:bounded}
We prove that the integrand of (\ref{eq:tmp_app_b_4}) converges in the
limit on any sequence of points on $\Omega_I$ which converges to a
point on $\bar\Omega_I$.  The sufficient condition is that
the integrand on any sequences of points satisfying the following
condition is convergent.
The sequence of points converges
to a point where some of $x_\sigma$ are equal to $i$ or $-i$.

Then, all we have to prove is that the equation
\begin{eqnarray}
   \lim\left[
     \sum_{\sigma\in\theta_+}\frac{N_\sigma}{x_\sigma-i}
     -\sum_{\sigma\in\theta_-}\frac{N_\sigma}{x_\sigma+i}
   \right]
&=&0
\label{eq:tmp_app_b_5}  
\end{eqnarray}
holds. Here, $\lim$ means a sequence on $\Omega_I$ converging to
a point where $x_{\sigma_\pm\in\theta_\pm}=\pm i$ and
$x_{{\sigma}\not\in\theta_\pm}\neq\pm i$ and $|\Im I_\sigma| \neq
\infty$.  This relation is proved in \ref{sec:limit_relation}.

\subsection{}
\label{sec:singurality}
We prove the following: In case $\{x_{\sigma}\},\{e^{I_\sigma}\}$
are in $\partial\Omega_I$, the condition that some variables satisfy
$x_\sigma=\pm i$ or $x_{\sigma'}-x_{\sigma''}=\pm 2i$ is equal to the
condition that there are subsets $\theta_\pm\subseteq\theta$ which
satisfies $x_{\sigma_\pm}=\pm i$, $x_{\sigma_\pm}-x_{\sigma'}\neq\pm
2i$, $x_{\sigma'}\neq\pm i,\infty$ and
$x_{\sigma'}-x_{\sigma''}\neq\pm 2i$ where $\sigma',\sigma''\in
\theta-\theta_+-\theta_-$ and $\sigma_\pm\in \theta_\pm$.

In case $x_{\sigma'}-x_{\sigma''}= 2i$, 
eq.(\ref{eq:statistical_Bethe}) with respect to $\sigma'$, $\sigma''$
and a condition $|\Im (I_{\sigma'})|,|\Im (I_{\sigma''})|<\infty$
require that $x_{\sigma'}$ is equal to $i$ or $x_{\sigma'''}-2i$ and
$x_{\sigma''}$ is equal to $-i$ or $x_{\sigma''''}+2i$. This result and
the fact that the number of the variables is finite lead to
$x_{\sigma'}=i$ and $x_{\sigma''}=-i$

\subsection{}
\label{sec:limit_relation}
We prove two relations,
\begin{eqnarray}
   \lim\left[
     \sum_{\sigma\in\theta_+}\frac{N_\sigma}{x_\sigma-i}
     -\sum_{\sigma\in\theta_-}\frac{N_\sigma}{x_\sigma+i}
   \right]
&=&0,
  \\
  \lim
  \frac{\prod_{\sigma\in\theta_+}\left(x_{\sigma}-i\right)^{N_\sigma}}
       {\prod_{\sigma\in\theta_-}\left(x_{\sigma}+i\right)^{N_\sigma}}
&=&1,
\end{eqnarray}
where $\lim$ means a limit on a sequence of points on $\Omega_I$
converging to a point where $x_{\sigma_\pm\in\theta_\pm}=\pm i$,
$x_{{\sigma}\not\in\theta_\pm}\neq\pm i$, $|\Im I_\sigma| \neq \infty$
and $\theta_\pm \subseteq \theta$.

To prove them, we define two topological spaces $\Omega_{Ip}$ and
$\partial\Omega_{Ip}$.  They are defined by subsets $\theta_\pm$.
The topological space $\Omega_{Ip}$ is a subspace
on which all the points can be written as $(\{x_\sigma\},$
$\{e^{I_\sigma}\},$ $\{\delta/\delta'\},$
$\{\delta^{-1}-\delta^{\prime -1}\})$ under the restrictions
$x_{\sigma}\neq \pm i,\infty,x_{\sigma'}\pm2i$, $|\Im I_\sigma|\neq
\infty$, and (\ref{eq:statistical_Bethe}), where
$\delta,\delta'=x_{\sigma\in\theta_\pm}\mp i$.  The topological space $\partial\Omega_{Ip}$ is
a subspace of the closure of $\Omega_{Ip}$, and a point in the closure
is in $\partial\Omega_{Ip}$ if and only if the point satisfies
$x_{\sigma_\pm\in\theta_\pm}=\pm i$,
$x_{\sigma\not\in\theta_\pm}\neq\pm i$ and $|\Im I_\sigma|\neq
\infty$.

A sufficient condition of these relation is that there is a set
$\zeta\in\Theta(\theta_++\theta_-)$ any element $\theta'$ of which
satisfies
\begin{eqnarray}
  \lim\frac1{x_{\sigma_+}-i}-\frac1{x_{\sigma_-}+i}
&=&
  0
\quad\quad \sigma_+\in\theta'_+,\sigma_-\in\theta'_-,
\nonumber\\
  \sum_{\sigma\in\theta'\cap\theta_+'}N_\sigma
&=&
  \sum_{\sigma\in\theta'\cap\theta_-'}N_\sigma,
\label{eq:definition_zeta}
\end{eqnarray}
where $\theta'_\pm=\theta'\cap\theta_\pm$ and $\lim$ means a limit on
an sequence of points on $\Omega_{Ip}$ which converges to a point on
$\partial\Omega_{Ip}$. The reason is that, any sequence on $\Omega_I$
converging to a point where $x_{\sigma_\pm\in\theta_\pm}=\pm i$ and
$x_{{\sigma}\not\in\theta_\pm}\neq\pm i$ and $|\Im I_\sigma| \neq
\infty$ corresponds to a sequence on $\Omega_{Ip}$ converging to a
subspace of a compact space in $\partial\Omega_{Ip}$.

First, we introduce an equivalence relation on a set $\theta_++\theta_-$,
\begin{eqnarray}
  \lim\frac1{x_{\sigma_\pm}\mp i}-\frac1{x_{\sigma'_\pm}\mp i}
\;=\;0,&{\rm then}\quad\sigma_\pm\sim\sigma'_\pm
 \nonumber\\
 \lim\frac1{x_{\sigma_\pm}\mp i}-\frac1{x_{\sigma'_\mp}\pm i}
\;=\;0,&{\rm then}\quad\sigma_\pm\sim\sigma'_\mp
\end{eqnarray}
where $\sigma_\pm,\sigma'_\pm\in\theta_\pm$.  Therefore, we define
$\zeta_0\in\theta_++\theta_-$ as a set any element of which belongs to the
equivalence class.  It is clear that $\zeta_0$ satisfies the first
condition of (\ref{eq:definition_zeta}).  In the following, we show
that $\zeta_0$ satisfies the second condition of
(\ref{eq:definition_zeta}).

From  (\ref{eq:statistical_Bethe}), it follows that
\begin{eqnarray}
  \lim
 \frac{\left(x_{\sigma_\pm}\mp i\right)^L}
       {\prod_{\sigma\in\theta_\mp}\left(x_{\sigma_\pm}-x_\sigma\mp 2i\right)^{N_\sigma}}
&\neq&0,\infty,
\label{eq:tmp_app_b_1001}
\end{eqnarray}
where $\sigma_\pm\in\theta_\pm\cap\theta'$, $\theta'\in\zeta_0$.  Now,
we define $\delta_{\theta'}$ as a variable on $\Omega_{Ip}$ which
satisfies $\lim 1/(x_{\sigma_\pm}\mp i)-1/\delta_{\theta'}=0$ , where
$\theta'\in\zeta$ and $\sigma_\pm\in\theta'\cap\theta_\pm$. From this
definition, $\lim \delta_{\theta'}/(x_{\sigma_\pm}\mp i)=1$, $\lim
(\delta_{\theta'}-x_{\sigma'_\pm}\pm
i)/(x_{\sigma_\mp}-x_{\sigma'_\pm}\pm 2i)=1$ and $\lim
\delta_{\theta'}^3/(\delta_{\theta'}-x_{\sigma_\pm'}\pm i)=0$ are shown to hold
where $\sigma'_\pm\in\theta_\pm$, $\not\in\theta'$ and
$\sigma_\pm\in\theta'$.  We raise both sides of the equations
(\ref{eq:tmp_app_b_1001}) to the power of $\pm N_{\sigma_\pm}$. Then,
the product of these equations is
\begin{eqnarray}
  \lim\delta_{\theta'}^{
    \left(N_+-N_-
    \right)L
  }
  \frac{\prod_{\sigma\in\theta_-,\not\in\theta'}\left(\delta_{\theta'}-x_\sigma+i\right)^{N_\sigma N_-}}
  {\prod_{\sigma\in\theta_+,\not\in\theta'}\left(\delta_{\theta'}-x_\sigma-i\right)^{N_\sigma N_+}}
&\neq&0,\infty,
\label{eq:tmp_app_b_22}  
\end{eqnarray}
where
\begin{eqnarray}
  N_+\;\equiv\;\sum_{\sigma\in\theta_+\cap\theta'}N_\sigma,
&&
  N_-\;\equiv\;\sum_{\sigma\in\theta_-\cap\theta'}N_\sigma.
\end{eqnarray}
Therefore, it holds that $N_+=N_-$. Here we have used that $L$ is much larger than $N_\sigma N_\pm$. 

\section{Definition of the integral path}
\label{app:definition_of_integral_path}
We show that (\ref{eq:Z_LM_integral_formulation(formal)}) is equal to
(\ref{eq:Z_LM_integral_formulation_1}), where $E_\theta$ is defined by
(\ref{eq:statistical_energy}), the relation between $\{x_\sigma\}$ and
$\{I_\sigma\}$ is defined by (\ref{eq:statistical_Bethe}) and the
integral path in (\ref{eq:Z_LM_integral_formulation(formal)}) is
defined by (\ref{eq:integral_path_Bethe})

In \ref{app:independency_of_integral_path} we have proved that the
integrals in the l.h.s. of (\ref{eq:integral_path_Bethe}) do
not depend on $A_\sigma$.  Then, we examine the case
$A_\sigma\leq\epsilon$.  In this case, all the points
on the integral path satisfy the condition,
for any
  $x_\sigma$,
there  exists
  $n_\sigma\in 2\mathbb{Z}_{\leq 0}+1$ such that
\begin{eqnarray}
  \left| x_\sigma -n_\sigma i\right| 
<
3\epsilon^{\frac1{L(M+1)}},
\end{eqnarray}
because eq.(\ref{eq:integral_path_Bethe}) indicates that one of
terms $\left|\frac{x_\sigma-i}{x_\sigma+i}\right|^L$
$\left|\frac{x_\sigma-x_{\sigma'}+2i}{x_\sigma-x_{\sigma'}-2i}\right|^{N_{\sigma'}}$
is less than $<\epsilon^{\frac1{M+1}}$.  In other words, the integral
path (\ref{eq:integral_path_Bethe}) is divided into several pieces,
which are classified by the vector $(n_\sigma i,n_{\sigma'}
i,\cdots),\; n_\sigma ,n_{\sigma'}\in\mathbb{Z}$.  In the following,
we consider a piece of the integral path characterized by the vector,
and regard $D$ as a set of variables $x_{\sigma_\pm1}\mp
i$,$x_{\sigma_n}-x_{\sigma_{n-2}}-2i$, where $\sigma_n\in\theta$ is a
set which satisfies $\lim_{\epsilon\rightarrow0}|x_\sigma-ni|=0$ on
the piece of the integral path.  Remark that the variables are linearly
dependent.  Before we change the integral path, we like to derive
several values from (\ref{eq:statistical_Bethe}), and prove several
facts.  For a while, we fix $\{n_\sigma\}$, equivalently $D$, and
consider the corresponding part of the integral path.

We define $A_\sigma$ which  depends on $\epsilon$,
\begin{eqnarray}
  A_\sigma&=&\epsilon^{m_\sigma},
\label{eq:def_limit_a}
\end{eqnarray}
where $m_\sigma\in\mathbb{R}_{\geq1}$ are defined as below.  We
choose an element $\mathbb{D}$ in $\Theta(D)$ which satisfies
$N_\mathbb{D}\leq N_\theta$. Using $\mathbb{D}$, we modify
eq.(\ref{eq:statistical_Bethe}) by replacing terms in the l.h.s.  of
equation as follows.  We replace all the terms which converge to
non-zero numbers in the limit $\epsilon\rightarrow 0$ with the
numbers, and replace all the terms which converge to zero in the limit
$\epsilon\rightarrow 0$, any of which is made of a value in $D$, with
variables $\delta_i$ under the restriction that all variables in any
set $D_i\in\mathbb{D}$ are substituted for the same variable.  We name
this modification the limitation of (\ref{eq:statistical_Bethe}) using
$\mathbb{D}$.  These equations can be solved with respect to
$\{\delta_i\}$.  Note that, $\{m_\sigma\}$ defines the $\{\Re
I_\sigma\}$.  Using these equations, we demand that $\{m_\sigma\}$
satisfy the following conditions: In case $N_\theta>N_{\mathbb{D}}$
there is no solution. In case $N_\theta=N_{\mathbb{D}}$ there are
finite number of solutions. The value of different variables as a
solution of the equations has different powers of $\epsilon$.  Then,
we fix $\{m_\sigma\}$ so that the above three conditions are satisfied
for all $\mathbb{D}\in\Theta(D)$ under the restriction of
$N_\mathbb{D}\leq N_\theta$.

We also define a subset $\mathcal{D}$ of $\Theta(D)$. Any element
$\mathbb{D}$ in $\Theta(D)$ which satisfies the following three
conditions is in $\mathcal{D}$.  First, $N_\mathbb{D}= N_\theta$.
Second, any value $\delta_i$ has the positive power of
$\epsilon$. Here, we regard the variables $\{\delta_i\}$ as a solution
of the limitation of (\ref{eq:statistical_Bethe}) using
$\mathbb{D}$. To write the third condition, we need some
preparations.  We denote elements in $\mathbb{D}$ by
$D_1,D_2,\cdots,D_{N_\theta}$ so that $m_i$ is not larger than
$m_{i+1}$.  Here, $m_i$ is the powers of $\epsilon$ with respect to
the value $\delta_i$ corresponding to $D_i$.  We regard any element in
$D$ as an oriented connection between two element in a set
$\{0\}$,$\sigma\in\theta$.  Here, $x_{\sigma_{\pm1}}\mp i$ is considered
as an oriented connection from $\sigma_{\pm1}$ to $\{0\}$ and
$x_{\sigma_n}-x_{\sigma_{n-2}}-2i$ is considered as an oriented
connection from $\sigma_{n}$ to $\sigma_{n-2}$.  Then, a subset of $D$
defines a division $\zeta\in\Theta(\{0\}+\theta)$ as follows:  Any two
elements in a set $\theta'\in\zeta$ are directly or indirectly
connected by connections in the subset, and any two elements in
different sets $\theta',\theta''\in \zeta$ are not connected by
connections in the subset.  We write $\zeta_n$ for a division defined
by $\bigcup_{m\geq n} D_m$.  The third condition is that any element
in $D_n$ connects two elements which are in different clusters in
$\zeta_{n+1}$.  Note that from the definition of $\zeta_{n}$, there
are only two clusters in $\zeta_{n+1}$ whose elements are linked by
connections in $D_n$.

Using $\mathbb{D}\in\mathcal{D}$, we shall define $\{m_\delta\}$ and
$\{r_\delta\}$ where $\delta\in D$. We modify the set $D$ by
multiplying some elements by $-1$. The modification causes a
reflection on $\mathbb{D}$. We have regarded an elements in $D$ as an
oriented connection; An element multiplied by $-1$ is the
oppositely oriented connection.  Then, $D$ is modified so that all the
connections in the modified $D_i$ connect elements in the same
cluster in $\zeta_{i+1}$ to other elements. We define $\{r_i\}$ and
$\{m_i\}$ so that $\{\delta_i=r_i\epsilon^{m_i}\}$ is a solution of
the limitation of (\ref{eq:statistical_Bethe}) using the modified
$\mathbb{D}$. Then, $m_\delta=m_i$ and $r_\delta=r_i$ where $\delta$
is in the modified $D_i$, and $m_{-\delta}=m_{\delta}$,
$r_{-\delta}=-r_{\delta}$.  We remark that while $\{m_\delta\}$ is
uniquely defined, $\{r_\delta\}$ is not uniquely defined, due to the
ununiqueness of the solutions for the limitation of (\ref{eq:statistical_Bethe}).

We define $\Omega_R$ as a topological space embedded in $\Omega$: Any
point on $\Omega_R$ is written as $(\{m_\delta\},\{r_\delta\})$ where
$\{m_\delta\}$,$\{r_\delta\}$ are given by the above procedure, and
$\mathbb{D}\in\mathcal{D}$ and $\{\Re I_\sigma\}$ are not fixed.  Note
that, $\Omega_R$ depends on $D$ and the number of elements in $\Omega$
in this case is $2N_D$.

The topological space $\Omega_R$ has the following properties.  Given sufficiently small $\epsilon'$ and $\epsilon$ which depends on
$\epsilon'$ (for simplicity, $\epsilon(\epsilon')$), there is a one-to-one correspondence between a point
on $\Omega_R$ and a point on a part of the integral path defined by
(\ref{eq:integral_path_Bethe}) and (\ref{eq:def_limit_a}) which
satisfies $|x_\sigma-n_\sigma i|<3\epsilon^{\frac1{L(M+1)}}$. We
denote by $C$ this piece of the integral path.  The correspondence
can be regarded as a homeomorphic mapping.  Furthermore, the relation
\begin{eqnarray}
  \left|\frac{\delta}{r_\delta\epsilon^{m_\delta}}-1\right|&<&\epsilon'
\label{eq:define_region}
\end{eqnarray}
holds where $\{\delta\}=D$ is the value on a point on $C$,
$(\{m_\delta\}, \{r_\delta\})$ is a point on $\Omega_R$ and the point
on $C$ corresponds to the point on $\Omega_R$.  These facts are proved
in \ref{sec:one_to_one_correspondence}.

Next, we shall define an integral path $C'$ which consists of
$N_\mathcal{D}$ connected integral paths.  There is a one-to-one
correspondence between elements in $\mathcal{D}$ and connected spaces
of $C'$.  From $\mathbb{D}\in\mathcal{D}$, we define the corresponding
connected space as follows. We select one variable from each element
$D_i\in\mathbb{D}$. Then, the space is $\{x_\sigma\}$ with the
restriction of $N_\theta$ conditions
$|\delta|=|r_\delta\epsilon^{m_\delta}|$ where $\delta$ is one of the
elements we choose.  The integral path $C'$ has the following
property.  Given $\epsilon'$ and a sufficiently small $\epsilon(\epsilon')$, there is a one-to-one
correspondence between a point on $\Omega_R$ and a point on the
integral path $C'$.  The correspondence can be regarded as
a homeomorphic mapping.  Furthermore, the relation
(\ref{eq:define_region}) holds.  The point on $C'$ corresponding to a
point $(\{m_\delta\},\{r_\delta\})$ on $\Omega_R$ is given by
$N_\theta$ conditions $\delta=r_\delta\epsilon^{m_\delta}$ where
$\delta$ is the element we choose from $D_i\in\mathbb{D}$.

From the previous two homeomorphic mappings, we make a homeomorphic
mapping from $C$ to $C'$.  This causes a continuous modification of
the integral path from $C$ to $C'$ by means of moving each point on
$C$ to the mapped point on $C'$.  For sufficiently small $\epsilon'$
and $\epsilon(\epsilon')$, this modification dose
not change the value of the integral in
(\ref{eq:Z_LM_integral_formulation(formal)}) . The reason is that $C'$
has no boundary, $C'$ is compact and the integrand in
(\ref{eq:Z_LM_integral_formulation(formal)}) is analytic on
(\ref{eq:define_region}), in which the corresponding points on $C$ and
$C'$ are.  Using $C'$, we can evaluate the integral
(\ref{eq:Z_LM_integral_formulation(formal)}), because the integral
path $C'$ is a sum of direct products of integral paths with respect
to a single integral.

Using these procedure, we can check the equality of
(\ref{eq:Z_LM_integral_formulation(formal)}) and
(\ref{eq:Z_LM_integral_formulation_1}) for any $M$ case.  Explicit
check is done in \ref{sec:explicite_check}.

\subsection{}
\label{sec:one_to_one_correspondence}

We prove the following: For sufficiently small $\epsilon'>0$ and
$\epsilon(\epsilon')>0$, there is a one-to-one
correspondence between a point on the piece $C$ of the integral path
(\ref{eq:integral_path_Bethe}) and a point on $\Omega_R$, where $C$ is
defined using a vector $(n_\sigma,\cdots)$ and $\Omega_R$ is defined using
$D$ made from the vector.  The correspondence can be considered as a
homeomorphic mapping.  Furthermore, the relation
(\ref{eq:define_region}) holds where $\{\delta\}=D$ is a value on a
point on $C$, $(\{m_\delta\}, \{r_\delta\})$ is a point on $\Omega_R$,
and the point on $C$ corresponds to the point on $\Omega_R$.

We define a subset $\Omega_{RI}$ of $\Omega_R$.  $\Omega_{RI}$ is a
set of points which correspond to $\{m_\delta\}$,$\{r_\delta\}$
generated by a fixed set of phases $\{\Re I_\sigma\}$ where
$\mathbb{D}$ are freely moved. There are a finite number of elements in
$\Omega_{RI}$. For a while, we fix a set of phases $\{\Re I_\sigma\}$.

First, we prove the following: lemma 1.  For $\epsilon'>0$ and a
sufficiently small $\epsilon(\epsilon')>0$, any
solution of (\ref{eq:statistical_Bethe}),
(\ref{eq:integral_path_Bethe}), (\ref{eq:def_limit_a}) which satisfies
$|x_\sigma-n_\sigma i|<3\epsilon^{\frac1{L(M-1)}}$ exists in
(\ref{eq:define_region}) generated by a point $(\{m_\delta\},
\{r_\delta\})$ on $\Omega_{RI}$.  We can choose $\epsilon$ independently
on ${\{\Re I_\sigma\}}$.

We define a phase space $\Omega_F(\epsilon)$ embedded in $\Omega$,
\begin{eqnarray}
  \left(\left\{\frac\delta{r_\delta\epsilon^{m_\delta}}\right\},\epsilon\right),\makebox[2cm]{}\delta\in D,\quad \epsilon > 0,
\label{eq:define_Omega_a}
\end{eqnarray}
where $(\{r_\delta\},\{m_\delta\})$ are all the points on
$\Omega_{RI}$.
For $\Omega$ in this case, note that the number of elements is
$N_{\Omega_{RI}}N_D+1$, and that $\{\delta\}$
and $\epsilon$ do not need to satisfy the conditions (\ref{eq:integral_path_Bethe})
and (\ref{eq:def_limit_a}). We define a subspace $\Omega_C(\epsilon)$
of $\Omega_F(\epsilon)$ where $\{\delta\}$ and $\epsilon$ satisfy the
conditions (\ref{eq:statistical_Bethe}),
(\ref{eq:integral_path_Bethe}), (\ref{eq:def_limit_a}),
$x_\sigma\neq\pm i$ and $|x_\sigma-n_\sigma
i|<3\epsilon^{\frac1{L(M+1)}}$.  We also define a subspace
$\Omega_B(\epsilon,\epsilon')$ of $\Omega_F(\epsilon)$ where
$\{r_\delta\}$ and $\{m_\delta\}$ satisfy (\ref{eq:define_region})
with respect to a $(\{r_\delta\}, \{m_\delta\})\in \Omega_{RI}$.
Next, we define a sequence $(P_{Bn})$ of points on $\Omega_C(\epsilon)$
as follows, where $\epsilon$ is freely moved.  The series of
$\epsilon$ which corresponds to $(P_{Bn})$ is decreasing and converges
to $0$. A necessary and sufficient condition of the lemma 1 is that
there is a subsequence of any $(P_{Bn})$ defined previously which
converges to a point on $\lim_{\epsilon'\rightarrow
  0}\lim_{\epsilon\rightarrow 0}\Omega_{B}(\epsilon,\epsilon')$. We
note that since the parameter ${\{\Re I_\sigma({\rm mod}\; 1)\} }$ is
compact we can choose $\epsilon$ independently on ${\{\Re
  I_\sigma\}}$.

We define a subspace $\Omega_D(\epsilon)$ of  $\Omega$ as 
\begin{eqnarray}
  \left(
    \left\{
      \frac{\delta}
           {\delta'}
    \right\},
   \left\{
           {\delta}
    \right\},\epsilon
   \right),\makebox[2cm]{}\delta,\delta'\in D,
\end{eqnarray}
where  the variables satisfy
(\ref{eq:statistical_Bethe}), (\ref{eq:integral_path_Bethe}),
(\ref{eq:def_limit_a}), $x_\sigma\neq \pm 2$ and $|x_\sigma-n_\sigma
i|<3\epsilon^{\frac1{Ln}}$. The number of variables in $\Omega$ is $
N_D^2+1$ in this case.  We denote the closure of $\Omega_D(\epsilon)$
by $\bar\Omega_D(\epsilon)$.  We define a sequence $(P_{Dn})$ of points
on $\Omega_D(\epsilon)$, where $\epsilon$ is freely moved: The series
of $\epsilon$ which corresponds to $(P_{Dn})$ is decreasing and
converges to $0$.  Then, we can choose a subsequence of $(P_{Dn})$
which converges to a point $P_D$ on $\bar\Omega_D(+0)$, because
$\bar\Omega_D(\epsilon)$ is a compact space.  Using this subsequence,
we define a partial order, $\prec$, for $D$,
\begin{eqnarray}
  \lim\frac \delta{\delta'}=0\rightarrow
 \delta\prec \delta'\makebox[2cm]{}{\rm for}\quad\delta,\delta'\in D,
\end{eqnarray}
where $\lim$ means a limit on the subsequence.  We define $D_1$ as the
set of maximal elements in $D$, and $D_2$ as the set of maximal
elements in $D-D_1$, and in the same way define $D_3,D_4,\cdots$.
Then, $\mathbb{D}$ is defined as
$\{D_1,D_2,D_3,\cdots\}$. Eqs.(\ref{eq:statistical_Bethe}),
(\ref{eq:integral_path_Bethe}), (\ref{eq:def_limit_a}) are solved in
the meaning that $\{r_\delta\neq0\}$ and $\{m_\delta\}$ which satisfy
\begin{eqnarray}
  \lim \frac\delta{\epsilon^{m_\delta}}=r_\delta
\label{eq:meaning_of_solution}
\end{eqnarray}
are given using $P_D$, and therefore using the limit of variable's
ratios.  From the definition of $\{m_\sigma\}$, $\{m_\delta\}$
corresponding to elements in different clusters of $\mathbb{D}$ are
distinct positive numbers, and $N_\mathbb{D}$ is equal to $N_\theta$.
Therefore, the limit of variable's ratios is either of $0,\pm
1,\infty$.  Then, $\{r_\delta\}$, $\{m_\delta\}$ are given without any
ambiguity.  It is clear that $(\{r_\delta\},\{m_\delta\})$ derived
here are in $\Omega_{RI}$.  In fact, the number of solutions
$(\{r_\delta\},\{m_\delta\})$ is not always one but a finite number.
 We can therefore choose a subsequence for the subsequence previously
defined which converges to a single solution in the meaning of
(\ref{eq:meaning_of_solution}). We note that there is a one-to-one
correspondence between $(P_{Bn})$ and $(P_{Dn})$. Therefore, there is
a subsequence of $(P_{Bn})$ corresponding to the subsequence of
$(P_{Dn})$ and the subsequence of $(P_{Bn})$ converges to a point on
$\lim_{\epsilon'\rightarrow 0}\lim_{\epsilon\rightarrow
  0}\Omega_{B}(\epsilon,\epsilon')$.  Thus, the lemma 1 is proved.

Second, the following lemma 2 holds: For $\epsilon'>0$ and a
sufficiently small $\epsilon(\epsilon')>0$, there
are one or more solutions of (\ref{eq:statistical_Bethe}),
(\ref{eq:integral_path_Bethe}), (\ref{eq:def_limit_a}) in
(\ref{eq:define_region}) generated by any element $(\{m_\delta\},
\{r_\delta\})$ on $\Omega_{RI}$.  In other words, $\{I_\sigma\}$ is
defined by means of $\epsilon$ and is in the image of any region
(\ref{eq:define_region}) where (\ref{eq:statistical_Bethe}) defines
$\{I_\sigma\}$ as a function of $\{x_\delta\}$.  We can choose $\epsilon$ independently on ${\{\Re
  I_\sigma\}}$.

Third, we prove the following lemma 3: For sufficiently small
$\epsilon'>0$ and $\epsilon(\epsilon')>0$, there
exists one or no solution of (\ref{eq:statistical_Bethe}),
(\ref{eq:integral_path_Bethe}), (\ref{eq:def_limit_a}) in
(\ref{eq:define_region}) generated by any element $(\{m_\delta\},
\{r_\delta\})$ on $\Omega_{RI}$.  We can choose $\epsilon$
independently on ${\{\Re I_\sigma\}}$.

We say that an analytic vector function $f(x)$ is ``monotonic'' on $A$
when $A$ is a convex domain and an analytic vector function $f(x)$
satisfies a relation; for any  $x\in A$ and $a\neq0$, there exists $b$ such that
\begin{eqnarray}
  \Re\left(b,\frac{df}{dx}\cdot a\right)>0 
\label{eq:define_tanchou}
\end{eqnarray}
where $(,)$ is the inner product on an $M$-dimensional complex linear
space, $a$ is an $N$-dimensional vector, $b$ is an $M$-dimensional
vector, $N$ and $M$ are dimensions of vector $x$ and $f$.  Then, an
equation $f(x)=b$ has one or no solution on $A$ when $f(x)$ is a
monotonic analytic vector function.

A sufficient condition of the lemma 3 is that (\ref{eq:statistical_Bethe})
is a monotonic analytic vector function on any region
(\ref{eq:define_region}) where (\ref{eq:statistical_Bethe}) defines
$\{I_\sigma\}$ as a function of $\{x_\sigma\}$.  It is evident that
(\ref{eq:define_region}) generated by any element on $\Omega_{RI}$ is
a convex domain.  In the following, we fix
$(\{r_\delta\},\{m_\delta\})$ to an element in $\Omega_{RI}$ and
$\mathbb{D}$ to an element in $\mathcal{D}$ which gives the element
$(\{r_\delta\},\{m_\delta\})$. Now, we express $b$ in terms of $a$ which
satisfies (\ref{eq:define_tanchou}).  We derive the limitation of
(\ref{eq:statistical_Bethe}) using $\mathbb{D}$.  We replace any
variable with an element in $D_i$ corresponding to the variable.
Then, the Jacobi matrix of this functions at
$\delta=r_\delta\epsilon^{m_\delta}$ is written as $\partial
I_\sigma/\partial \delta_0$ where $\delta$ are selected variables in
the previous modification.  We choose $b$ to be $\partial I_\sigma/\partial
\delta_0\cdot a$ for any $a$, and then (\ref{eq:define_tanchou}) holds
when $\epsilon'$ and $\epsilon$ are sufficiently small.  The
$\epsilon'$ and $\epsilon$ can be given independently on $\{\Re
I_\sigma\}$ and the element of $\Omega_{RI}$.

Fourth, we point out the following: Two regions
(\ref{eq:define_region}) corresponding to different elements on
$\Omega_{RI}$ have no common element when $\epsilon$ and $\epsilon'$
are sufficiently small.

Three lemmas and the above fact indicate the following.
Given a sufficiently small $\epsilon$ and a sufficiently small
$\epsilon'(\epsilon)$, there is only one solution of
(\ref{eq:statistical_Bethe}), (\ref{eq:integral_path_Bethe}),
(\ref{eq:def_limit_a}) in (\ref{eq:define_region}) generated by any
element on $\Omega_{RI}$.  Then, the correspondence between
$\Omega_{RI}$ and solutions of (\ref{eq:statistical_Bethe}),
(\ref{eq:def_limit_a}) which satisfy $|x_\sigma-n_\sigma
i|<3\epsilon^{\frac1{Ln}}$ is  one-to-one.  This
correspondence can be regard as a one-to-one correspondence between
$\Omega_R$ and solutions of (\ref{eq:integral_path_Bethe}) which
satisfy $|x_\sigma-n_\sigma i|<3\epsilon^{\frac1{L(M+1)}}$, that is,
$C$.  And, any point on $C$ and the corresponding element
$(\{m_\delta\},\{r_\delta\})$ on $\Omega_R$ satisfy
(\ref{eq:define_region}).

Finally, we prove that this correspondence is a homeomorphic mapping.
This is due to the following three facts: $\Omega_R$ is a compact
space.  The correspondence is a bijection.  And the correspondence is
a continuous mapping from to $\Omega_R$ to $C$.

\subsection{An example}
\label{sec:explicite_check}

In the case $M$ of $\Theta_M$ is equal to $1$, a set ${\Theta_1}$ has
one element $\theta=\{1\}$.  There is only a case that $n_1=1$, and
therefore $D$ has only one element $x_1-i=\delta_1$.  We can choose
$m_1$ to be $1$.  $\mathcal{D}$ has only one element $\mathbb{D}$, and
$\mathbb{D}$ has only one element $D$.  Then, $\Omega_R$ is a space
$(|r_{\delta_1}|=2,m_{\delta_1}=1/L)$.  Therefore, the integral path
$C'$ becomes $\left|{x_1-i}\right|=2\epsilon^{1/L}$.  Then, the value
of (\ref{eq:Z_LM_integral_formulation(formal)}) is equal to the value
of (\ref{eq:Z_LM_integral_formulation_1}) in the case $M=1$.

In case that $\theta$ is arbitrary, there is a case $n_{\sigma}=1$ for
all $\sigma\in\theta$.  For that case, all the elements in $D$ is
$x_\sigma-i=\delta_\sigma$.  We can choose $m_\sigma$ arbitrarily.
$\mathcal{D}$ has only one element $\mathbb{D}$, and all the elements
in $\mathbb{D}$ is $\{\delta_\sigma\}$.  Then, $\Omega_R$ is a space
$(|r_{\delta_\sigma}|=2,m_{\delta_\sigma}=m_{\sigma}/L)$.  Therefore,
the integral path $C'$ becomes
$\left|{x_\sigma-i}\right|=2\epsilon^{m_\sigma/L}$.  Then, if a sum of
 the integrals using the paths corresponding to the
case $n_{\sigma}\neq 1$ for some $\sigma\in\theta$ is zero, the value
of (\ref{eq:Z_LM_integral_formulation(formal)}) is equal to the value
of (\ref{eq:Z_LM_integral_formulation_1}).

In the case $M$ of $\Theta_M$ is equal to $2$, it is only possible
that $n_{\sigma}$ is not equal to $1$ for some $\sigma\in\theta$.
Note that, we judge whether the case $\{n_{\sigma}\}$ occurs or not by
checking the number of $\mathcal{D}$ corresponding to $\{n_{\sigma}\}$
is non-zero or not.  Then, the values of
(\ref{eq:Z_LM_integral_formulation(formal)}) and
(\ref{eq:Z_LM_integral_formulation_1}) are the same in the case $M=2$.
This procedure can be summarized in the following table:
\begin{center}
\begin{tabular}[t]{|c||c|c|}
\hline
$\theta$        &    $1\!-\!2$     & $1, 2$\\
\hline
$\{n_\sigma\}$  &$n_{1\!-\!2}=1$   &$n_1=1,n_2=1$\\
\hline
$D$             &$x_{1\!-\!2}-i$   &$x_1-i$, $x_2-i$\\
\hline
$\{m_\sigma\}$  &   $m_{1\!-\!2}$  & $m_1$, $m_2$\\
\hline
$\mathbb{D}$    &    $\{\delta_{1\!-\!2}\}$       &$\{\delta_1\}$, $\{\delta_2\}$\\
\hline
$\Omega_R$            &
$|r_{\delta_{1\!-\!2}}|=2$&
 $|r_{\delta_1}|=|r_{\delta_2}|=2$
\\
&
 $m_{\delta_{1\!-\!2}}=m_{1\!-\!2}/L$&
$m_{\delta_1}=m_1/L$
\\
                &
 &
$m_{\delta_2}=m_2/L$\\
\hline
\end{tabular}
\end{center}
where $x_\sigma-i$ is equal to $\delta_\sigma$.

In the same way, the procedure in case of $M=3$ becomes
\begin{center}
\begin{tabular}[t]{|c||c|c|c|c|c|}
\hline
$\theta$        &
$1\!-\!2\!-\!3$   &
\multicolumn{2}{c|}{  $1\!-\!2,3$}&
\multicolumn{2}{c|}{  $1,2,3$}\\
\hline
$\{n_\sigma\}$    &
$n_{1\!-\!2\!-\!3}=1$   &
$n_{1\!-\!2}=1$   &
$n_{1\!-\!2}=1$   &
$n_{1}=1      $   &
{ $n_1=1$}\\
   &
   &
$n_{3}=1      $   &
$n_{3}=-1     $   &
$n_{2}=1      $   &
{ $n_2=1$}\\
   &
   &
   &
   &
$n_{3}=1      $   &
{ $n_3=-1$}\\
\hline
$D$             &
$x_{1\!-\!2\!-\!3}-i$   &
$x_{1\!-\!2}-i$                 &
$x_{1\!-\!2}-i$                 &
$x_1-i$                 &
$x_1-i$                 
\\
&
&
$x_3-i$   &
$x_{1\!-\!2}-x_3-2i$                 &
$x_2-i$                 &
$x_2-i$                 
\\
&
&
&
$x_{3}+i$&
$x_3-i$                 &
$x_1-x_3-2i$                 
\\
&
&
&
&
               &
$x_2-x_3-2i$                 
\\
&
&
&
&
               &
$x_3+i$                 
\\
\hline
$\{m_\sigma\}$  &
  $m_{1\!-\!2\!-\!3}$  &
 $m_{1\!-\!2}$ &
 $m_{1\!-\!2}=1$ &
 $m_1$ &
 $m_1=3$ 
\\
 &
 &
 $m_3$ &
 $m_3=1$ &
 $m_2$ &
 $m_2=2$ 
\\
 &
 &
  &
 &
 $m_3$ &
 $m_3=1$ 
\\
\hline
$\mathbb{D}$    &
$\{\delta_{1\!-\!2\!-\!3}\}$       &
$\{\delta_{1\!-\!2}\}$       &
$\{\delta_{1\!-\!2},\delta_{3}\}$       &
$\{\delta_1\}$&
$\{\delta_1, \delta_2,\delta_3\}$
\\
    &
    &
$\{\delta_3\}$&
$\{\delta_{1\!-\!2,3}\}$&
$\{\delta_2\}$&
$\{\delta_{1\!-\!3}\}$
\\
    &
    &
&
&
$\{\delta_3\}$&
$\{\delta_{2\!-\!3}\}$
\\
\hline
$\Omega_R$            &
 $|r_{\delta_{1\!-\!2\!-\!3}}|=2$&
 $|r_{\delta_{1\!-\!2}}|=2$&
 $|r_{\delta_{1\!-\!2}}|=2$&
 $|r_{\delta_1}|=2$&
 $|r_{\delta_1}|=2$
\\
                &
&
 $|r_{\delta_3}|=2$&
 $|r_{\delta_{1\!-\!2,3}}|=4$&
 $|r_{\delta_2}|=2$&
 $|r_{\delta_2}|=2$
\\
                &
&
&
&
 $|r_{\delta_3}|=2$&
 $|r_{\delta_3}|=2$
\\
                &
&
&
&
&
 $|r_{\delta_{1\!-\!3}}|=4$
\\
                &
&
&
&
&
 $|r_{\delta_{2\!-\!3}}|=4$
\\
&
 $m_{\delta_{1\!-\!2\!-\!3}}=m_{1\!-\!2\!-\!3}/L$&
 $m_{\delta_{1\!-\!2}}=m_{1\!-\!2}/L$&
 $m_{\delta_{1\!-\!2}}=3/L$&
$m_{\delta_1}=m_1/L$&
$m_{\delta_1}=6/L$
\\
         &
&
 $m_{\delta_3}=m_3/L$&
 $m_{\delta_{1\!-\!2,3}}=2$&
$m_{\delta_2}=m_2/L$&
$m_{\delta_2}=6/L$
\\
         &
&
&
&
$m_{\delta_3}=m_3/L$&
$m_{\delta_3}=6/L$
\\
         &
&
&
&
&
$m_{\delta_{1,3}}=3$
\\
         &
&
&
&
&
$m_{\delta_{2,3}}=4$
\\
\hline
\end{tabular}
\end{center}
where $x_\sigma-x_{\sigma'}-2i$ is equal to $\delta_{\sigma,\sigma'}$.
Note that we have other choices of values $\{m_\sigma\}$.
In the table, we have omitted the cases
which correspond to ($\theta=2\!-\!3,1$), ($\theta=3\!-\!1,2$),
($n_{2}=n_{3}=1,n_{1}=-1$) and ($n_{3}=n_{1}=1,n_{2}=-1$).  The
columns for ($\theta=2\!-\!3,1$) and ($\theta=3\!-\!1,2$)
 are given by replacing $1,2,3$ of the subscripts in the column  ($\theta=1\!-\!2,3$)
with $2,3,1$ or $3,1,2$ respectively.  Similarly, the columns
for ($n_{2}=n_{3}=1,n_{1}=-1$) and
($n_{3}=n_{1}=1,n_{2}=-1$) are given by replacing
$1,2,3$ of the subscripts in the column ($n_{1}=n_{2}=1,n_{3}=-1$) with $2,3,1$ or $3,1,2$
respectively.  Then, the integral path corresponding to
$\theta=1\!-\!2,3$ becomes a union of two connected integral paths;
\begin{eqnarray}
&&
|x_{1\!-\!2}-i|\quad=\quad 2\epsilon^{m_{1\!-\!2}/L},\quad\quad
|x_{3}-i|\quad=\quad 2\epsilon^{m_{3}/L} ,
\\&{\rm and}&
 |x_{1\!-\!2}-i|\quad=\quad 2\epsilon^{3/L},\quad\quad
|x_{1\!-\!2}-x_{3}-2i|\quad=\quad4\epsilon^2. 
\end{eqnarray}
And, the integral path corresponding to $\theta=1,2,3$ becomes a union
of four connected integral paths;
\begin{eqnarray}
&&
|x_{1}-i|\:=\: 2\epsilon^{m_{1}/L},\quad
|x_{2}-i|\:=\: 2\epsilon^{m_{2}/L},\quad
|x_{3}-i|\:=\: 2\epsilon^{m_{3}/L},
\\&&
|x_{1}-i|\:=\: 2\epsilon^{6/L},\quad
|x_{1}-x_{3}-2i|\:=\: 4\epsilon^3,\quad
|x_{2}-x_{3}-2i|\:=\: 4\epsilon^4
\label{eq:tmp_appb_1000001}
\end{eqnarray}
and the integral paths which are given by replacing $1,2,3$ of the
subscripts in the second integral path (\ref{eq:tmp_appb_1000001})
with $2,3,1$ or $3,1,2$ respectively.

Using the part of the integral path corresponding to $n_{1\!-\!2}=1$,
$n_{3}=-1$, we evaluate the integral which is a term in
(\ref{eq:Z_LM_integral_formulation(formal)}) corresponding to
$\theta=\{1\!-\!2,3\}$,
\begin{eqnarray}
&&\int \left|\frac{\partial I}{\partial x}\right|
\exp\left(\frac2{x_{1\!-\!2}^2+1}+\frac1{x_3^2+1}\right)
d(x_{1\!-\!2}-i)
d(x_{1\!-\!2}-x_{3}-2i)
\nonumber\\&=&
-\int_{|x-i|=+0}
L\left(
\frac{2}{x+i}
-\frac{1}{x-i}
-\frac{1}{x-3i}
\right)
e^{-\beta\left(
\frac{2i}{x+i}
-\frac{i}{x-i}
-\frac{i}{x-3i}
\right)}
\frac{d(x-i)}{2\pi i}.{}\makebox[1cm]{}
\end{eqnarray}
The orientation of the path is defined so that the value of
the integral is a positive integer when $e^{-\beta E}$ is replaced with 1.
This definition is consistent with the orientation of
 integrals in (\ref{eq:Z_LM_integral_formulation(formal)}).

Using the part of the integral path corresponding to $n_{1}=n_{2}=1$,
$n_{3}=-1$, we evaluate the integral which is a term in
(\ref{eq:Z_LM_integral_formulation(formal)}) corresponding to
$\theta=\{1,2,3\}$,
\begin{eqnarray}
&&\int \left|\frac{\partial n}{\partial x}\right|
\exp\left(\frac1{x_1^2+1}+\frac1{x_2^2+1}+\frac1{x_3^2+1}\right)
d(x_{1}-i)
d(x_{1}-x_{3}-2i)
d(x_{2}-x_{3}-2i)
\nonumber\\&=&
-\!\!\int_{|x-i|=+0}
\!\!\!\!\!\!L\left(
\frac{2}{x+i}
-\frac{1}{x-i}
-\frac{1}{x-3i}
\right)
e^{-\beta\left(
\frac{2i}{x+i}
-\frac{i}{x-i}
-\frac{i}{x-3i}
\right)}
\frac{d(x-i)}{2\pi i}.
\end{eqnarray}
The coefficient of the integrals in
(\ref{eq:Z_LM_integral_formulation(formal)}) are
$\mu\left(\hat0_n,\{1\!-\!2,3\}\right)=-1$ and
$\mu\left(\hat0_n,\{1,2,3\}\right)=1$.  Therefore, in
(\ref{eq:Z_LM_integral_formulation(formal)}), these terms cancel out.
Then,  (\ref{eq:Z_LM_integral_formulation(formal)}) is
equal to (\ref{eq:Z_LM_integral_formulation_1}) in the
case $M=3$.  We have checked that all the values of integrals
corresponding to $n_{\sigma}\neq 1$ for some $\sigma\in\theta$ cancel
out in the case of $M=4$.  It is adequate to suppose that the value of
(\ref{eq:Z_LM_integral_formulation(formal)}) is equal to the value of
(\ref{eq:Z_LM_integral_formulation_1}) in case of $M>4$.

\section{Modifications of integral paths 1}
\label{app:modification_of_integral_path_1}
We prove that the value
of (\ref{eq:modification_of_integral_path_1_tmp_1}) does not change when
$m$ is decreased by 1.

First, we change the integral path into $(-\infty,\infty)$ with respect
to $x_{\theta_m}$.  Then
(\ref{eq:modification_of_integral_path_1_tmp_1}) becomes
\begin{eqnarray}
&&(\ref{eq:modification_of_integral_path_1_tmp_1})\nonumber\\
&=&  \left[\prod_{\sigma\in\theta}N_\sigma\right]^{-1}
  \left[
    \prod_{\{\sigma,\sigma'\}\in\lambda}
    N_{\sigma} N_{\sigma'}
  \right]
  \sum_{\zeta\in\Theta\left(\theta|\lambda|\sigma_1,\cdots,\sigma_m\right)}
  \sum_{\{\sigma^{(\theta_k)}\}\in\{\theta_k\in\zeta\}_{k>m}}
  \left[
    \prod_{k=1}^{m-1}
    \int_{\left|i-x_{\theta_k}\right|_+=\bar\delta_{\sigma_k}}
    \frac{dx_{\theta_k}}{2\pi}
  \right]
\nonumber\\&&
  \left[
    \prod_{k=m}^{N_\zeta}
    \int_{-\infty+i\delta_{\sigma^{(\theta_k)}}}
        ^{+\infty+i\delta_{\sigma^{(\theta_k)}}}
    \frac{dx_{\theta_k}}{2\pi}
  \right]
    \left[
      \prod_{\theta'\in G_\theta\left(\lambda\right)}
      \left(
      \sum_{\sigma\in\theta'}
      \frac
        {2N_{\sigma}L}
        {
          \left(
            x_{\theta[\zeta,\sigma]}
            +2l_\lambda\left(\sigma,\sigma^{(\theta[\zeta,\sigma])}\right)i\right)^2+1}
\right)
    \right]
    \nonumber\\&&
    \left[
      \prod_{\{\sigma,\sigma'\}\in\lambda,\;\sigma\not\stackrel{\zeta}{\sim}\sigma'}
        \frac
          {-4}{
            \left(
               x_{\theta[\zeta,\sigma ]}
              -x_{\theta[\zeta,\sigma']}
              +2l_\lambda\left(\sigma ,\sigma^{(\theta[\zeta,\sigma ])}\right)i
              -2l_\lambda\left(\sigma',\sigma^{(\theta[\zeta,\sigma'])}\right)i\right)^2+4}
    \right]
    \nonumber\\&&
    \exp
    \left[-\beta
      \sum_{\sigma\in\theta}
      \frac
        {2N_{\sigma}}
        {
          \left(
            x_{\theta[\zeta,\sigma]}
            +2l_\lambda\left(\sigma,\sigma^{(\theta[\zeta,\sigma])}\right)i\right)^2+1}
    \right]
\nonumber\\&&{}+
  \left[\prod_{\sigma\in\theta}N_\sigma\right]^{-1}
  \left[
    \prod_{\{\sigma,\sigma'\}\in\lambda}
    N_{\sigma} N_{\sigma'}
  \right]
  \sum_{\zeta\in\Theta\left(\theta|\lambda|\sigma_1,\cdots,\sigma_m\right)}
  \sum_{\{\sigma^{(\theta_k)}\}\in\{\theta_k\in\zeta\}_{k>m}}
  \left[
    \prod_{k=1}^{m-1}
    \int_{\left|i-x_{\theta_k}\right|_+=\bar\delta_{\sigma_k}}
    \frac{dx_{\theta_k}}{2\pi}
  \right]
\nonumber\\&&
  \int
  \frac{dx_{\theta_m}}{2\pi}
  \left[
    \prod_{k=m+1}^{N_\zeta}
    \int_{-\infty+i\delta_{\sigma^{(\theta_k)}}}
        ^{+\infty+i\delta_{\sigma^{(\theta_k)}}}
    \frac{dx_{\theta_k}}{2\pi}
  \right]
    \left[
      \prod_{\theta'\in G_\theta\left(\lambda\right)}
      \left(
      \sum_{\sigma\in\theta'}
      \frac
        {2N_{\sigma}L}
        {
          \left(
            x_{\theta[\zeta,\sigma]}
            +2l_\lambda\left(\sigma,\sigma^{(\theta[\zeta,\sigma])}\right)i\right)^2+1}
\right)
    \right]
    \nonumber\\&&
    \left[
      \prod_{\{\sigma,\sigma'\}\in\lambda,\;\sigma\not\stackrel{\zeta}{\sim}\sigma'\not\stackrel{\zeta}{\sim}\sigma_m}
        \frac
          {-4}{
            \left(
               x_{\theta[\zeta,\sigma ]}
              -x_{\theta[\zeta,\sigma']}
              +2l_\lambda\left(\sigma ,\sigma^{(\theta[\zeta,\sigma ])}\right)i
              -2l_\lambda\left(\sigma',\sigma^{(\theta[\zeta,\sigma'])}\right)i\right)^2+4}
    \right]
    \nonumber\\&&
    \left[
      \sum_{\{\sigma,\sigma'\}\in\lambda,\;\sigma\stackrel{\zeta}{\sim}\sigma_m\not\stackrel{\zeta}{\sim}\sigma'}
      \sum_{s=\pm1}
      2\pi s 
      \theta\left(
        l_\lambda\left(\sigma',\sigma^{(\theta[\zeta,\sigma'])}\right)
        -l_\lambda\left(\sigma,\sigma_m\right)
        -\frac12+s
      \right)
    \right.
    \nonumber\\&&
    \left.
      \delta\left(
        x_{\theta_m}
        -x_{\theta[\zeta,\sigma']}
        +2l_\lambda\left(\sigma ,\sigma_m\right)i
        -2l_\lambda\left(\sigma',\sigma^{(\theta[\zeta,\sigma'])}\right)i
        -2si
      \right)
    \right.
    \nonumber\\&&
    \left.
      \prod_{
        \{\sigma'',\sigma'''\}\in\lambda,\;
        \sigma''\stackrel{\zeta}{\sim}\sigma\not\stackrel{\zeta}{\sim}
        \sigma'''\not\stackrel{\zeta}{\sim}\sigma'
      }
        \frac
          {-4}{
            \left(
              x_{\theta_m}
              -x_{\theta[\zeta,\sigma''']}
              +2l_\lambda\left(\sigma '',\sigma_m\right)i
              -2l_\lambda\left(\sigma''',\sigma^{(\theta[\zeta,\sigma'''])}\right)i\right)^2+4}
    \right]
    \nonumber\\&&
    \exp
    \left[-\beta
      \sum_{\sigma\in\theta}
      \frac
        {2N_{\sigma}}
        {
          \left(
            x_{\theta[\zeta,\sigma]}
            +2l_\lambda\left(\sigma,\sigma^{(\theta[\zeta,\sigma])}\right)i\right)^2+1}
    \right].
\label{eq:modification_of_integral_path_1_tmp_2}
\end{eqnarray}
The second term comes from the residues.  In the second term, the
$\delta$-function and the integral with respect to $x_{\theta_m}$ mean
the following procedure:  Before integrations with respect to the
other variables, the variable $x_{\theta_m}$ is replaced with the
other variables and constants so that a value of the expression which
is the independent variable of the $\delta$-function becomes $0$. And,
$\frac12$ in the expression which is the independent variable of the
$\theta$-function is caused by the relation
(\ref{eq:large_small_relation}).

There occur  cancellations in the second term. We
recognize the term as a series with respect to sets
$(\zeta,\{\sigma^{(\theta_k)}\},\sigma,\sigma',s)$,
\begin{eqnarray}
  \sum_{\zeta\in\Theta\left(\theta|\lambda|\sigma_1,\cdots,\sigma_m\right)}
  \sum_{\{\sigma^{(\theta_k)}\}\in\{\theta_k\in\zeta\}_{k>m}}
  \sum_{\{\sigma,\sigma'\}\in\lambda,\;\sigma\stackrel{\zeta}{\sim}\sigma_m\not\stackrel{\zeta}{\sim}\sigma'}
  \sum_{s=\pm1}.
\label{eq:modification_of_integral_path_1_tmp_3}
\end{eqnarray}
Then, there are cancellations between any term corresponding to
$(\zeta ,\{\sigma^{(\theta_k)}\},\sigma,\sigma',-)$ which has a non-zero
value and a term corresponding to
$(\zeta',\{\sigma^{(\theta_k)}\},\sigma'',\sigma',+)$ which is defined
as
\begin{eqnarray}
&\theta'''\equiv\{\sigma'''|\sigma'''\in\theta[\lambda,\sigma'],
l_\lambda\left(\sigma''',\sigma^{(\theta[\zeta,\sigma'])}\right) =
l_\lambda\left(\sigma',\sigma^{(\theta[\zeta,\sigma'])}\right)+
l_\lambda\left(\sigma',\sigma'''\right)
\}
\nonumber\\
&\theta'\equiv\theta_m+\theta''',\quad
\theta''\equiv\theta[\zeta,\sigma']-\theta'''
\nonumber\\
&\zeta'\equiv\zeta-\theta_m-\theta[\zeta,\sigma']+\theta'+\theta''
\nonumber\\
&l_\lambda\left(\sigma',\sigma''\right)=1,\quad
l_\lambda\left(\sigma',\sigma^{(\theta[\zeta,\sigma'])}\right) =
l_\lambda\left(\sigma'',\sigma^{(\theta[\zeta,\sigma'])}\right)+1.
\end{eqnarray}
We note that the term
$(\zeta',\{\sigma^{(\theta_k)}\},\sigma'',\sigma',+)$ exists for the condition
$\sigma''\neq\sigma^{(\theta[\zeta\sigma''])}$. This condition is
satisfied automatically when the term $(\zeta ,
\{\sigma^{(\theta_k)}\},\sigma,\sigma',-)$ has a non-zero value, in
other words, the $\theta$-function is not equal to $0$ in the term $(\zeta ,
\{\sigma^{(\theta_k)}\},\sigma,\sigma',-)$.

Then, the second term in
(\ref{eq:modification_of_integral_path_1_tmp_2}) becomes
\begin{eqnarray}
&&  \left[\prod_{\sigma\in\theta}N_\sigma\right]^{-1}
  \left[
    \prod_{\{\sigma,\sigma'\}\in\lambda}
    N_{\sigma} N_{\sigma'}
  \right]
  \sum_{\zeta\in\Theta\left(\theta|\lambda|\sigma_1,\cdots,\sigma_m\right)}
  \sum_{\{\sigma^{(\theta_k)}\}\in\{\theta_k\in\zeta\}_{k>m}}
  \left[
    \prod_{k=1}^{m-1}
    \int_{\left|i-x_{\theta_k}\right|_+=\bar\delta_{\sigma_k}}
    \frac{dx_{\theta_k}}{2\pi}
  \right]
\nonumber\\&&
  \int
  \frac{dx_{\theta_m}}{2\pi}
  \left[
    \prod_{k=m+1}^{N_\zeta}
    \int_{-\infty+i\delta_{\sigma^{(\theta_k)}}}
        ^{+\infty+i\delta_{\sigma^{(\theta_k)}}}
    \frac{dx_{\theta_k}}{2\pi}
  \right]
    \left[
      \prod_{\theta'\in G_\theta\left(\lambda\right)}
      \left(
      \sum_{\sigma\in\theta'}
      \frac
        {2N_{\sigma}L}
        {
          \left(
            x_{\theta[\zeta,\sigma]}
            +2l_\lambda\left(\sigma,\sigma^{(\theta[\zeta,\sigma])}\right)i\right)^2+1}
\right)
    \right]
    \nonumber\\&&
    \left[
      \prod_{\{\sigma,\sigma'\}\in\lambda,\;\sigma\not\stackrel{\zeta}{\sim}\sigma'\not\stackrel{\zeta}{\sim}\sigma_m}
        \frac
          {-4}{
            \left(
               x_{\theta[\zeta,\sigma ]}
              -x_{\theta[\zeta,\sigma']}
              +2l_\lambda\left(\sigma ,\sigma^{(\theta[\zeta,\sigma ])}\right)i
              -2l_\lambda\left(\sigma',\sigma^{(\theta[\zeta,\sigma'])}\right)i\right)^2+4}
    \right]
    \nonumber\\&&
    \left[
      \sum_{\{\sigma_m,\sigma'\}\in\lambda,\;\sigma_m\not\stackrel{\zeta}{\sim}\sigma'}
      2\pi      \delta\left(
        x_{\theta_m}
        -x_{\theta[\zeta,\sigma']}
        +2l_\lambda\left(\sigma ,\sigma_m\right)i
        -2l_\lambda\left(\sigma',\sigma^{(\theta[\zeta,\sigma'])}\right)i
        -2si
      \right)
    \right.
    \nonumber\\&&
    \left.
      \prod_{
        \{\sigma'',\sigma'''\}\in\lambda,\;
        \sigma''\stackrel{\zeta}{\sim}\sigma_m\not\stackrel{\zeta}{\sim}
        \sigma'''\not\stackrel{\zeta}{\sim}\sigma'
      }
        \frac
          {-4}{
            \left(
              x_{\theta_m}
              -x_{\theta[\zeta,\sigma''']}
              +2l_\lambda\left(\sigma '',\sigma_m\right)i
              -2l_\lambda\left(\sigma''',\sigma^{(\theta[\zeta,\sigma'''])}\right)i\right)^2+4}
    \right]
    \nonumber\\&&
    \exp
    \left[-\beta
      \sum_{\sigma\in\theta}
      \frac
        {2N_{\sigma}}
        {
          \left(
            x_{\theta[\zeta,\sigma]}
            +2l_\lambda\left(\sigma,\sigma^{(\theta[\zeta,\sigma])}\right)i\right)^2+1}
    \right].
\label{eq:modification_of_integral_path_1_tmp_4}
\end{eqnarray}
This expression is the sum with respect to
terms corresponding
$(\zeta,\{\sigma^{(\theta_k)}\},\sigma,\sigma',+)$ which avoids the
cancellation, equivalently, which satisfies $\sigma=\sigma_m$

Next, we execute the replacement which is indicated by the
$\delta$-function and the integral with respect to $x_{\sigma_m}$ and
change an expression of summations with respect to permutations.
Then, (\ref{eq:modification_of_integral_path_1_tmp_4}) is modified as
\begin{eqnarray}
&&  \left[\prod_{\sigma\in\theta}N_\sigma\right]^{-1}
  \left[
    \prod_{\{\sigma,\sigma'\}\in\lambda}
    N_{\sigma} N_{\sigma'}
  \right]
  \sum_{\zeta\in\Theta\left(\theta|\lambda|\sigma_1,\cdots,\sigma_{m-1}\right)}
  \sum_{\{\sigma^{(\theta_k)}\}\in\{\theta_k\in\zeta\}_{k>{m-1}},\;\sigma_m\neq\sigma^{(\theta_k)}}
  \left[
    \prod_{k=1}^{m-1}
    \int_{\left|i-x_{\theta_k}\right|_+=\bar\delta_{\sigma_k}}
    \frac{dx_{\theta_k}}{2\pi}
  \right]
\nonumber\\&&
  \left[
    \prod_{k=m}^{N_\zeta}
    \int_{-\infty+i\delta_{\sigma^{(\theta_k)}}}
        ^{+\infty+i\delta_{\sigma^{(\theta_k)}}}
    \frac{dx_{\theta_k}}{2\pi}
  \right]
    \left[
      \prod_{\theta'\in G_\theta\left(\lambda\right)}
      \left(
      \sum_{\sigma\in\theta'}
      \frac
        {2N_{\sigma}L}
        {
          \left(
            x_{\theta[\zeta,\sigma]}
            +2l_\lambda\left(\sigma,\sigma^{(\theta[\zeta,\sigma])}\right)i\right)^2+1}
\right)
    \right]
    \nonumber\\&&
    \left[
      \prod_{\{\sigma,\sigma'\}\in\lambda,\;\sigma\not\stackrel{\zeta}{\sim}\sigma'\not\stackrel{\zeta}{\sim}\sigma_m}
        \frac
          {-4}{
            \left(
               x_{\theta[\zeta,\sigma ]}
              -x_{\theta[\zeta,\sigma']}
              +2l_\lambda\left(\sigma ,\sigma^{(\theta[\zeta,\sigma ])}\right)i
              -2l_\lambda\left(\sigma',\sigma^{(\theta[\zeta,\sigma'])}\right)i\right)^2+4}
    \right]
    \nonumber\\&&
    \exp
    \left[-\beta
      \sum_{\sigma\in\theta}
      \frac
        {2N_{\sigma}}
        {
          \left(
            x_{\theta[\zeta,\sigma]}
            +2l_\lambda\left(\sigma,\sigma^{(\theta[\zeta,\sigma])}\right)i\right)^2+1}
    \right].
\label{eq:modification_of_integral_path_1_tmp_5}
\end{eqnarray}
Here, we have used a simple relation,
\begin{eqnarray}
l_\lambda\left(\sigma^{(\theta[\zeta,\sigma'])},\sigma\right)
&=&l_\lambda\left(\sigma^{(\theta[\zeta,\sigma'])},\sigma'\right)+1+
l_\lambda\left(\sigma_m,\sigma\right)
\nonumber\\
&&{\rm for }\quad
  \zeta\in\Theta\left(\theta|\lambda\right),
\quad
  \{\sigma_m,\sigma'\}\in\lambda,\;\sigma_m\not\stackrel{\zeta}{\sim}\sigma',
\quad
  \sigma\stackrel{\zeta}{\sim}\sigma_m
\end{eqnarray}
In this change, there is a one-to-one correspondence between
$\{\zeta,\{\sigma^{(\theta_k)}\},\sigma'\}$ and
$\{\zeta,\{\sigma^{(\theta_k)}\}\}$
associated with
\begin{eqnarray}
  \sum_{\zeta\in\Theta\left(\theta|\lambda|\sigma_1,\cdots,\sigma_m\right)}
  \sum_{\{\sigma^{(\theta_k)}\}\in\{\theta_k\in\zeta\}_{k>m}}
      \sum_{\{\sigma_m,\sigma'\}\in\lambda,\;\sigma_m\not\stackrel{\zeta}{\sim}\sigma'}  
\leftrightarrow
  \sum_{\zeta\in\Theta\left(\theta|\lambda|\sigma_1,\cdots,\sigma_{m-1}\right)}
  \sum_{\{\sigma^{(\theta_k)}\}\in\{\theta_k\in\zeta\}_{k>{m-1}},\;\sigma_m\neq\sigma^{(\theta_k)}}\!\!\!\!\!\!\!\!\!\!\!
\end{eqnarray}
for $\zeta,\{\sigma^{(\theta_k)}\},\sigma'$ and
$\zeta',\{\sigma^{\prime(\theta'_k)}\}$ which satisfy
conditions,
\begin{eqnarray}
  \theta_m+\theta[\zeta,\sigma']
=
  \theta[\zeta',\sigma_m]
,\;
  \zeta-\theta_m-\theta[\zeta,\sigma']
=
  \zeta-\theta[\zeta',\sigma_m]  
,\;
  \{\sigma^{(\theta_k)}\}-\sigma^{(\theta_m)}
=
  \{\sigma^{\prime(\theta'_k)}\},
\end{eqnarray}
where $\zeta,\{\sigma^{(\theta_k)}\},\sigma'$ correspond to 
$\zeta',\{\sigma^{\prime(\theta'_k)}\}$.
This correspondence is a ``one-to-one correspondence''.

In this way, the second term in (\ref{eq:modification_of_integral_path_1_tmp_2}) becomes the expression (\ref{eq:modification_of_integral_path_1_tmp_5}).
Using this expression (\ref{eq:modification_of_integral_path_1_tmp_2})
is
\begin{eqnarray}
&&
  \left[\prod_{\sigma\in\theta}N_\sigma\right]^{-1}
  \left[
    \prod_{\{\sigma,\sigma'\}\in\lambda}
    N_{\sigma} N_{\sigma'}
  \right]
  \sum_{\zeta\in\Theta\left(\theta|\lambda|\sigma_1,\cdots,\sigma_{m-1}\right)}
  \sum_{\{\sigma^{(\theta_k)}\}\in\{\theta_k\in\zeta\}_{k>{m-1}}}
  \left[
    \prod_{k=1}^{m-1}
    \int_{\left|i-x_{\theta_k}\right|_+=\bar\delta_{\sigma_k}}
    \frac{dx_{\theta_k}}{2\pi}
  \right]
\nonumber\\&&
  \left[
    \prod_{k=m}^{N_\zeta}
    \int_{-\infty+i\delta_{\sigma^{(\theta_k)}}}
        ^{+\infty+i\delta_{\sigma^{(\theta_k)}}}
    \frac{dx_{\theta_k}}{2\pi}
  \right]
    \left[
      \prod_{\theta'\in G_\theta\left(\lambda\right)}
      \left(
      \sum_{\sigma\in\theta'}
      \frac
        {2N_{\sigma}L}
        {
          \left(
            x_{\theta[\zeta,\sigma]}
            +2l_\lambda\left(\sigma,\sigma^{(\theta[\zeta,\sigma])}\right)i\right)^2+1}
\right)
    \right]
    \nonumber\\&&
    \left[
      \prod_{\{\sigma,\sigma'\}\in\lambda,\;\sigma\not\stackrel{\zeta}{\sim}\sigma'\not\stackrel{\zeta}{\sim}\sigma_m}
        \frac
          {-4}{
            \left(
               x_{\theta[\zeta,\sigma ]}
              -x_{\theta[\zeta,\sigma']}
              +2l_\lambda\left(\sigma ,\sigma^{(\theta[\zeta,\sigma ])}\right)i
              -2l_\lambda\left(\sigma',\sigma^{(\theta[\zeta,\sigma'])}\right)i\right)^2+4}
    \right]
    \nonumber\\&&
    \exp
    \left[-\beta
      \sum_{\sigma\in\theta}
      \frac
        {2N_{\sigma}}
        {
          \left(
            x_{\theta[\zeta,\sigma]}
            +2l_\lambda\left(\sigma,\sigma^{(\theta[\zeta,\sigma])}\right)i\right)^2+1}
    \right].
\label{eq:modification_of_integral_path_1_tmp_6}
\end{eqnarray}
Here, we have used 
a simple relation,
\begin{eqnarray}
&&
  \sum_{\zeta\in\Theta\left(\theta|\lambda|\sigma_1,\cdots,\sigma_{m}\right)}
  \sum_{\{\sigma^{(\theta_k)}\}\in\{\theta_k\in\zeta\}_{k>m}}
 +\sum_{\zeta\in\Theta\left(\theta|\lambda|\sigma_1,\cdots,\sigma_{m-1}\right)}
  \sum_{\{\sigma^{(\theta_k)}\}\in\{\theta_k\in\zeta\}_{k>{m-1}},\;\sigma_m\neq\sigma^{(\theta_k)}}
\nonumber\\&=&
  \sum_{\zeta\in\Theta\left(\theta|\lambda|\sigma_1,\cdots,\sigma_{m-1}\right)}
  \sum_{\{\sigma^{(\theta_k)}\}\in\{\theta_k\in\zeta\}_{k>{m-1}}}.
\end{eqnarray}

When $m$ is replaced with $m-1$,
(\ref{eq:modification_of_integral_path_1_tmp_1}) becomes
(\ref{eq:modification_of_integral_path_1_tmp_6}). And (\ref{eq:modification_of_integral_path_1_tmp_1}) has been shown to be
equivalent to (\ref{eq:modification_of_integral_path_1_tmp_6}).
Thus, we have proved that the value of
(\ref{eq:modification_of_integral_path_1_tmp_1}) does not depend on
 $m$

\section{A proof with respect to a permutation 1}
\label{sec:proof_w.r.t_set_1}
We prove
\begin{eqnarray}
  \sum_{\theta\in\Theta_n}
  \left(
    \prod_{\sigma\in\theta}
    \left(-\right)^{N_\sigma-1}
    \left(N_\sigma-1\right)!
  \right)
  x^{N_\theta}
&=&
  x\left(x-1\right)\cdots\left(x-n+1\right).
  \label{eq:hodai_5}
\end{eqnarray}
For the purpose,
we introduce an expression
\begin{eqnarray}
  \prod_{i=2}^{n}
  \left(1-\sum_{j=1}^{i-1}y_{i,j}\right),
\end{eqnarray}
and expand this product. We interpret $y_{i,j}$ as a connection
between elements $i$ and $j$, and a product of $y_{i,j}$
as a set of connections.  Then, the expansion can be
considered as a linear combination with respect to a set $\lambda$ of
connections, and the coefficient of each term is $(-1)^{N_\lambda}$. In
fact, all the set $\lambda$ is an element of
$\Lambda\left(\{1,\cdots,n\}\right)$.  Any two
elements in $\{1,\cdots,n\}$ are connected or not, and there is no
closed path in the connections.  That is to say, there is no term like
$y_{2,1}y_{2,1}$ or $y_{2,1}y_{3,2}y_{3,1}$.

By replacing of $\lambda$ with $\theta=G_{\{1,\cdots,n\}}(\lambda)$,
the expansion can be considered as a linear combination with
respect to $\theta\in\Theta_n$.  And, it is easily shown that the
coefficient of $\theta\in\Theta_n$ is $ \prod_{\sigma\in\theta}
\left(-\right)^{N_\sigma} \left(N_\sigma-1\right)!$.  The reason is
that $ \prod_{\sigma\in\theta} \left(-\right)^{N_\sigma}$
is caused by $(-1)^{N_\lambda}$ and $ \prod_{\sigma\in\theta}
\left(N_\sigma-1\right)!$ counts the number of terms when $\lambda$ is
replaced with the $\theta$.  Due to this fact and a relation
$N_\theta=n-N_\lambda$ where $\theta=G_{\{1,\cdots,n\}}(\lambda)$, the l.h.s. of (\ref{eq:hodai_5}) is equal to
\begin{eqnarray}
  x^n\prod_{i=2}^{n}
  \left(1-\sum_{j=1}^{i-1}x^{-1}\right).
\end{eqnarray}
It is easy to see that this expression is equivalent to the r.h.s. of 
(\ref{eq:hodai_5}).

\section{A proof with respect to a permutation 2}
\label{sec:proof_w.r.t_set_2}
\label{sec:moveing_of_integral_path}
Here, we prove
eq.(\ref{eq:moveing_of_integral_path_coodinate_relation}).

Using a relation (\ref{eq:hodai_5}), the function with respect to $\sigma_m$
and $\sigma_{m-1}$
\begin{eqnarray}
        \sum_{\theta'\in\Theta\left(\sigma_m\right)}
      N_{\sigma_{m-1}}^{N_{\theta'}}
      \prod_{\sigma'\in\theta'}
      \left(-\right)^{N_{\sigma'}-1}\left(N_{\sigma'}-1\right)!
\end{eqnarray}
in the l.h.s of (\ref{eq:moveing_of_integral_path_coodinate_relation})
is simplifies as 
\begin{eqnarray}
\theta\left(N_{\sigma_m-1}- N_{\sigma_{m}}+\frac12\right)
      \frac{N_{\sigma_{m-1}}!}{\left(N_{\sigma_{m-1}}-N_{\sigma_m}\right)!},
\label{eq:moveing_of_integral_path_tmp_01}
\end{eqnarray}
and therefore the l.h.s of (\ref{eq:moveing_of_integral_path_coodinate_relation}) becomes
\begin{eqnarray}
  \theta_>\left(\tilde\theta\right)
    \left(-\right)^{N_{\sigma_1}-1}\left(N_{\sigma_1}-1\right)!
    N_{\sigma_1}^{-1}
    \prod_{\sigma_{m>1}\in\tilde\theta}
      \frac{N_{\sigma_{m-1}}!}{\left(N_{\sigma_{m-1}}-N_{\sigma_m}\right)!}.
\end{eqnarray}

A remaining task is to prove the relation
\begin{eqnarray}
&&
  \sum_{\tilde\zeta'\in D\left(\tilde\theta\right)}
  \sum_{\lambda'\in\Lambda_c\left(\tilde\zeta'\right)}
  \delta\left(\tilde\zeta',\lambda',l\right)
  \left[
    \prod_{\tilde\theta'\in\tilde\zeta'}
    \left(
      \left(-\right)^{M_{\tilde\theta'}-1}\left(M_{\tilde\theta'}-1\right)!
      M_{\tilde\theta'}^{-1}
      \left(
        M_{\tilde\theta'}!
      \right)^{N_{\tilde\theta'}-1}
    \right)
  \right]
\nonumber\\&&
  \left[
    \prod_{\{\tilde\theta',\tilde\theta''\}\in\lambda'}
    \left(-M_{\tilde\theta'}M_{\tilde\theta''}\right)
  \right]
\nonumber\\&=&
  \left(-\right)^{N_{\sigma_1}-1}\left(N_{\sigma_1}-1\right)!
  N_{\sigma_1}^{-1}
  \prod_{\sigma_{k>1}\in\tilde\theta}
  \frac{N_{\sigma_{k-1}}!}{\left(N_{\sigma_{k-1}}-N_{\sigma_{k}}\right)!},
\label{eq:proof_w.r.t_set_2_tmp_1}
\end{eqnarray}
where $\tilde\theta$ is written as
$(\sigma_1,\sigma_2,\cdots,\sigma_m)$, and the number of elements in
each set $\sigma_k$ is decreasing in the order of the sequence
$\tilde\theta$, and $l$ is an element of $\sigma_1$.

We use a symbol
$\sum_{\tilde\theta'\leftarrow\tilde\theta}$ defined as follows for a
convenience' sake.  The 
 summation $\sum_{\tilde\theta'\leftarrow\tilde\theta}$  with respect to $\tilde\theta'$ satisfies
conditions  related to $\tilde\theta$,
\begin{eqnarray}
  \sigma'_k\;\subset\;\sigma_k
\quad
  l\;\in\;\sigma'_1
\quad
  N_{\sigma'_k}\;=\;N_{\sigma_k}-N_{\sigma_ms}
\end{eqnarray}
where $(\sigma_1,\cdots,\sigma_m)=\tilde\theta$ and
$(\sigma'_1,\cdots,\sigma_{m-1})=\tilde\theta'$.  We change an
expression of summations in the l.h.s. of
(\ref{eq:proof_w.r.t_set_2_tmp_1}) with respect to permutations.
The l.h.s. of
(\ref{eq:proof_w.r.t_set_2_tmp_1}) becomes
\begin{eqnarray}
&&
  \sum_{\tilde\theta'\leftarrow\tilde\theta}
  \left[
    \left(-\right)^{N_{\sigma_m}}
    \sum_{\theta\in\Theta\left(\sigma_m\right)}
    \left(N_{\sigma_1}-N_{\sigma_m}\right)^{N_{\theta}}
    \left[\frac{N_{\sigma_m}!}{\prod_{\sigma\in\theta}N_\sigma!}\right]^{m-1}
    \prod_{\sigma\in\theta}
    \left(N_\sigma-1\right)!\left(N_\sigma!\right)^{m-1}
  \right]
  \sum_{\tilde\zeta'\in D\left(\tilde\theta'\right)}
  \sum_{\lambda'\in\Lambda_c\left(\tilde\zeta'\right)}
\nonumber\\&&
  \delta\left(\tilde\zeta',\lambda',m\right)
  \left[
    \prod_{\tilde\theta''\in\tilde\zeta'}
    \left(
      \left(-\right)^{M_{\tilde\theta''}-1}\left(M_{\tilde\theta''}-1\right)!
      M_{\tilde\theta''}^{-1}
      \left(
        M_{\tilde\theta''}!
      \right)^{N_{\tilde\theta''}-1}
    \right)
  \right]
  \left[
    \prod_{\{\tilde\theta'',\tilde\theta'''\}\in\lambda'}
    \left(-M_{\tilde\theta''}M_{\tilde\theta'''}\right)
  \right].
\label{eq:proof_w.r.t_set_2_tmp_2}
\end{eqnarray}
In this change, there is a one-to-many correspondence between
$\{\tilde\zeta',\lambda'\}$ and $\{\theta,\tilde\zeta',\lambda'\}$
associated with
\begin{eqnarray}
  \sum_{\tilde\zeta'\in D\left(\tilde\theta\right)}
  \sum_{\lambda'\in\Lambda_c\left(\tilde\zeta'\right)}
&\leftrightarrow&
  \sum_{\tilde\theta'\leftarrow\tilde\theta}
  \sum_{\theta\in\Theta\left(\sigma_m\right)}
  \sum_{\tilde\zeta'\in D\left(\tilde\theta'\right)}
  \sum_{\lambda'\in\Lambda_c\left(\tilde\zeta'\right)}
\end{eqnarray}
When $\tilde\zeta',\lambda'$ correspond to a subset $A$ of
$\{\theta,\tilde\zeta',\lambda'\}$,  $\tilde\zeta',\lambda'$ and
$\theta'',\tilde\zeta'',\lambda''$ in $A$ satisfy the following
conditions.  There is a bijection $f$ from a set
$\{\tilde\theta\in\tilde\zeta|N_{\tilde\theta}=m\}$ to the set
$\theta''$ which satisfies
 $\sigma'_m=\sigma''$; $f(\tilde\theta'=(\sigma'_1,\cdots))=\sigma''$.
The conditions for $\tilde\zeta''$ and $\lambda''$ are
\begin{eqnarray}
  \tilde\zeta''
\;=\;
  \left\{
    \tilde\theta'|\tilde\theta'\in\tilde\zeta,N_{\tilde\theta'}\neq m
  \right\},
\quad
  \lambda''
\;=\;
\lambda\left[\lambda',\tilde\zeta''\right].
\end{eqnarray}
This correspondence is a ``one-to-many correspondence''.

In fact, the l.h.s. of (\ref{eq:proof_w.r.t_set_2_tmp_1}) depends only
on a decreasing sequence
$(N_{\sigma_1},N_{\sigma_2},\cdots,N_{\sigma_m})$ where $\tilde\theta$
is written as $({\sigma_1},{\sigma_2},\cdots,{\sigma_m})$ which is a
sequence defining the expression (\ref{eq:proof_w.r.t_set_2_tmp_1}).
Then, we can regard the l.h.s. as a function
$F(N_{\sigma_1},N_{\sigma_2},\cdots,N_{\sigma_m})$.  This fact
indicates that elements which are summed up by means of
$\sum_{\tilde\theta'\leftarrow\tilde\theta}$ in
(\ref{eq:proof_w.r.t_set_2_tmp_2}) have the same value.  We note that
the function $F$ is defined for any number of independent variables,
e.g. $F(n_1)$, $F(n_1,n_2)$.  Using the function $F$, the equivalence
between the l.h.s. of (\ref{eq:proof_w.r.t_set_2_tmp_1}) and
(\ref{eq:proof_w.r.t_set_2_tmp_2}) is written as
\begin{eqnarray}
&&
F\left(n_1,n_2,\cdots,n_m\right)
\nonumber\\&=&
  (-)^{n_m}
  \sum_{\theta\in\Theta_{n_m}}
  \left(n_1-n_m\right)^{N_\theta}
  \left[\frac{n_m!}{\prod_{\sigma\in\theta}N_\sigma!}\right]^{m-1}
  \left[
    \prod_{\sigma\in\theta}
    \left(N_\sigma-1\right)!\left(N_\sigma!\right)^{m-1}
  \right]
\nonumber\\&&
  F\left(n_1-n_m,n_2-n_m,\cdots,n_{m-1}-n_m\right)
\nonumber\\&&
  \left[
    \prod_{k=2}^{m-1}
    \frac{n_k!}{\left(n_k-n_m\right)!n_m!}
  \right]
  \frac{\left(n_1-1\right)!}{\left(n_1-n_m-1\right)!n_m!}.
\label{eq:proof_w.r.t_set_2_tmp_3}
\end{eqnarray}
The factor $ \left[ \prod_{k=2}^{m-1}
  \frac{n_k!}{\left(n_k-n_m\right)!n_m!} \right]
\frac{\left(n_1-1\right)!}{\left(n_1-n_m-1\right)!n_m!}$ is the number
of elements which is summed up with respect to
$\sum_{\tilde\theta'\leftarrow\tilde\theta}$. Using the relation
(\ref{eq:hodai_5}), we simplify (\ref{eq:proof_w.r.t_set_2_tmp_3}) as
\begin{eqnarray}
&&
F\left(n_1,n_2,\cdots,n_m\right)
\nonumber\\
&=&
(-)^{n_m}F\left(n_1-n_m,\cdots,n_{m-1}-n_m\right)
\left[\frac{
\left(n_1-1\right)!
}{
  \left(n_1-n_m-1\right)!
}
\right]^2
\left[\prod_{k=2}^{m-1}
  \frac{n_k!}{\left(n_k-n_m\right)!}
\right]
\label{eq:proof_w.r.t_set_2_tmp_6}
\end{eqnarray}

In case $m=1$, 
eq.(\ref{eq:proof_w.r.t_set_2_tmp_1}) obviously holds, which gives
\begin{eqnarray}
  F\left(n_1\right)
&=&
  \left(-\right)^{n_1-1}\left(n_1-1\right)!
  n_1^{-1}.
\label{eq:proof_w.r.t_set_2_tmp_4}
\end{eqnarray}
On the other hand, eq.(\ref{eq:proof_w.r.t_set_2_tmp_6}) can
be regarded as an inductive relation of $F$.  Therefore, all we have to
do for a proof of the relation (\ref{eq:proof_w.r.t_set_2_tmp_1}) is
to show
\begin{eqnarray}
  F\left(n_1,n_2,\cdots,n_m\right)
&=&
  \left(-\right)^{n_1-1}\left(n_1-1\right)!
  n_1^{-1}
  \prod_{k=2}^{m}
  \frac{n_{k-1}!}{\left(n_{k-1}-n_{k}\right)!}
\label{eq:proof_w.r.t_set_2_tmp_5}
\end{eqnarray}
from (\ref{eq:proof_w.r.t_set_2_tmp_6}) and
(\ref{eq:proof_w.r.t_set_2_tmp_4}). It is sufficient to derive (\ref{eq:proof_w.r.t_set_2_tmp_5}) from
(\ref{eq:proof_w.r.t_set_2_tmp_6}) by supposing an expression of
$F\left(n_1-n_m,\cdots,n_{m-1}-n_m\right)$ like
(\ref{eq:proof_w.r.t_set_2_tmp_5}), i.e.
\begin{eqnarray}
  \left(-\right)^{n_1-n_m-1}
\frac{\left(n_1-n_m-1\right)!}{n_1-n_m}
  \prod_{k=2}^{m-1}
  \frac{\left(n_{k-1}-n_m\right)!}{\left(n_{k-1}-n_{k}\right)!}.
\label{eq:proof_w.r.t_set_2_tmp_7}
\end{eqnarray}
It is clear that (\ref{eq:proof_w.r.t_set_2_tmp_5}) is given when we
substitute (\ref{eq:proof_w.r.t_set_2_tmp_7}) for
$F\left(n_1-n_m,n_2-n_m,\cdots,n_{m-1}\right)$ in
(\ref{eq:proof_w.r.t_set_2_tmp_6}).

\section{Modifications of integral paths 2}
\label{sec:proof_of_integral_path}
We prove the equivalence of 
(\ref{eq:moveing_of_integral_path_tmp_2})
 and (\ref{eq:moveing_of_integral_path_tmp_3}).

A sufficient condition for the equivalence is 
\begin{eqnarray}
&&
  \sum_{
    \lambda'\subseteq \lambda,\;
    \xi\equiv G_{\tilde\zeta}\left(\lambda'\right)
  }
  \left[
    \prod_{\tilde\zeta'\in \xi}
    \delta\left(
      \tilde\zeta',
      \lambda\left[\lambda',\tilde\zeta'\right],
      {\min}_1\left(\tilde\zeta'\right)
    \right)
    \int_{-\infty+i\min_1\left(\tilde\zeta'\right)\delta}
        ^{+\infty+i\min_1\left(\tilde\zeta'\right)\delta}
    dx_{\tilde\zeta'}
  \right]
\nonumber\\&&
  \left[
    \prod_{
      \{\tilde\theta,\tilde\theta'\}\in\lambda-\lambda'
    }
    f_{N_{\tilde\theta},N_{\tilde\theta'}}\left(
      x_{\tilde\zeta\left[\xi,\tilde\theta \right]}-
      x_{\tilde\zeta\left[\xi,\tilde\theta'\right]}
    \right)
  \right]
  g\left(
    x_{
      \tilde\zeta\left[
        \xi,
        \tilde\theta_1
      \right]
    },    
    x_{
      \tilde\zeta\left[
        \xi,
        \tilde\theta_2
      \right]
    },    
    \cdots
  \right)
\nonumber\\&=&
  \left[
    \prod_{\tilde\theta\in\tilde\zeta}
    \int_{-\infty-iN_{\tilde\theta}\delta}
        ^{+\infty-iN_{\tilde\theta}\delta}
    dx_{\tilde\theta}
  \right]
  \left[
    \prod_{\{\tilde\theta,\tilde\theta'\}\in\lambda}
    f_{N_{\tilde\theta},N_{\tilde\theta'}}\left(
       x_{\tilde\theta }
      -x_{\tilde\theta'}
    \right)
  \right]
g\left(x_{\tilde\theta_1},x_{\tilde\theta'_2},\cdots\right)
\label{eq:proof_of_integral_path_tmp_1}
\end{eqnarray}
where $\theta\in\Theta(\sigma\subset\mathbb{N})$,
$\tilde\zeta\in\tilde{\bar\Theta}\left(\theta\right)$,
$\lambda\in\Lambda\left(\tilde\zeta\right)$ and
$\{\tilde\theta_1,\tilde\theta_2,\cdots\}=\tilde\zeta$.  Both sides of
the equation are functions of $\tilde\zeta$ and
$\lambda$.  Here, we have used several functions defined below.
The definition of $\tilde\zeta[\xi,\tilde\theta]$ is the
same as $\theta\left[\zeta,\sigma\right]$.

  Let $\xi$ be a set with a finite number of sets as elements,
  and $\tilde\theta$ be an element in a set which is an element of
  $\xi$. Then, $\tilde\zeta[\xi,\tilde\theta]$ is a set which is in $\xi$
  and includes $\tilde\theta$,
  \begin{eqnarray}
    \tilde\theta\in\tilde\zeta[\xi,\tilde\theta]\in\xi.
  \end{eqnarray}
The definition of the function $f_{n,m}$ for natural numbers $n,m$ is
\begin{eqnarray}
  f_{n,m}\left(x\right)
&\equiv&
  \frac{2\theta\left(n-m\right)-1}{2\pi i x}
  +
  \tilde f_{n,m}\left(x\right)
\label{eq:proof_of_integral_path_tmp_01}
\end{eqnarray}
where $\tilde f_{n,m}(x)$ is a function of $x$ which satisfies
the following conditions. First, the function is analytic on neighborhood
of the real axis. Second, the function does not diverge in the limit
$x\rightarrow \infty$. Finally, the function satisfies $\tilde
f_{n,m}(x)=\tilde f_{m,n}(-x)$.   In
(\ref{eq:proof_of_integral_path_tmp_1}), $g(\cdots)$ is an analytic function on
neighborhood of the real axis. And, the multiple integral of $g$ from
$-\infty$ to $\infty$ with respect to the all variables has a finite
value.  These definitions imply that the relation such as
\begin{eqnarray*}
&&
\int_{-\infty}^{+\infty}dy
\int_{-\infty-i\delta}^{+\infty-i\delta}dx
f_{n,m}\left(x-y\right)g\left(x,y\right)
\\
&=&
\int_{-\infty}^{+\infty}dy
\int_{-\infty+i\delta}^{+\infty+i\delta}dx
f_{n,m}\left(x-y\right)g\left(x,x\right)
-
\int_{-\infty}^{+\infty}dx
g\left(x,x\right),
\end{eqnarray*}
holds for $n<m$.

The reason why the condition (\ref{eq:proof_of_integral_path_tmp_1}) is a sufficient condition of the equivalence between
(\ref{eq:moveing_of_integral_path_tmp_2})
 and (\ref{eq:moveing_of_integral_path_tmp_3}) is as follows.
When   we apply
$\sum_{\theta\in\Theta_M}$,
$\sum_{\tilde\zeta\in\tilde{\bar\Theta}\left(\theta\right)}$ and
$\sum_{\lambda\in\Lambda\left(\tilde\zeta\right)}$
to both sides of (\ref{eq:proof_of_integral_path_tmp_1}) and
 define functions $f$ and $g$ as
\begin{eqnarray}
  f_{n,m}\left(x\right)
&\equiv&
  \frac1{2\pi}K_{n,m}\left(x+\left(n-m\right)i\right)
\nonumber\\
  g\left(x_{\tilde\theta_1},x_{\tilde\theta'_2},\cdots\right)
&\equiv&
\frac1{2\pi}
  \left[
    \prod_{\tilde\zeta'\in G_{\tilde\zeta}\left(\lambda\right)}
    \left(
      \sum_{\tilde\theta\in\tilde\zeta}
      \frac
        {2M_{\tilde\theta}N_{\tilde\theta}}
        {
          \left(x_{\tilde\theta}+\left(N_{\tilde\theta}-1\right)i\right)^2
          +N_{\tilde\theta}^2
        }
    \right)
  \right]
  \left[
    \prod_{\{\tilde\theta,\tilde\theta'\}\in\lambda}
    \frac{-M_{\tilde\theta}M_{\tilde\theta'}}{2\pi}
  \right]
\nonumber\\&&
  e^{
    -\beta
    \sum_{\tilde\theta\in\tilde\zeta}
    \frac
      {2M_{\tilde\theta}N_{\tilde\theta}}
      {
        \left(
          x_{\tilde\theta}+\left(N_{\tilde\theta}-1\right)i
        \right)^2
        +N_{\bar\theta}^2
      }
    }
\prod_{\tilde\theta\in\tilde\zeta}
\left[
  \left(-\right)^{M_{\tilde\theta}-1}\left(M_{\tilde\theta}-1\right)!
  M_{\tilde\theta}^{-1}
  \left(
    M_{\tilde\theta}!
  \right)^{N_{\tilde\theta}-1}
\right],
\end{eqnarray}
which satisfy the previous conditions, both sides of this equation
become (\ref{eq:moveing_of_integral_path_tmp_2}) and
(\ref{eq:moveing_of_integral_path_tmp_3}). Therefore, we may prove
(\ref{eq:proof_of_integral_path_tmp_1}) by means of the mathematical
induction with respect to $N_{\tilde\zeta}$.

It is clear that (\ref{eq:proof_of_integral_path_tmp_1}) holds in case
of $N_{\tilde\zeta}=1$.  We suppose that
(\ref{eq:proof_of_integral_path_tmp_1}) holds for
$\tilde\zeta,\lambda$ which satisfy
\begin{eqnarray}
&
 \theta_+\;\in\;\Theta(\sigma_+\subset\mathbb{N}),
 \quad 
\tilde\zeta_+\;\in\;\tilde{\bar\Theta}(\theta_+),
\quad
\lambda_+\;\in\;\Lambda(\tilde\zeta_+),
\label{eq:proof_of_integral_path_tmp_02}
\\&
  \tilde\theta_0\;\in\;\tilde\zeta_+,
\quad
  \tilde\zeta\;=\;\tilde\zeta_+-\tilde\theta_0,
\quad
  \lambda\;=\;\lambda[\lambda_+,\tilde\zeta],
\quad
N_{\lambda_+}-N_\lambda\;=\;0\;{\rm or}\; 1.
\label{eq:proof_of_integral_path_tmp_03}
\end{eqnarray}
We note that for any $\tilde\zeta_+$ and $\lambda_+$ satisfying
(\ref{eq:proof_of_integral_path_tmp_02}), there are $\tilde\zeta$ and
$\lambda$ which satisfy (\ref{eq:proof_of_integral_path_tmp_03}).
From now on we prove that (\ref{eq:proof_of_integral_path_tmp_1})
holds for the sets $\tilde\zeta_+,\lambda_+$.  Hereafter in this
appendix, (\ref{eq:proof_of_integral_path_tmp_1}) for the sets
$\tilde\zeta,\lambda$ is just referred to as
(\ref{eq:proof_of_integral_path_tmp_1}), while
(\ref{eq:proof_of_integral_path_tmp_1}) for the sets
$\tilde\zeta_+,\lambda_+$ is referred to as
(\ref{eq:proof_of_integral_path_tmp_1})${}_+$.

In case of $N_\lambda=N_{\lambda_+}$, we replace
$g\left(x_1,x_2\cdots\right)$ in
(\ref{eq:proof_of_integral_path_tmp_1}) with
\begin{eqnarray}
  g\left(x_1,x_2\cdots\right)
&\leftarrow&
  \int_{-\infty}
      ^{+\infty}
      dx_{\tilde\theta_0}
  g\left(x_0,x_1,\cdots\right).
\end{eqnarray}
 Then, (\ref{eq:proof_of_integral_path_tmp_1}) becomes
(\ref{eq:proof_of_integral_path_tmp_1})${}_+$.  Therefore,
(\ref{eq:proof_of_integral_path_tmp_1})${}_+$ holds in case (\ref{eq:proof_of_integral_path_tmp_1}) does.

Next we analyze (\ref{eq:proof_of_integral_path_tmp_1})${}_+$ in the
case of $N_\lambda+1=N_{\lambda_+}$, which means that
$\tilde\theta_0=(\sigma_1^{(0)},\cdots)$ connects a sequence
$\tilde\theta_1=(\sigma_1^{(1)},\cdots)\in\tilde\zeta_+$ by
$\lambda_+$.  We subdivide the case into three,
\begin{eqnarray}
  \makebox{Case}1
\quad
  N_{\tilde\theta_0}\;=\;N_{\tilde\theta_1},
&
  \makebox{Case}2
\quad
  N_{\tilde\theta_0}\;>\;N_{\tilde\theta_1},
&
  \makebox{Case}3
\quad
  N_{\tilde\theta_0}\;<\;N_{\tilde\theta_1}.
\end{eqnarray}

Case 1: $N_{\tilde\theta_0}=N_{\tilde\theta_1}$ \\
We replace $g\left(x_1,x_2\cdots\right)$ in
(\ref{eq:proof_of_integral_path_tmp_1}) with
\begin{eqnarray}
  g\left(x_1,x_2\cdots\right)
  &\leftarrow&
  \int_{-\infty}
  ^{+\infty}
  dx_{\tilde\theta_0}
  g\left(x_0,x_1,\cdots\right)
  f_{N_{\tilde\theta_0},N_{\tilde\theta_1}}\left(
    x_{\tilde\theta_0}-x_{\tilde\theta_1}
  \right).    
\end{eqnarray}
Then, (\ref{eq:proof_of_integral_path_tmp_1}) becomes
(\ref{eq:proof_of_integral_path_tmp_1})${}_+$.  Therefore,
(\ref{eq:proof_of_integral_path_tmp_1})${}_+$ holds in case
(\ref{eq:proof_of_integral_path_tmp_1}) does.

Case 2: $ N_{\tilde\theta_0}>N_{\tilde\theta_1}$
\\
We replace $g\left(x_1,x_2\cdots\right)$ in
(\ref{eq:proof_of_integral_path_tmp_1}) with
\begin{eqnarray}
  g\left(x_1,x_2\cdots\right)
&\leftarrow&
  \int_{-\infty-iN_{\tilde\theta_0}\delta}
      ^{+\infty-iN_{\tilde\theta_0}\delta}
      dx_{\tilde\theta_0}
  g\left(x_0,x_1,\cdots\right)
  f_{N_{\tilde\theta_0},N_{\tilde\theta_1}}\left(
    x_{\tilde\theta_0}-x_{\tilde\theta_1}
  \right).
\end{eqnarray}
We apply this relation to the r.h.s. of
(\ref{eq:proof_of_integral_path_tmp_1})${}_+$. The
integral path with respect to $x_{\tilde\theta_0}$  is changed into
$(-\infty+i\min(\sigma_1^{(0)})\delta,+\infty+i\min(\sigma_1^{(0)})\delta)$. Then,
the r.h.s. of (\ref{eq:proof_of_integral_path_tmp_1})${}_+$ becomes
\begin{eqnarray}
&&  \sum_{
    \lambda'\subseteq \lambda,\;
    \xi\equiv G_{\tilde\zeta_+}\left(\lambda'\right)
  }
  \left[
    \prod_{\tilde\zeta'\in \xi}
    \delta\left(
      \tilde\zeta',
      \lambda\left[\lambda',\tilde\zeta'\right],
      {\min}_1\left(\tilde\zeta'\right)
    \right)
    \int_{-\infty+i\min_1\left(\tilde\zeta'\right)\delta}
        ^{+\infty+i\min_1\left(\tilde\zeta'\right)\delta}
    dx_{\tilde\zeta'}
  \right]
\nonumber\\&&
  \left[
    \prod_{
      \{\tilde\theta,\tilde\theta'\}\in\lambda_+-\lambda'
    }
    f_{N_{\tilde\theta},N_{\tilde\theta'}}\left(
      x_{\tilde\zeta\left[\xi,\tilde\theta \right]}-
      x_{\tilde\zeta\left[\xi,\tilde\theta'\right]}
    \right)
  \right]
  g\left(
    x_{
      \tilde\zeta\left[
        \xi,
        \tilde\theta_0
      \right]
    },    
    x_{
      \tilde\zeta\left[
        \xi,
        \tilde\theta_1
      \right]
    },    
    \cdots
  \right)
\nonumber\\&&{}+
  \sum_{
    \lambda'\subseteq \lambda,\;
    \lambda_+'\equiv\lambda'+\{\tilde\theta_0,\tilde\theta_1\},\;
    \xi_+\equiv G_{\tilde\zeta_+}\left(\lambda'_+\right)
  }
  \left[
    \prod_{
      \tilde\zeta'
      \in
      \xi_+
    }
    \delta\left(
      \tilde\zeta',
      \lambda\left[
        \lambda'_+,
        \tilde\zeta'\right],
      {\min}_1\left(\tilde\zeta'\right)
    \right)
    \int_{-\infty+i\min_1\left(\tilde\zeta'\right)\delta}
        ^{+\infty+i\min_1\left(\tilde\zeta'\right)\delta}
    dx_{\tilde\zeta'}
  \right]
\nonumber\\&&
  \left[
    \prod_{
      \{\tilde\theta,\tilde\theta'\}\in\lambda_+
      -\lambda'_+
    }
    f_{N_{\tilde\theta},N_{\tilde\theta'}}\left(
      x_{\tilde\zeta\left[\xi_+,\tilde\theta \right]}-
      x_{\tilde\zeta\left[\xi_+,\tilde\theta'\right]}
    \right)
  \right]
  g\left(
    x_{
      \tilde\zeta\left[
        \xi_+,
        \tilde\theta_0
      \right]
    },    
    x_{
      \tilde\zeta\left[
        \xi_+,
        \tilde\theta_1
      \right]
    },    
    \cdots
  \right).
\label{eq:proof_of_integral_path_tmp_4}
\end{eqnarray}
The second term is a sum of the residues generated when the integral
path with respect to $x_{\{\tilde\theta_0\}}$ is moved from $\Im
x_{\{\tilde\theta_0\}}=-N_{\tilde\theta_0}\delta$ to
$\min(\sigma^{(1)}_0)\delta$.  The residues are due to poles at $ x_
{ \tilde\zeta\left[ G_{\tilde\zeta_+}\left(\lambda'\right),
    \tilde\theta_1 \right]} =x_{\{\tilde\theta_0\}}$. Note that a
residue characterized by a set $\lambda'$ in
$\sum_{\lambda'\subseteq\lambda}$ of the second term is generated by a
term characterized by the same set $\lambda'$ in
$\sum_{\lambda'\subseteq\lambda}$ of the first term, and, the
following fact enable us to write these residues like the second
terms.  In case
\begin{eqnarray}
  \delta\left(\tilde\zeta_0,
      \lambda\left[
        \lambda',
        \tilde\zeta_0\right],
      {\min}_1\left(\tilde\zeta_0\right)
    \right)
&=&1
\end{eqnarray}
and 
\begin{eqnarray}
  \min\left(
    \sigma_1^{(0)}
    \right)
&>&
  {\min}_1\left(
    \tilde\zeta\left[
G_{\tilde\zeta}\left(\lambda'\right)      
,
      \tilde\theta_1
    \right]
  \right)
\label{eq:proof_of_integral_path_tmp_5},
\end{eqnarray}
which means that the pole is in a region surrounded by
integral paths $\Im x_{\{\tilde\theta_0\}}=-N_{\tilde\theta_0}\delta$
and $\min(\sigma^{(1)}_0)\delta$ , it follows that
\begin{eqnarray}
    \delta\left(
      \tilde\zeta_1,
      \lambda\left[
        \lambda',
        \tilde\zeta_1\right],
      {\min}_1\left(\tilde\zeta_1\right)
    \right)
\;=\;1
&\Leftrightarrow&
    \delta\left(
      \tilde\zeta_1',
      \lambda\left[
        \lambda'_+,
        \tilde\zeta_1'\right],
      {\min}_1\left(\tilde\zeta_1'\right)
    \right)
\;=\;1,
\end{eqnarray}
and in the other case,
\begin{eqnarray}
    \delta\left(
      \tilde\zeta_1',
      \lambda\left[
        \lambda'_+,
        \tilde\zeta_1'\right],
      {\min}_1\left(\tilde\zeta_1'\right)
    \right)
&=&0,
\end{eqnarray}
where we suppose $ N_{\tilde\theta_0}>N_{\tilde\theta_1}$,
$\lambda'\subseteq\lambda$,
$\lambda_+'\equiv\lambda'+\{\tilde\theta_0,\tilde\theta_1\}$, $\xi
\equiv G_{\tilde\zeta_+}\left( \lambda '\right)$ $\xi_+\equiv
G_{\tilde\zeta_+}\left( \lambda_+'\right)$,
$\tilde\zeta_1=\tilde\zeta\left[ \xi,\tilde\theta_1 \right]$,
$\tilde\zeta_0=\tilde\zeta\left[ \xi,\tilde\theta_0 \right]$ and
$\tilde\zeta_1'=\tilde\zeta\left[\xi_+,\tilde\theta_1 \right]$.

It is clear that (\ref{eq:proof_of_integral_path_tmp_4}) is equal to
the l.h.s. of (\ref{eq:proof_of_integral_path_tmp_1})${}_+$.  A term
in the l.h.s. of (\ref{eq:proof_of_integral_path_tmp_1})${}_+$  characterized by $\lambda'_1$ in
$\sum_{\lambda'\subseteq\lambda_+}$ which
$\{\tilde\theta_0,\tilde\theta_1\}$ belongs to is the same as a term
in the second term of (\ref{eq:proof_of_integral_path_tmp_4})
characterized by
$\lambda'_2=\lambda'_1-\{\tilde\theta_0,\tilde\theta_1\}$ in
$\sum_{\lambda'\subseteq\lambda}$. And, a term in
(\ref{eq:proof_of_integral_path_tmp_1})${}_+$ characterized by
$\lambda'_1$ which $\{\tilde\theta_0,\tilde\theta_1\}$ does not belong
to is the same as a term in the first term of
(\ref{eq:proof_of_integral_path_tmp_4}) characterized by
$\lambda'_2=\lambda'_1$.

Case 3: $N_{\tilde\theta_0}<N_{\tilde\theta_1}$\\ 
We shall modify the r.h.s. of
(\ref{eq:proof_of_integral_path_tmp_1})${}_+$.  First, we move the
integral path with respect to $x_{\{\tilde\theta_0\}}$ from $\Im
x_{\{\tilde\theta_0\}}=-N_{\tilde\theta_0}\delta$ to
$(\max(\sigma_+)+1)\delta$.  Next, we apply the relation
(\ref{eq:proof_of_integral_path_tmp_1}) in the same way that we have
done in the case 1 and 2. Finally, we move the integral path with
respect to $x_{\{\tilde\theta_0\}}$ from $\Im
x_{\{\tilde\theta_0\}}=(\max(\sigma_+)+1)\delta$ to
$\min(\sigma_1^{(0)})\delta$.  Then, the r.h.s. of
(\ref{eq:proof_of_integral_path_tmp_1})${}_+$ becomes
\begin{eqnarray}
&=&
  \sum_{
    \lambda'\subseteq \lambda,\;
    \xi\equiv G_{\tilde\zeta_+}\left(\lambda'\right)
  }
  \left[
    \prod_{\tilde\zeta'\in\xi}
    \delta\left(
      \tilde\zeta',
      \lambda\left[\lambda',\tilde\zeta'\right],
      {\min}_1\left(\tilde\zeta'\right)
    \right)
    \int_{-\infty+i\min_1\left(\tilde\zeta'\right)\delta}
        ^{+\infty+i\min_1\left(\tilde\zeta'\right)\delta}
    dx_{\tilde\zeta'}
  \right]
\nonumber\\&&
  \left[
    \prod_{
      \{\tilde\theta,\tilde\theta'\}\in\lambda_+-\lambda'
    }
    f_{N_{\tilde\theta},N_{\tilde\theta'}}\left(
      x_{\tilde\zeta\left[\xi,\theta \right]}-
      x_{\tilde\zeta\left[\xi,\theta'\right]}
    \right)
  \right]
  g\left(
    x_{
      \tilde\zeta\left[
        \xi,
        \tilde\theta_0
      \right]
    },    
    x_{
      \tilde\zeta\left[
        \xi,
        \tilde\theta_1
      \right]
    },    
    \cdots
  \right)
\nonumber\\&&{}+
    \sum_{
      \lambda'\subseteq \lambda,\;
      \lambda'_+\equiv\lambda'+\left\{\tilde\theta_0,\tilde\theta_1\right\},\;
      \xi_+\equiv G_{\tilde\zeta}\left(\lambda'_+\right),\;
      \tilde\zeta_1'\equiv\tilde\zeta\left[\xi_+,\tilde\theta_1\right]
    }
\nonumber\\&&
    \left[
      \theta\left(
        {\min}_1\left(\tilde\zeta_1'\right)
        -\min\left(\sigma_1^{(0)}\right)
        +\frac12
      \right)
      \sum_{
        \lambda''
        \subseteq
        \lambda\left[\lambda',\tilde\zeta_1'\right],\;
        \xi'\equiv G_{\tilde\zeta'_1}\left(\lambda''\right),\;
      }
  \left[
    \prod_{
      \tilde\zeta'\in\xi',\;
      }
    \delta\left(
      \tilde\zeta',
      \tilde\lambda\left[\lambda'',\tilde\zeta'\right],
      {\min}_1\left(
        \tilde\zeta'
      \right)
    \right)
  \right]
\right.
\nonumber\\&&\left.
      \left[
        \prod_{\{\tilde\theta,\tilde\theta'\}\in\lambda\left[\lambda',\tilde\zeta_1'\right]-\lambda''}
        Sgn\left[
          \tilde\zeta\left[\xi',\tilde\theta \right],
          \tilde\zeta\left[\xi',\tilde\theta'\right],
          l_{\lambda'}\left[\tilde\theta,\tilde\theta_1\right]
          -l_{\lambda'}\left[\tilde\theta',\tilde\theta_1\right]
        \right]
      \right]
\right]
\nonumber\\&&
  \left[
    \prod_{
      \tilde\zeta'
      \in
      \xi_+
      ,\;\neq\tilde\zeta_1'
    }
    \delta\left(
      \tilde\zeta',
      \lambda\left[
        \lambda'_+,
        \tilde\zeta'\right],
      {\min}_1\left(\tilde\zeta'\right)
    \right)
    \int_{-\infty+i\min_1\left(\tilde\zeta'\right)\delta}
        ^{+\infty+i\min_1\left(\tilde\zeta'\right)\delta}
    dx_{\tilde\zeta'}
  \right]
    \int_{-\infty+i\min_1\left(\tilde\zeta'_1\right)\delta}
        ^{+\infty+i\min_1\left(\tilde\zeta'_1\right)\delta}
    dx_{\tilde\zeta'_1}
\nonumber\\&&
  \left[
    \prod_{
      \{\tilde\theta,\tilde\theta'\}\in\lambda_+-\lambda'_+
    }
    f_{N_{\tilde\theta},N_{\tilde\theta'}}\left(
      x_{\tilde\zeta\left[\xi_+,\theta \right]}-
      x_{\tilde\zeta\left[\xi_+,\theta'\right]}
    \right)
  \right]
  g\left(
    x_{
      \tilde\zeta\left[
        \xi_+,
        \tilde\theta_0
      \right]
    },    
    x_{
      \tilde\zeta\left[
        \xi_+,
        \tilde\theta_1
      \right]
    },    
    \cdots
  \right),
\label{eq:proof_of_integral_path_tmp_6}
\end{eqnarray}
where $Sgn[\tilde\zeta,\tilde\zeta',1]$ is defined as
\begin{eqnarray}
  Sgn[\tilde\zeta,\tilde\zeta',1]&=&\left\{
    \begin{array}{lll}
       1&\makebox{In Case}&
       m<m'
       \quad{\rm and}\quad
       \min_{\tilde\theta\in \tilde\zeta}N_{\tilde\theta}
       >
       \min_{\tilde\theta\in \tilde\zeta'}N_{\tilde\theta}
       \\
       -1&\makebox{In Case}& m<m'
       \quad{\rm and}\quad
       \min_{\tilde\theta\in \tilde\zeta}N_{\tilde\theta}
       <
       \min_{\tilde\theta\in \tilde\zeta'}N_{\tilde\theta}
       \\
       0&\makebox{Otherwise}
    \end{array}
  \right.
\nonumber
\end{eqnarray}
\\[-2.2cm]\begin{eqnarray}
\end{eqnarray}
and $Sgn[\tilde\zeta',\tilde\zeta,-1]\equiv
Sgn[\tilde\zeta,\tilde\zeta',1]$ in case of $\theta\in\Theta_n$¡¢
$\tilde\zeta,\tilde\zeta'\subseteq\tilde{\bar\Theta}(\theta)$,
$m={\min}_1\left(\tilde\zeta\right)$ and
$m'={\min}_1\left(\tilde\zeta'\right)$.  The second term in
(\ref{eq:proof_of_integral_path_tmp_6}) is a sum of residues
characterized by sets $\lambda'_1$ and $\lambda''_1$ in
$\sum_{\lambda'\subseteq\lambda}\sum_{\lambda''\subseteq\lambda\left[\lambda',\tilde\zeta'_1\right]}$.
Each residue is generated by a term which is
characterized by
$\lambda'_2=\lambda'_1-\lambda\left[\lambda'_1,\tilde\zeta'_1\right]+\lambda''_1$
in $\sum_{\lambda'\subseteq\lambda}$ $\lambda'$ of the first
term.  We note two facts. First, we move the integral path with respect to
$\tilde\zeta_1'$ in the second term; as a result, a term in the first
term as a sum generates not one but several residues in the second
term.  Second, a condition that a pole in the first term corresponding
to a residue is in a region where the integral path passes is the same as a
condition that $\theta$-function and all the $Sgn$-function in the
residue are not zero.

Using a relation proved in \ref{sec:proof_w.r.t_set_3}, the second
term of (\ref{eq:proof_of_integral_path_tmp_6}) becomes
\begin{eqnarray}
&&
  \sum_{
    \lambda'\subseteq \lambda,\;
    \lambda'_+\equiv\lambda'+\left\{\tilde\theta_0,\tilde\theta_1\right\},\;
    \xi_+\equiv
    G_{\tilde\zeta_+}\left(\lambda'+\{\tilde\theta_0,\tilde\theta_1\}\right)
  }
  \left[
    \prod_{
      \tilde\zeta'
      \in
      \xi_+
    }
    \delta\left(
      \tilde\zeta',
      \lambda\left[
        \lambda'_+,
        \tilde\zeta'\right],
      {\min}_1\left(\tilde\zeta'\right)
    \right)
    \int_{-\infty+i\min_1\left(\tilde\zeta'\right)\delta}
        ^{+\infty+i\min_1\left(\tilde\zeta'\right)\delta}
    dx_{\tilde\zeta'}
  \right]
\nonumber\\&&
  \left[
    \prod_{
      \{\tilde\theta,\tilde\theta'\}\in\lambda_+-\lambda'_+
    }
    f_{N_{\tilde\theta},N_{\tilde\theta'}}\left(
      x_{\tilde\zeta\left[\xi_+,\tilde\theta \right]}-
      x_{\tilde\zeta\left[\xi_+,\tilde\theta'\right]}
    \right)
  \right]
  g\left(
    x_{
      \tilde\zeta\left[
        \xi_+,
        \tilde\theta_0
      \right]
    },    
    x_{
      \tilde\zeta\left[
        \xi_+,
        \tilde\theta_1
      \right]
    },    
    \cdots
  \right)
\label{eq:proof_of_integral_path_tmp_8}.
\end{eqnarray}
By the same reason as the case 2, the first term of
(\ref{eq:proof_of_integral_path_tmp_6}) plus
(\ref{eq:proof_of_integral_path_tmp_8}) equals the l.h.s. of
(\ref{eq:proof_of_integral_path_tmp_1})${}_+$.

\section{A proof with respect to a permutation 3}
\label{sec:proof_w.r.t_set_3}
We prove that
\begin{eqnarray}
&&
        \theta\left(
        {\min}_1\left(\tilde\zeta_1'\right)
        -\min\left(\sigma_1^{(0)}\right)
        +\frac12
      \right)
      \sum_{
        \lambda''
        \subseteq
        \lambda',\;
        \xi'\equiv G_{\tilde\zeta'_1}\left(\lambda''\right),\;
      }
  \left[
    \prod_{
      \tilde\zeta'\in\xi',\;
      }
    \delta\left(
      \tilde\zeta',
      \tilde\lambda\left[\lambda'',\tilde\zeta'\right],
      {\min}_1\left(
        \tilde\zeta'
      \right)
    \right)
  \right]
\nonumber\\&&
      \left[
        \prod_{
          \{\tilde\theta,\tilde\theta'\}
          \in
          \lambda'-\lambda''
        }
        Sgn\left[
          \tilde\zeta\left[\xi',\tilde\theta \right],
          \tilde\zeta\left[\xi',\tilde\theta'\right],
          l_{\lambda'}\left[\tilde\theta,\tilde\theta_1\right]
          -l_{\lambda'}\left[\tilde\theta',\tilde\theta_1\right]
        \right]
      \right]
\nonumber\\&=&
  \delta\left(\tilde\zeta'_1,\lambda'_+,{\min}_1\left(\tilde\zeta'_1\right)\right),
\label{eq:proof_w.r.t_set_3_tmp_1}
\end{eqnarray}
where $Sgn$-function is defined in \ref{sec:proof_of_integral_path}, and
$\theta\in\Theta\left(\sigma\subset\mathbb{N}\right)$,
$\tilde\zeta'_1\in\tilde{\bar\Theta}\left(\theta\right)$,
$\tilde\theta_0,\tilde\theta_1\in\tilde\zeta'_1$,
$N_{\tilde\theta_0}<N_{\tilde\theta_1}$,
$\lambda'\in\Lambda_c\left(\tilde\zeta'_1-\tilde\theta_0\right)$ and
$\lambda'_+=\lambda'+\{\tilde\theta_0,\tilde\theta_1\}$.  Note that
the assumption $\tilde\theta_0\in\tilde\zeta'_1$ leads to
$\min_1(\tilde\zeta_1)\leq\min(\sigma_1^{(0)})$.

It is evident that in the following two case, both sides of
(\ref{eq:proof_w.r.t_set_3_tmp_1}) are equal to zero. One is a case of
$\min_1(\tilde\zeta'_1)<\min(\sigma_1^{(0)})$. The other is a case
that there are two elements $\tilde\theta$, $\tilde\theta'$ which
satisfy $N_{\tilde\theta}=N_{\tilde\theta'}$ and
$\{\tilde\theta,\tilde\theta'\}\in\lambda'$.
Therefore, we shall suppose
 $\min_1(\tilde\zeta)=\min(\sigma_1^{(0)})$ and $N_{\tilde\theta}\neq N_{\tilde\theta'}$ in case of $\{\tilde\theta,\tilde\theta'\}\in\lambda'$.

Case 1: the r.h.s. of (\ref{eq:proof_w.r.t_set_3_tmp_1}) is equal to 1\\
We define a subset of $\lambda'$ as
\begin{eqnarray}
  \lambda''_1
  &\equiv&
  \left\{
    \left\{
      \tilde\theta ,
      \tilde\theta'=\left(\sigma'_1,\cdots\right),
    \right\}\in\lambda'
    \left|
    N_{\tilde\theta}<N_{\tilde\theta'},\quad
    {\min}_1\left(
      \left\{
        \tilde\theta_1,
        \sim,
        \tilde\theta
      \right\}
    \right)
    <
    \min{\sigma_1'}
  \right.\right\}
\end{eqnarray}
where
\begin{eqnarray}
  \left\{
    \tilde\theta,
    \sim,
    \tilde\theta'
  \right\}
&\equiv&
  \left\{
    \tilde\theta''\in\tilde\zeta
    \left|
    l_{\lambda}\left(\tilde\theta,\tilde\theta'\right)
    =
    l_{\lambda}\left(\tilde\theta  ,\tilde\theta''\right)+
    l_{\lambda}\left(\tilde\theta'',\tilde\theta' \right)
  \right.\right\}.
\end{eqnarray}
In the l.h.s. of (\ref{eq:proof_w.r.t_set_3_tmp_1}), any term
corresponding to $\lambda''_2$ which satisfies $\lambda''_2\subseteq
\lambda''_1$ is equal to 0, because $Sgn$-function with respect to
$\{\tilde\theta,\tilde\theta'\}\in\lambda''_1-\lambda''_2$, where
$\tilde\theta''' \not\in\{\tilde\theta_1,\sim,\tilde\theta\}$ in case
$\{\tilde\theta'',\tilde\theta'''\}\in\lambda''_1-\lambda''_2$,
$l[\tilde\theta_0,\tilde\theta]<l[\tilde\theta_0,\tilde\theta']$ and
$l[\tilde\theta_0,\tilde\theta'']<l[\tilde\theta_0,\tilde\theta''']$ ,
is equal to 0.  Also, any term corresponding to $\lambda''_2$ which
satisfies $\lambda''_2\subseteq\lambda'$ and
$(\lambda'-\lambda''_1)\cap\lambda''_2\neq\emptyset$ is equal to 0,
because $\delta$-function with respect to
$\tilde\zeta[\xi',\tilde\theta]$, where
$\{\tilde\theta,\tilde\theta'\}\in(\lambda'-\lambda''_1)\cap\lambda''_2$,
is equal to 0. Therefore, only one term, which corresponds to
$\lambda''_1$, survives.  Each function in the term satisfies
\begin{eqnarray}
    \delta\left(
      \tilde\zeta',
      \tilde\lambda\left[\lambda''_1,\tilde\zeta'\right],
      {\min}_1\left(
        \tilde\zeta'
      \right)
    \right)
&=&1
\nonumber\\
    Sgn\left[
      \tilde\zeta\left[
        \xi',
        \tilde\theta
      \right],
      \tilde\zeta\left[
        \xi',
        \tilde\theta'
      \right],
      l_{\lambda'}\left(\tilde\theta ,\tilde\theta_1\right)-
      l_{\lambda'}\left(\tilde\theta',\tilde\theta_1\right)
    \right]
&=&1
\end{eqnarray}
where ${\{\tilde\theta,\tilde\theta'\}\in
  \lambda\left[\lambda',\tilde\zeta_1'\right]-\lambda''_1}$,
$\xi'=G_{\tilde\zeta'_1}(\lambda''_1)$ and
$\tilde\zeta'\in\xi'$. Thus, the relation
(\ref{eq:proof_w.r.t_set_3_tmp_1}) holds in the case 1.

Case 2: the r.h.s. of (\ref{eq:proof_w.r.t_set_3_tmp_1}) is equal to 0\\ 
In this case, we prove that the l.h.s. of
(\ref{eq:proof_w.r.t_set_3_tmp_1}) equals to 0.  First, we introduce
an equivalence relation on a set which consists of
 subsets of $\lambda'$ corresponding to non-zero terms in the l.h.s.
of (\ref{eq:proof_w.r.t_set_3_tmp_1}).  Next, we prove that there are
two or no terms in an equivalence class.  Finally, two terms in any
equivalence classes cancel out each other.

To define the equivalence relation, we define several sets by use of
$\lambda,\tilde\zeta,\tilde\theta_1$.  First, a set $\tilde\zeta_*$
 consists of elements $\tilde\theta\in\tilde\zeta$ which satisfy
the following two conditions.  There is $\tilde\theta'\in\zeta$ which
satisfies
\begin{eqnarray}
  \left\{\tilde\theta,\tilde\theta'\right\}
\;\in\;
\lambda'
,\quad
  N_{\tilde\theta}
\;>\;
  N_{\tilde\theta'}
,\quad
  l_{\lambda'}\left(
    \tilde\theta_1,
    \tilde\theta
  \right)
  +1
\;=\;
  l_{\lambda'}\left(
    \tilde\theta_1,
    \tilde\theta'
  \right).
\end{eqnarray}
And,
\begin{eqnarray}
\left\{\tilde\theta'',\tilde\theta'''\right\}\in\lambda'
,\;
  N_{\tilde\theta''}
<
  N_{\tilde\theta'''}
,\;
  \tilde\theta'',\tilde\theta'''
\in
  \left\{
    \tilde\theta_1,
    \sim,
    \tilde\theta
  \right\}
\;\Rightarrow\;
  l_{\lambda'}\left(
    \tilde\theta_1,
    \tilde\theta''
  \right)
  +1=
  l_{\lambda'}\left(
    \tilde\theta_1,
    \tilde\theta'''
  \right).
\end{eqnarray}
  It is assured that there are elements in $\tilde\zeta_*$ by
the assumption that the r.h.s. of (\ref{eq:proof_w.r.t_set_3_tmp_1})
is equal to 0.  Next, we define $\tilde\theta_*$ as one of sets in
$\tilde\zeta_*$ which satisfy
\begin{eqnarray}
  \tilde\theta\in\tilde\zeta, \quad \tilde\theta\neq\tilde\theta_*
&\Rightarrow& \theta_*\not\in\left\{\tilde\theta_1,\sim,\tilde\theta\right\}.
\end{eqnarray}
We also define sets $\lambda_{*0}$, $\tilde\theta_{*0}$, $\lambda_{*-}$,
$\tilde\zeta_{*-}$, $\lambda_{*+}$ and $\lambda_e$
\begin{eqnarray}
  \lambda_{*0}
&\equiv&
  \left\{\
    \left\{\tilde\theta,\tilde\theta_*\right\}
    \in
    \lambda
    \left|
    \tilde\theta
    \in
    \left\{
      \tilde\theta_1,
      \sim,
      \tilde\theta_*
    \right\}
  \right.\right\}
\nonumber\\
&=&
  \left\{\left\{\tilde\theta_{*0},\tilde\theta_*\right\}\right\}
\nonumber\\
  \lambda_{*-}
&\equiv&
  \left\{
    \left\{\tilde\theta,\tilde\theta_*\right\}
    \in
    \lambda
    \left|
    N_{\tilde\theta}<N_{\tilde\theta_*},
    \tilde\theta
    \not\in
    \left\{
      \tilde\theta_1,
      \sim,
      \tilde\theta_*
    \right\}
  \right.\right\}
\nonumber\\
  \tilde\zeta_{*-}
&\equiv&
  \left\{
    \tilde\theta\in\tilde\zeta_1'
    \left|
    \left\{\tilde\theta,\tilde\theta_*\right\}\in\lambda_{*-}
  \right.\right\}
\nonumber\\
  \lambda_{*+}
&\equiv&
  \lambda\left[
    \lambda',
    \left\{
      \tilde\theta\in\tilde\zeta'_1
      \left|
      \tilde\theta_*
      \in
      \left\{
        \tilde\theta_1,
        \sim,
        \tilde\theta
      \right\},\;
      \left\{
        \tilde\theta_1,
        \sim,
        \tilde\theta
      \right\}
      \cap
      \tilde\zeta_{*-}=\emptyset
    \right.\right\}
  \right]
\nonumber\\
  \lambda_e
&=&
\lambda'-
\lambda_{*0}
-
\lambda_{*+}
-
\lambda_{*-}.
\end{eqnarray}
Note that 
$\lambda_{*0}$ contains only one  element.
We introduce an equivalence relation on  a set which consists of
subsets of  $\lambda'$ corresponding to non-zero terms in the l.h.s.
of (\ref{eq:proof_w.r.t_set_3_tmp_1})
as
\begin{eqnarray}
  \lambda''\sim\lambda'''&\Leftrightarrow&\lambda''\cap\lambda_e=\lambda'''\cap\lambda_e
\end{eqnarray}
Then, $\lambda''\cap\lambda_e$ is a character of an equivalence class.
Hereafter, we shall study in the case of a term corresponding to
$\lambda''$ in an equivalence class $\lambda''_e$. We define
$\xi''_e$ and $\tilde\theta_{*-}$ as
\begin{eqnarray}
  \xi''_e 
\equiv
  G_{\tilde\zeta_1'}\left[\lambda''_e\right]
,\quad
  \tilde\theta_{*-}=\left(\sigma_1^{(*-)},\cdots\right)
\in\tilde\zeta_{*-}
,\quad
  \min\left(\sigma_1^{(*-)}\right)
 =
  \max_{\tilde\theta\in\tilde\zeta_{*-}}\left(
    {\min}_1\left(
      \tilde\zeta\left[\xi_e'',\tilde\theta\right]
    \right)
  \right).
\end{eqnarray}

We prove that there are two or no terms in an equivalence class.

Case 2.1 $\min(\sigma_1^{(*)})<\min_1(\tilde\zeta[\xi''_e,\tilde\theta_{*0}])$\\
In this case, the assumptions that all $\delta$-functions and $Sgn$-functions in the term are non-zero request that
\begin{eqnarray}
  \lambda''\cap\left(\lambda_{*0}\cup\lambda_{*-}\right) &=&
  \left\{\left\{\tilde\theta_{*-},\tilde\theta_*\right\}\right\}
  \quad{\rm or}\quad \emptyset,
\nonumber\\
  \min\left(\sigma^{(*)}_1\right)&>&{\min}_1\left(\tilde\zeta\left[\xi''_0,\tilde\theta_{am}\right]\right)
\label{eq:proof_w.r.t_set_3_tmp_12}
\end{eqnarray}
In case 
$\lambda''\cap\left(\lambda_{*0}\cap\lambda_{*-}\right) =
\left\{\left\{\tilde\theta_{*-},\tilde\theta_*\right\}\right\} $,
a condition that the term corresponding to $\lambda''$ is non-zero restricts 
$\lambda''$ to be
\begin{eqnarray}
  \lambda''
  &=&
  \lambda''_e
  +
  \left\{\tilde\theta_{*-},\tilde\theta_*\right\}
  +
  \left\{
    \left\{
      \tilde\theta ,
      \tilde\theta'=\left(\sigma'_1,\cdots\right),
    \right\}\in\lambda_{*+}
    |
    N_{\tilde\theta}<N_{\tilde\theta'},\quad
\right.
\nonumber\\&&{}
\left.
\min\left(
    {\min}_1\left(
        \tilde\zeta\left[\xi''_e,\tilde\theta_{*-}\right],
    \right),    {\min}_1\left(
      \left\{
        \tilde\theta_{*},
        \sim,
        \tilde\theta
      \right\}
    \right)\right)
    <
    \min\left(\sigma_1'\right)
  \right\}.
\label{eq:proof_w.r.t_set_3_tmp_101}
\end{eqnarray}
In case 
$\lambda''\cap\left(\lambda_{*0}\cap\lambda_{*-}\right) =
\emptyset$,
$\lambda''$ is restricted as
\begin{eqnarray}
  \lambda''
  &=&
  \lambda''_o
  +
  \left\{
    \left\{
      \tilde\theta ,
      \tilde\theta'=\left(\sigma'_1,\cdots\right),
    \right\}\in\lambda_{a+}
    |
    N_{\tilde\theta}<N_{\tilde\theta'},\quad
\right.
\nonumber\\&&{}
\left.
    {\min}_1\left(
      \left\{
        \tilde\theta_*,
        \sim,
        \tilde\theta
      \right\}
    \right)
    <
    \min\left(\sigma_1'\right)
  \right\}.
\label{eq:proof_w.r.t_set_3_tmp_102}
\end{eqnarray}
The condition that the term corresponding to
(\ref{eq:proof_w.r.t_set_3_tmp_101}) is non-zero and the condition
(\ref{eq:proof_w.r.t_set_3_tmp_102}) is non-zero request the same
condition with respect to $\lambda''_e$

Case 2.2 $\min(\sigma_1^{(*)})>\min_1(\tilde\zeta[\xi''_e,\tilde\theta_n])$\\
In this case, the assumptions that all $\delta$-functions and $Sgn$-functions in the term are non-zero request that
\begin{eqnarray}
  \lambda''\cap\left(\lambda_{*0}\cup\lambda_{*-}\right)
&=&
\left\{\left\{\tilde\theta_{*-},\tilde\theta_*\right\}\right\}
\quad{\rm or}\quad
\left\{\left\{\tilde\theta_{*0},\tilde\theta_*\right\}\right\},
\nonumber\\
  \min\left(\sigma^{(*)}_1\right)&>&{\min}_1\left(\tilde\zeta\left[\xi''_0,\tilde\theta_{am}\right]\right).
\end{eqnarray}
In case 
$\lambda''\cap\left(\lambda_{*0}\cap\lambda_{*-}\right) =
\left\{\left\{\tilde\theta_{*-},\tilde\theta_*\right\}\right\} $,
$\lambda''$ is restricted as
\begin{eqnarray}
  \lambda''
  &=&
  \lambda''_e
  +
  \left\{\tilde\theta_{*-},\tilde\theta_*\right\}
  +
  \left\{
    \left\{
      \tilde\theta ,
      \tilde\theta'=\left(\sigma'_1,\cdots\right),
    \right\}\in\lambda_{*+}
    |
    N_{\tilde\theta}<N_{\tilde\theta'},\quad
\right.
\nonumber\\&&{}
\left.
\min\left(
    {\min}_1\left(
        \tilde\zeta\left[\xi''_e,\tilde\theta_{*-}\right],
    \right),    {\min}_1\left(
      \left\{
        \tilde\theta_{*},
        \sim,
        \tilde\theta
      \right\}
    \right)\right)
    <
    \min\left(\sigma_1'\right)
  \right\}.
\label{eq:proof_w.r.t_set_3_tmp_103}
\end{eqnarray}
In case 
$\lambda''\cap\left(\lambda_{*0}\cap\lambda_{*-}\right) =
\left\{\left\{\tilde\theta_{*+},\tilde\theta_*\right\}\right\} $,
$\lambda''$ is restricted as
\begin{eqnarray}
  \lambda''
  &=&
  \lambda''_e
  +
  \left\{\tilde\theta_{*0},\tilde\theta_*\right\}
  +
  \left\{
    \left\{
      \tilde\theta ,
      \tilde\theta'=\left(\sigma'_1,\cdots\right),
    \right\}\in\lambda_{*+}
    |
    N_{\tilde\theta}<N_{\tilde\theta'},\quad
\right.
\nonumber\\&&{}
\left.
\min\left(
    {\min}_1\left(
        \tilde\zeta\left[\xi''_e,\tilde\theta_{*0}\right],
    \right),    {\min}_1\left(
      \left\{
        \tilde\theta_{*},
        \sim,
        \tilde\theta
      \right\}
    \right)\right)
    <
    \min\left(\sigma_1'\right)
  \right\}.
\label{eq:proof_w.r.t_set_3_tmp_104}
\end{eqnarray}
The condition that the term corresponding to
(\ref{eq:proof_w.r.t_set_3_tmp_103}) is non-zero and the condition
(\ref{eq:proof_w.r.t_set_3_tmp_104}) is non-zero requests the same
condition with respect to $\lambda''_e$

Thus, we have proved that there are two or no terms in an equivalence class.

Finally, we consider about a sign of the term corresponding to
$\lambda''$.  The sign depends on only $Sgn$-functions. And, each
$Sgn$-function depends on a connection in $\lambda'-\lambda''$ when
$\lambda'$, $\tilde\zeta'_1$ and $\tilde\theta_1$ are fixed.
Moreover, a sign of $Sgn$-function corresponding to any connection in
$\lambda_{*+}$ and $\lambda_{*0}$ is positive, and a sign of
$Sgn$-function corresponding to any connection in $\lambda_{*-}$ is
negative.  Therefore, in both the case 2.1 and the case 2.2, terms
corresponding to the two $\lambda''$ classes cancel out each other.

\end{document}